\documentclass{aa}
\usepackage{graphicx}

\usepackage{subfigure}
\usepackage{txfonts}
\usepackage{amsmath}
\usepackage{amssymb}
\usepackage{multirow}
\usepackage{mathtools}
\usepackage{esint}
\usepackage{bm}
\usepackage[colorlinks=true,allcolors=blue]{hyperref}
\usepackage{float}
\newcommand{\figref}[1]{\figurename~\ref{#1}}
\usepackage{makecell}
\usepackage{tabularx}

\begin{document}

   \title{Dynamical friction in Bose-Einstein condensed self-interacting dark matter at finite temperatures, and the Fornax dwarf spheroidal}
   
   \author{S. T. H. Hartman
          \and
          H. A. Winther
          \and
          D. F. Mota
          }

   \institute{Institute of Theoretical Astrophysics, University of Oslo, PO Box 1029, Blindern 0315, Oslo, Norway
}
   \date{Received ; accepted }
   
  \abstract
   {}
   {The aim of the present work is to better understand the gravitational drag forces, also referred to as dynamical friction, acting on massive objects moving through a self-interacting Bose-Einstein condensate, also known as a superfluid, at finite temperatures. This is relevant for models of dark matter consisting of light scalar particles with weak self-interactions that require nonzero temperatures, or that have been heated inside galaxies.}
   {We derived expressions for dynamical friction using linear perturbation theory, and compared these to numerical simulations in which nonlinear effects are included. After testing the linear result, it was applied to the Fornax dwarf spheroidal galaxy, and two of its gravitationally bound globular clusters. Dwarf spheroidals are well-suited for indirectly probing properties of dark matter, and so by estimating the rate at which these globular clusters are expected to sink into their host halo due to dynamical friction, we inferred limits on the superfluid dark matter parameter space.}
   {The dynamical friction in a finite-temperature superfluid is found to behave very similarly to the zero-temperature limit, even when the thermal contributions are large. However, when a critical velocity for the superfluid flow is included, the friction force can transition from the zero-temperature value to the value in a conventional thermal fluid. Increasing the mass of the perturbing object induces a similar transition to when lowering the critical velocity. When applied to two of Fornax's globular clusters, we find that the parameter space preferred in the literature for a zero-temperature superfluid yields decay times that are in agreement with observations. However, the present work suggests that increasing the temperature, which is expected to change the preferred parameter space, may lead to very small decay times, and therefore pose a problem for finite-temperature superfluid models of dark matter.}
   {}

   \keywords{cosmology: dark matter - theory}

   \maketitle
%

\section{Introduction}
\label{sec:introduction}

When a massive object moves through a background medium, its gravitational field can cause the background to form an overdensity that trails it, and in turn exerts a gravitational force on the object that produced it. This is known as dynamical friction, and is a purely gravitational phenomenon. It can therefore also arise in systems in which the constituent components otherwise have very weak or no coupling to one another, or behave as collisionless particles, such as dark matter (DM) and stars. Many important processes in the formation of structure, the evolution of galaxies, and the dynamics of astrophysical systems, such as mergers \citep{Jiang2007,Kolchin2008}, the sinking of satellites into their host halos \citep{Colpi1999,Cowsik2009,Cole2012,Tamfal2020}, the decay of orbiting black holes and binaries \citep{Just2011,Pani2015,Dosopoulou2017,Gomez2017}, and bar--halo interactions in disk galaxies \citep{Weinberg1985,Debattista2000,Sellwood2014}, therefore depend on the nature of this drag force.

The first detailed calculation of dynamical friction was carried out by \citet{Chandrasekhar1943} in the context of stellar dynamics. Chandrasekhar considered the varying gravitational forces acting on a star as it moves through its stellar neighborhood, and found that it experiences a net average force opposite to its direction of motion, that is, a gravitational drag force. He treated the background of stars as an infinite homogeneous gas of collisionless particles following a Maxwell-Boltzmann velocity distribution, an approach that can also be used for collisionless DM \citep{Mulder1983,Colpi1999,Binney2008}. However, for a collisional medium, pressure forces must be taken into account when computing the dynamical friction, and this has been done both analytically and numerically for various types of gases, such as ideal \citep{Ostriker1999,Salcedo1999,Lee2011,Lee2014,Thun2016}, relativistic \citep{Barausse2007,Katz2019}, and magnetized gases \citep{Salcedo2012, Shadmehri2012}.

The nature of the dynamical friction due to DM is related to the nature of DM itself. The standard model of the universe, $\Lambda$CDM, includes cold and collisionless DM as the predominant matter component, making up about 80\% of all matter. While extremely successful at explaining observables such as the microwave background radiation, large-scale structure, the expansion history of the Universe, and important properties of galaxies \citep{Davis1985,Percival2001,Tegmark2004,Trujillo-Gomez2011,Vogelsberger2014,Planck2015,Riess2016}, the identity of DM has remained elusive. Furthermore, there are discrepancies between simulations of structure formation at small scales, and observations (for reviews, see e.g., \citet{Weinberg2015,DelPopolo2017,Bullock2017}). These discrepancies may have their solution within $\Lambda$CDM by including more realistic models of baryonic physics in simulations \citep{Santos-Santos2015,Sales2016,Zhu2016,Sawala2016}, but the solution may also be in alternative models of DM \citep{Hu2000,Spergel2000,Shao2013,Lovell2014,Schive2014,Elbert2015,Berezhiani2015,Khoury2016,Schwabe2016,Mocz2017,Tulin2018,Clesse2018,Boldrini2020}. For these reasons, studies have also been carried out on dynamical friction in various DM models, such as fuzzy DM \citep{Hui2017,Or2019,Lancaster2020}, and self-interacting Bose-Einstein condensed (SIBEC) DM \citep{Berezhiani2019}, also known as superfluid DM.
A number of studies have considered finite-temperature effects of interacting superfluid DM \citep{Harko2012, Slepian2012, Harko2015, Sharma2019}. Of particular note is the one presented by \citet{Berezhiani2015}, who suggested that superfluid DM, when provided with a special Lagrangian structure and coupling to the visible sector, can give rise to modified Newtonian dynamics (MOND) \citep{Milgrom1983b,Milgrom1983c,Milgrom1983a,Famaey2012} between baryons at galactic scales. This MONDian force is mediated by superfluid phonons, which cease to be coherent on scales larger than galaxies, resulting in the vanishing of the extra force and the preservation of the large-scale success of CDM. For the fifth force to be MONDian, the DM particles need exotic three-body self-interactions, and the DM fluid has to be above a certain temperature to be well-behaved. Finite-temperature DM might arise through processes inside galaxies that transfer energy to the DM halo \citep{Goerdt2010,Pontzen2012,Read2019}, possibly heating up the DM fluid.

Because the form of the dynamical friction experienced by visible matter embedded in DM halos depends on the properties of DM, observations of galaxies can be used to constrain DM. Dwarf spheroidal galaxies (dSph) are particularly well-suited for this purpose. Being poor in visible matter, their dynamical behavior is dominated by their DM component and they therefore provide a testing ground for DM models \citep{Battaglia2013,Walker2013,Strigari2018}. One such system is the Fornax dSph and its five gravitationally bound globular clusters (GCs) \citep{Mackey2003}, with a sixth one recently found to likely be a genuine, albeit dim, GC \citep{Wang2019b}. The orbital decay times of these GCs, in particular the inner two (not counting the recently discovered sixth GC), due to dynamical friction from a CDM background, have been estimated to $\tau_{\text{DF}} \lesssim 1$ Gyr \citep{Oh2000,Cole2012,Hui2017,Arca-Sedda2017}, much shorter than the supposed age of the host system, $\tau_{\text{age}} \sim 10$ Gyr \citep{delPino2013, Wang2019a}. Furthermore, there is no bright stellar nucleus at the center of Fornax dSph that would suggest the sinking of other GCs in the past. This apparent mismatch between theoretical prediction and observation suggests one of two scenarios; that we are witnessing these GCs just as they are about to fall into their host, implying a fine-tuning of their initial positions, which seems unlikely; or that there is some mechanism, or property of DM that stops the GCs from migrating towards the center of the Fornax dSph. This discrepancy between CDM estimates and observations is the so-called timing-problem, and a number of solutions have been proposed, such as massive black holes heating the system \citep{Oh2000}; assuming the CDM profile of Fornax to be cored instead of cuspy \citep{Goerdt2006,Cole2012}; inaccurate modeling of the Fornax system and the rate of the orbital decay \citep{Cowsik2009,Kaur2018,Boldrini2019,Leung2020,Meadows2020}; or some exotic property of DM \citep{Hui2017,Lancaster2020}.

In this work we extend the analysis of dynamical friction in a zero-temperature superfluid to finite temperatures, where the fluid is in a mixed state of normal fluid---made up of thermal excitations---and superfluid. This type of system has pressure terms coming from both thermal excitations and self-interactions, and can exhibit unique features due to the separate flow of the superfluid and normal fluid components. With an expression for the dynamical friction in SIBEC-DM, we estimate the time it takes for two of the GCs orbiting the Fornax dSph to sink into their host halo due to this gravitational drag, thereby inferring constraints on finite-temperature superfluid DM. The paper is organized as follows: In Section \ref{sec:superfluid_basics} the superfluid equations at both zero and finite temperatures are introduced, as well as some basic notions related to superfluidity. In Sections \ref{sec:dynfric_linpert_steady_state} and \ref{sec:dynfric_linpert_finite_time} these equations are used to derive the dynamical friction at linear order, both in a steady-state and a finite-time scenario. The dynamical friction is also found using numerical simulations of the full superfluid hydrodynamic equations in Section \ref{sec:dynfric_numerical}, which is compared to the linear result in Section \ref{sec:comparison_pert_theory_numeric}. The tools developed in the preceding sections are used in Section \ref{sec:application_fornax} to estimate the decay times of two of the GCs orbiting the Fornax dSph, and constraints on SIBEC-DM are inferred. In Section \ref{sec:Conclusion} a summary of this work and the main results are presented.
Natural units are used throughout.

\section{Hydrodynamics of finite-temperature superfluids}
\label{sec:superfluid_basics}

In the standard treatment, superfluids are often related to Bose-Einstein condensates (BEC), which form when the temperature is sufficiently low and the particle density high enough that the de Broglie wavelengths of identical bosons overlap, creating a coherent state that can be described by a single-particle wave-function. This wave-function is usually associated with the superfluid, and can therefore be regarded as a quantum mechanical effect at macroscopic scales. The wave-function $\psi$ at the mean-field level is governed by the Gross-Pitaevskii equation, a non-linear Schrödinger equation with effective contact interactions parameterized by $g$;
\begin{equation}
    i\frac{\partial \psi}{\partial t} = \Bigg[\frac{-\nabla^2}{2m} + g|\psi|^2 + mV_{\text{ext}}\Bigg]\psi.
\end{equation}
The external potential $V_{\text{ext}}$ can be a trapping potential, as is often used in cold atomic experiments, or a gravitational potential. The amplitude of $\psi$ is related to the particle number density by $n=|\psi|^2$, and mass density $\rho = m|\psi|^2$.

By inserting for the wave-function
\begin{equation}
    \psi = \sqrt{n} e^{iS} = \sqrt{\frac{\rho}{m}} e^{iS},
\end{equation}
and defining the velocity field $\bm{v} = \bm{\nabla}S/m$, the nonlinear Schrödinger equation can be reformulated in a hydrodynamic form. The real and imaginary parts of the Schrödinger equation give the set of equations
\begin{equation}
    \frac{\partial \rho}{\partial t} + \bm{\nabla}\cdot(\rho \bm{v}) = 0,
\end{equation}
\begin{equation}
    \frac{\partial \bm{v}}{\partial t} + (\bm{v}\cdot \bm{\nabla}) \bm{v} + \nabla\Bigg(\frac{g\rho}{m^2} + Q + V_{\text{ext}}\Bigg) = 0.
\end{equation}
These are the so-called Madelung equations \citep{Madelung1926}. The first is a continuity equation for mass, and the second is a quantum variant of the momentum equation, with the quantum potential 
\begin{equation}
    Q = -\frac{1}{2m^2}\frac{\nabla^2 \sqrt{\rho}}{\sqrt{\rho}},
\end{equation}
coming from the kinetic part of the Schrödinger equation that is present even in the absence of interactions. From the definition of the velocity field, we see that it is irrotational, because the curl of a gradient is zero. However, there can arise defects in the superfluid, around which the circulation is quantized as
\begin{equation}
    m\oint \bm{v}\cdot \text{d}\bm{l} = 2\pi N, \quad N\in\mathbb{Z},
\end{equation}
because the complex wave-function must be single-valued. These special structures in superfluids are called quantum vorticies.
Both the Schrödinger and Madelung formulations have been used in cosmology as models for DM in order to explain the absence of small-scale structure that is predicted in $N$-body simulations of CDM \citep{Schive2014,Mocz2017,Nori2018,Nori2020,Mina2020,Mina2020b}.

At finite temperatures, the hydrodynamic formulation of a superfluid must take into account that the fluid is no longer completely superfluid. There is a thermal cloud of excitations in addition to the coherent superfluid state that carries entropy, gives a thermal contribution to the fluid pressure, and can be viscous and rotational. To complicate matters further, as the temperature of the fluid changes, the fraction of the fluid in this thermal cloud changes as well.
This property of superfluids, to behave both as a superfluid (in the sense that we usually understand the term, as a fluid with zero viscosity, quantized circulation, and carrying no entropy) and a conventional fluid, has led to the development of a two-fluid picture of superfluids. The hydrodynamic equations for a finite-temperature superfluid are (neglecting the quantum potential) \citep{Taylor2005, Chapman2014}:

\begin{equation}
\label{eq:SF_mass_conservation}
    \frac{\partial \rho}{\partial t} + \bm{\nabla}\cdot\bm{j} = 0,
\end{equation}

\begin{equation}
\label{eq:SF_entropy_conservation}
    \frac{\partial S}{\partial t} + \bm{\nabla}\cdot(S \bm{u}_n) = 0,
\end{equation}

\begin{equation}
\label{eq:SF_velocity}
    \frac{\partial \bm{u}_s}{\partial t} + \bm{\nabla}(\mu + \frac{1}{2}\bm{u}^2_s) = -\bm{\nabla}\Phi,
\end{equation}

\begin{equation}
\label{eq:SF_momentum_conservation}
\begin{split}
    \frac{\partial \bm{j}}{\partial t} &+ \bm{\nabla}P + \rho_s(\bm{u}_s\cdot\bm{\nabla})\bm{u}_s + \rho_n(\bm{u}_n\cdot\bm{\nabla})\bm{u}_n \\
    &+ \bm{u}_s[\bm{\nabla}\cdot(\rho_s\bm{u}_s)] + \bm{u}_n[\bm{\nabla}\cdot(\rho_n\bm{u}_n)] = -\rho\bm{\nabla}\Phi.
\end{split}
\end{equation}
The thermal cloud, which we refer to as the "normal fluid", has density $\rho_n$, velocity $\bm{u}_n$, and transports both mass and thermal energy. The second component is the "superfluid", with density $\rho_s$, a velocity field $\bm{u}_s$, and carries no entropy. The total mass density is the sum of the two components, $\rho = \rho_n + \rho_s$, and likewise for momentum, $\bm{j} = \rho_n \bm{u}_n + \rho_s \bm{u}_s$. The fluid pressure is $P$, the entropy density $S$, temperature $T$, and $\mu = [P+U-ST - \frac{1}{2}\rho_{n}(\bm{u}_s-\bm{u}_n)^2]/\rho$.

As previously mentioned, superfluids and BECs are related phenomena, but it is important to stress that they are not equivalent. The formation of a BEC does not automatically imply a superfluid. To see this we must consider the co-called Landau criterion. Landau, in his seminal paper on superfluid liquid helium 4 \citep{Landau1941}, made the following argument: Assume that dissipation and heating in a fluid takes place via the creation of elementary excitations. If these excitations become energetically unfavorable and cannot spontaneously appear, then dissipation and heating ceases, and the fluid becomes superfluid. The criterion for such a condition is for the relative velocity $v$ between the superfluid and a scattering potential, such as an impurity or a container wall, to be smaller than a critical value,
\begin{equation}
\label{eq:landau_criterion}
    v < v_c = \min_{\bm{p}} \frac{\epsilon(\bm{p})}{p},
\end{equation}
where $\epsilon(\bm{p})$ is the energy of an elementary excitation with momentum $\bm{p}$ \citep{Pitaevskii2016}. This criterion shows that an ideal BEC, for which the excitation spectrum is $\epsilon(\bm{p}) = p^2/2m$, has $v_c=0$ and is therefore not a superfluid. On the other hand, a Bose gas with weak interactions has---upon the formation of a BEC---an energy spectrum that is linear at small momentum, $\epsilon(\bm{p}) = c_s p$. Hence $v_c = c_s$, and weakly interacting BECs are superfluids.

The Landau criterion is usually derived with the velocity relative to an external scatterer in mind, but it also applies to the thermal excitations that make up the normal fluid. The critical value for the relative velocity $\bm{w} = \bm{u}_s - \bm{u}_n$ of the normal fluid and superfluid is smaller than the one determined by Eq. \eqref{eq:landau_criterion}, but the difference is small at low temperatures and weak self-interactions \citep{Navez2006}.

The presence of the relative velocity $\bm{w}$, because of the partially independent motion of the superfluid and normal fluid components in a finite-temperature superfluid, has important consequences for its behavior. The superfluid part does not carry heat, while the normal fluid does, allowing mass and entropy to flow separately. This becomes clear if we define the velocity field for the mass flux, $\bm{v} = \bm{j}/\rho$, and express the equations for mass and entropy conservation in terms of $\bm{w}$ and $\bm{v}$;

\begin{equation}
\label{eq:mass_conservation_variant}
\frac{\partial \rho}{\partial t} + \bm{\nabla}\cdot (\rho \bm{v}) = 0,
\end{equation}

\begin{equation}
\label{eq:entropy_conservation_variant}
\frac{\partial S}{\partial t} + \bm{\nabla}\cdot (S \bm{v}) - \bm{\nabla}\cdot \Bigg(\frac{S\rho_s}{\rho} \bm{w} \Bigg) = 0.
\end{equation}
For a finite superfluid fraction, the entropy has an additional flux term, and therefore entropy and mass can have different flow patterns. This property is called thermal counterflow. The equation for $\partial\bm{w}/\partial t$ contains a driving term $S \bm{\nabla}T/\rho_n$, and so the counterflow $\bm{w}$ tends to be directed towards regions of higher temperature, washing out thermal differences in the superfluid. As we see below, it is this property that makes the dynamical friction in a superfluid different from a corresponding fully normal fluid (i.e., a conventional fluid, $\rho_s = 0$, with the same pressure forces).

When the Landau criterion is broken, with $w$ approaching and passing the critical velocity, the superfluid flow starts to decay as a tangle of quantum vortices form, and causes a mutual friction between the superfluid and normal fluid components \citep{Skrbek2011, Skrbek2012, Barenghi2014}. Such a dissipative effect is not present in the superfluid equations, but can be included with additional terms, as has been done in numerical studies of superfluid helium \citep{Doi2008, Darve2012, Soulaine2017}. However, to circumvent the need for extra parameters and the need to assume the functional form of the mutual friction, we instead follow the same approach used in a previous work \citep{Hartman2020}; the dissipative processes are assumed to take place instantaneously when the relative velocity $w$ exceeds the critical velocity. The velocity field $\bm{v}_s$ is changed in such a way that the fluid momentum is conserved, and that only the magnitude of $\bm{w}$ is altered, not its direction, bringing it to $w=v_c$. In other words, we assume the mutual friction to be directed along $\bm{w}$.

The critical temperature $T_c$ is a central quantity in BEC superfluids. For $T>T_c$, a gas of identical bosons is a normal fluid, but for $T<T_c$, the particles begin accumulating in the ground state, forming a BEC, which in turn can form a superfluid. In the three-dimensional, homogeneous, ideal Bose gas, this critical temperature is
\begin{equation}
    T_c = \frac{2\pi\hbar^2}{m^{5/3}}\left(\frac{\rho}{\zeta(3/2)}\right)^{2/3},
\end{equation}
where $\zeta(x)$ is the Riemann Zeta-function, and holds approximately for weakly interacting gases as well \citep{Andersen2004,Sharma2019}.

For the thermodynamic quantities of a weakly interacting Bose gas, we again follow the approach used in a previous work \citep{Hartman2020}. The equation of state is approximated by an ideal gas with contributions from two-body interactions,
\begin{equation}
    P = \frac{1}{2}\frac{g}{m^2} \rho^2 + \zeta(5/2) \left(\frac{m}{2\pi}\right)^{3/2}T^{5/2},
\end{equation}
\begin{equation}
    S = \frac{5}{2}\zeta(5/2) \left(\frac{m}{2\pi}\right)^{3/2}T^{3/2},
\end{equation}
valid only for $T<T_c$.
The fraction of particles in the condensate $f_0$ and the superfluid $f_s = \rho_s/\rho$ are both taken to be equal to the condensate fraction in the ideal case;
\begin{equation}
    f_s = f_0 = 1 - \Bigg(\frac{T}{T_c}\Bigg)^{3/2}.
\end{equation}
The critical velocity is approximated as
\begin{equation}
\label{eq:vc}
    v_c = \sqrt{\frac{gnf_0}{m}}.
\end{equation}
As long as the temperature is not too close to the transition point, and the interactions are sufficiently weak, these approximations work well.

\section{Dynamical friction from steady-state linear perturbation theory}
\label{sec:dynfric_linpert_steady_state}

The starting point for computing the dynamical friction acting on an object, or a "perturber", moving through the superfluid are Eqs. \eqref{eq:SF_mass_conservation}-\eqref{eq:SF_momentum_conservation}. The gravitational potential is sourced by both the background mass density $\rho$, and the mass distribution $\rho_{\text{pert}}$ of the  perturber:

\begin{equation}
    \label{eq_grav_pot}
    \nabla^2 \Phi = 4\pi G [\rho + \rho_{\text{pert}}].
\end{equation}
The superfluid is assumed to be homogeneous, and so the fluid variables are expanded  to linear order, $\rho = \rho_0 + \delta \rho$, $S = S_0 + \delta S$, $\bm{u}_s = \delta \bm{u}_s$, and so on.
The linear equations are

\begin{equation}
\label{eq:SF_mass_conservation_lin}
    \frac{\partial \delta\rho}{\partial t} + \bm{\nabla}\cdot\delta\bm{j} = 0,
\end{equation}

\begin{equation}
\label{eq:SF_entropy_conservation_lin}
    \frac{\partial \delta S}{\partial t} + S_0\bm{\nabla}\cdot\delta\bm{u}_n = 0,
\end{equation}

\begin{equation}
\label{eq:SF_velocity_lin}
    \frac{\partial \delta \bm{u}_s}{\partial t} + \frac{1}{\rho_0}\bm{\nabla}\delta P - \frac{S_0}{\rho_0}\bm{\nabla}\delta T = -\bm{\nabla}\delta\Phi,
\end{equation}

\begin{equation}
\label{eq:SF_momentum_conservation_lin}
    \frac{\partial \delta \bm{j}}{\partial t} + \bm{\nabla}\delta P = -\rho_0\bm{\nabla}\delta\Phi,
\end{equation}

\begin{equation}
    \delta\bm{u}_n = \frac{1}{\rho_0}\delta\bm{j} - \frac{\rho_{s0}}{\rho_0}\delta\bm{u}_s,
\end{equation}

\begin{equation}
    \label{eq_grav_pot_lin}
    \nabla^2 \delta\Phi = 4\pi G [\delta\rho + \rho_{\text{pert}}].
\end{equation}
These can be combined into two coupled equations for $\delta \rho$ and $\delta S$;

\begin{equation}
\label{eq:linear_pert_rho}
    \begin{split}
        \frac{\partial^2 \delta \rho}{\partial t^2} - \left[ \left(\frac{\partial P}{\partial \rho} \right)_0\nabla^2 + 4\pi G\rho_0 \right]\delta\rho - \left(\frac{\partial P}{\partial S} \right)_0\nabla^2\delta S = 4\pi G \rho_0 \rho_{\text{pert}},
    \end{split}
\end{equation}

\begin{equation}
\label{eq:linear_pert_s}
    \begin{split}
        \frac{\partial^2 \delta S}{\partial t^2} & - \frac{S_0}{\rho_0}\left[ \left(\frac{\partial P}{\partial \rho} \right)_0\nabla^2 + S_0\frac{\rho_{s0}}{\rho_{n0}}\left(\frac{\partial T}{\partial \rho} \right)_0\nabla^2 + 4\pi G\rho_0 \right]\delta\rho \\
        & - \frac{S_0}{\rho_0}\left[\left(\frac{\partial P}{\partial S}\right)_0\nabla^2 + S_0\frac{\rho_{s0}}{\rho_{n0}}\left(\frac{\partial T}{\partial S} \right)_0\nabla^2\right]\delta S = 4\pi G S_0 \rho_{\text{pert}}.
    \end{split}
\end{equation}
The "0" subscript indicates that the quantity is evaluated at the background level. As expected, there are scale-dependent pressure terms that inhibit the growth of mass density and entropy perturbations, but in the entropy equation there are additional effective pressure terms that further reduce entropy perturbations. These are due to thermal counterflow and depend on the superfluid fraction, vanishing in the fully normal fluid limit. It must be noted that the critical velocity $v_c$ is not included in the present approach, but the effect of this on linear theory is considered further in Section \ref{sec:dynfric_linpert_finite_time}, as well as in Section $\ref{sec:dynfric_numerical}$ using numerical simulations.

Writing $\delta\rho = \alpha\rho_0$, and Fourier transforming into momentum ($\bm{k}$) and frequency ($k_0$) space, the solutions of the $k$-modes $\alpha_k$ are found:

\begin{equation}
    \alpha_k = -4\pi G \rho_{\text{pert},k} \frac{k_0^2 - A k^2}{\left(k_0^2 - \omega_{k+}^2\right)\left(k_0^2 - \omega_{k-}^2\right)},
\end{equation}
where the dispersion relation is
\begin{equation}
    \omega_{k\pm}^2 = C_4k^2 - C_2 \pm\sqrt{C_3 k^4 - 2C_1C_2k^2 + C_2^2},
\end{equation}
and
\begin{equation}
    A = \frac{ S_0^{2}}{\rho_0} \frac{\rho_{s0}}{\rho_{n0}}\left(\frac{\partial T}{\partial S}\right)_0,
\end{equation}

\begin{equation}
    C_1 = \frac{1}{2}\left(\frac{\partial P}{\partial \rho}\right)_0 + \frac{S_0}{2\rho_0}\left(\frac{\partial P}{\partial S}\right)_0 - \frac{S_0^2}{2\rho_0}\frac{\rho_{s0}}{\rho_{n0}}\left(\frac{\partial T}{\partial S}\right)_0,
\end{equation}

\begin{equation}
    C_2 = 2\pi G \rho_0,
\end{equation}

\begin{equation}
    C_3 = C_4^2 + \frac{S_0^2}{\rho_0}\frac{\rho_{s0}}{\rho_{n0}}\left[\left(\frac{\partial P}{\partial S}\right)_0\left(\frac{\partial T}{\partial \rho}\right)_0 - \left(\frac{\partial P}{\partial \rho}\right)_0\left(\frac{\partial T}{\partial S}\right)_0\right],
\end{equation}

\begin{equation}
    C_4 = \frac{1}{2}\left(\frac{\partial P}{\partial \rho}\right)_0 + \frac{S_0}{2\rho_0}\left(\frac{\partial P}{\partial S}\right)_0 + \frac{S_0^2}{2\rho_0}\frac{\rho_{s0}}{\rho_{n0}}\left(\frac{\partial T}{\partial S}\right)_0.
\end{equation}
The dynamical friction is given by the change in the energy of the perturber,
\begin{equation}
    F_{\text{DF}} = -\frac{M}{V}\frac{\partial \Phi_{\alpha}}{\partial t},
\end{equation}
where $M$ and $V$ are the mass and velocity of the perturber, and $\Phi_{\alpha}$ is the gravitational potential of the background fluid,
\begin{equation}
    \nabla^2 \Phi_{\alpha} = 4\pi G \rho_0 \alpha.
\end{equation}
This is readily found in $k$-space,
\begin{equation}
    \Phi_{\alpha, k} = -\frac{4\pi G \rho_0 \alpha_{k}}{k^2},
\end{equation}
which can be Fourier transformed back into position-space to give the dynamical friction,
\begin{equation}
    \begin{split}
        F_{\text{DF}} &= \frac{M}{V}\frac{\partial}{\partial t}\int \frac{\text{d}k^4}{(2\pi)^4}e^{ik_0t - i\bm{k}\cdot\bm{x}}\Phi_{\alpha, k} \\
        & = -\frac{4\pi G M^2\rho_0}{V}\int \frac{\text{d}k^4}{(2\pi)^4}\frac{ik_0}{k^2}e^{ik_0t - i\bm{k}\cdot\bm{x}}\alpha_k.
    \end{split}
\end{equation}
Approximating the perturber as a point particle moving along the $z$-axis with constant velocity $V$,
\begin{equation}
    \rho_{\text{pert}}(\bm{x},t) = M\delta(x)\delta(y)\delta(z-Vt),
\end{equation}
or in $k$-space
\begin{equation}
    \rho_{\text{pert}, k} = 2\pi M\delta(k_0-Vk_z),
\end{equation}
yields the expression for the dynamical friction as
\begin{equation}
\label{eq:df_steady_integral_form}
    F_{\text{DF}} = \frac{32\pi^3 G^2 M^2 \rho}{V} \int \frac{\text{d}k^4}{(2\pi)^4}\frac{ik_0}{k^2}e^{ik_0t - i\bm{k}\cdot\bm{x}} \frac{(k_0^2 - Ak^2)\delta(k_0-Vk_z)}{(k_0^2 - \omega_{k+}^2)(k_0^2 - \omega_{k-}^2)}.
\end{equation}

Equation \eqref{eq:df_steady_integral_form} can be tackled by extending the $k_0$-integral into the complex place and closing it in the upper half plane (assuming $t>0$), meaning that contour integration can be used. The poles are pushed slightly off the real line by the prescription $\omega_{k\pm} \rightarrow \omega_{k\pm} + i\epsilon$, and only the residual of the poles inside the contour contribute to the integral. Taking the limit $\epsilon \rightarrow 0^+$ after integrating gives the dynamical friction as
\begin{equation}
\label{eq:F_k3_int_remaining}
    \begin{split}
        F_{\text{DF}} = & -\frac{16\pi^3G^2M^2 \rho_0}{V} \int \frac{\text{d}k^3}{(2\pi)^3}\frac{1}{k^2}\frac{1}{\omega_{k+}^2 - \omega_{k-}^2} \\
        & \quad\times\Bigg[ e^{i\omega_{k+}t - i\bm{k}\cdot\bm{x}}(\omega_{k+}^2 - Ak^2)\delta(\omega_{k+} - Vk_z) \\
        & \quad\quad- e^{i\omega_{k-}t  - i\bm{k}\cdot\bm{x}}(\omega_{k-}^2 - Ak^2)\delta(\omega_{k-} - Vk_z) \Bigg].
    \end{split}
\end{equation}
Spherical polar coordinates are adopted for the integral over momentum, with the polar angle $\theta$ defined as the angle relative to the direction of propagation, the $z$-axis, and the force is evaluated at the position of the perturber, $\bm{x}=Vt\hat{\bm{z}}$. The integrand is independent of the azimuthal angle, but depends on the polar angle through $k_z = k\cos\theta$. Integrating over the azimuthal angle therefore gives a factor $2\pi$, while the polar angle in combination with the $\delta$-function fixes the exponentials to one and places upper limits on the momentum, $k<k^{*,\pm}_{\text{max}}$, where $k^{*,\pm}_{\text{max}}$ satisfies $kV = \omega_{k\pm}$. Further constraints are placed on $k$: The remaining $k$-integral is bounded by the finite sizes of the perturber and the cloud it moves through, $R_{\text{max}}=R_{\text{cloud}}$ and $R_{\text{min}}=R_{\text{pert}}$, otherwise both ultraviolet (UV)- and infrared (IR) divergences may be encountered, because the perturber is modeled as a point particle, and the background fluid as infinite and uniform. We must also have $k>k^{*,\pm}_{\text{min}}$, where $k^{*,\pm}_{\text{min}}$ is the minimum momentum for which $\omega_{k\pm}$ are real. At small $k$, or, equivalently, large scales, where $\omega_{k\pm}$ become complex or imaginary, the background cloud will be gravitationally unstable and deform. We denote as a general measure the upper limits in $k$ for the two terms in Eq. \eqref{eq:F_k3_int_remaining} by $k_{\text{max}}^{\pm}$, and the lower limits by $k_{\text{min}}^{\pm}$. Inserting the expression for $\omega_{k\pm}$ and using that $C_4 - A = C_1$, the dynamical friction becomes
\begin{equation}
\label{eq:F_k_int_remaining}
    \begin{split}
        F_{\text{DF}} = & -\frac{4\pi G^2M^2 \rho_0}{V^2}\Bigg[\int_{k_{\text{min}}^{+}}^{k_{\text{max}}^{+}} \frac{\text{d}k}{2k}\Bigg(\frac{C_1 k^2 - C_2}{\sqrt{C_3 k^4 - 2C_1 C_2 k^2 + C_2^2}} + 1 \Bigg) \\
        & \quad - \int_{k_{\text{min}}^{-}}^{k_{\text{max}}^{-}} \frac{\text{d}k}{2k}\Bigg(\frac{C_1 k^2 - C_2}{\sqrt{C_3 k^4 - 2C_1 C_2 k^2 + C_2^2}} - 1 \Bigg)\Bigg].
    \end{split}
\end{equation}
There is an implicit criterion that $k_{\text{max}}^{\pm} > k_{\text{min}}^{\pm} > 0$, otherwise the integral is zero. 

Equation~\eqref{eq:F_k_int_remaining} can be solved analytically, but its expression is not particularly enlightening. Instead, we focus on a few limiting cases for which the force reduces to a simplified form; zero temperature, the fully normal fluid, small velocities, and no self-gravitation.

\subsection{Zero-temperature limit}
Taking the limit $T\rightarrow0$ (under the assumption that terms such as $S^2\rho_s/\rho_n$ go to zero as well) yields one band for the dispersion relation,
\begin{equation}
    \omega^2_k = c_{T=0}^2 k^2 - 4\pi G\rho_0,
\end{equation}
where the sound speed at zero temperature is
\begin{equation}
    c_{T=0}^2 = \left(\frac{\partial P}{\partial \rho}\right)_0.
\end{equation}
The dynamical friction becomes
\begin{equation}
\label{eq:F_simple_limit}
    F_{\text{DF}} = - \frac{4\pi G^2 M^2 \rho_0}{V^2}\ln\Bigg( \frac{k_{\text{max}}}{k_{\text{min}}} \Bigg),
\end{equation}
with
\begin{equation}
    k_{\text{max}} = \min\left(2\pi R_{\text{min}}^{-1},\,\sqrt{\frac{4\pi G\rho_0}{c_{T=0}^2 - V^2}}\right),
\end{equation}
\begin{equation}
    k_{\text{min}} = \max\left(2\pi R_{\text{max}}^{-1},\,\sqrt{\frac{4\pi G\rho_0}{c_{T=0}^2}}\right).
\end{equation}

\subsection{Normal fluid limit}
Taking the fully normal fluid limit $\rho_s\rightarrow0$ also gives one band for the dispersion relation,
\begin{equation}
    \omega^2_k = c_{n}^2 k^2 - 4\pi G\rho_0,
\end{equation}
with the sound speed in the fully normal fluid
\begin{equation}
    c_{n}^2 = \left(\frac{\partial P}{\partial \rho}\right)_0 + \frac{S_0}{\rho_0}\left(\frac{\partial P}{\partial S}\right)_0.
\end{equation}
The dynamical friction is again given by Eq. \eqref{eq:F_simple_limit}, but with
\begin{equation}
    k_{\text{max}} = \min\left(2\pi R_{\text{max}}^{-1},\,\sqrt{\frac{4\pi G\rho_0}{c_{n}^2 - V^2}}\right),
\end{equation}
\begin{equation}
    k_{\text{min}} = \max\left(2\pi R_{\text{min}}^{-1},\,\sqrt{\frac{4\pi G\rho_0}{c_{n}^2}}\right).
\end{equation}
This is the same as the zero-temperature case, but with a different sound speed.

\subsection{Small-velocity limit}
At sufficiently small velocities, $V^2 \ll C_4-\sqrt{C_3}$, assuming that the finite sizes of the background cloud and perturber do not set the integral limits in Eq. \eqref{eq:F_k_int_remaining}, the dynamical friction becomes
\begin{equation}
    F_{\text{DF}} = -\frac{2\pi G^2 M^2 \rho_0}{c_{T=0}^2}.
\end{equation}
This is equal to the friction force at $T=0$ in the same limit, as opposed to when $\rho_s = 0$;
\begin{equation}
    F_{\text{DF}} = -\frac{2\pi G^2 M^2 \rho_0}{c_{n}^2}.
\end{equation}
The dynamical friction of a superfluid therefore approaches the zero-temperature limit even when there is a significant thermal contribution. This happens because counterflow in the superfluid conspires against thermal perturbations, allowing the mass over-density to behave similarly to a zero-temperature fluid. With only the interaction pressure that is present at zero temperature effectively damping density perturbations, the density contrast of the superfluid can grow larger (compared to a normal fluid at the same temperature) and hence produce a stronger net gravitational force acting on the perturber. However, we recall that this result does not include the effect of the critical velocity which would limit this thermal counterflow. In Section \ref{sec:dynfric_linpert_finite_time} we propose a scheme to include the critical velocity in linear perturbation theory, and then test the scheme using hydrodynamic simulations in Section \ref{sec:comparison_pert_theory_numeric}.

\subsection{Neglecting self-gravitation}
The numerical results presented in Section \ref{sec:comparison_pert_theory_numeric}, as well as the decay times of globular clusters in Section \ref{sec:application_fornax}, are obtained when self-gravitation is neglected. It is therefore of interest to see what the steady-state linear theory predicts in this case as well.

Neglecting self-gravitation amounts to setting $C_2 = 0$. The dispersion relation becomes
\begin{equation}
    \label{eq:dispersion_no_self-gravity}
    \omega_{k\pm}^2 = (C_4 \pm \sqrt{C_3})k^2 = c_{\pm}^2k^2.
\end{equation}
For the equation of state used throughout this work, and $T/T_c \lesssim 0.2$, the above superfluid sound speeds can be accurately approximated by
\begin{equation}
    \label{eq:cs_p}
    c_{+} = \sqrt{\frac{c_{n}^2 - c_{T=0}^2}{f_n}},
\end{equation}
\begin{equation}
    \label{eq:cs_m}
    c_{-} = c_{T=0}.
\end{equation}
We note that for $c_{n}\gg c_{T=0}$, we have $c_{+} \approx c_{n}/\sqrt{f_n} \gg c_n$. The dynamical friction takes the form
\begin{equation}
\begin{split}
    \label{eq:df_steady_state_no_self_grav}
    F_{\text{DF}} = &- \frac{4\pi G^2M^2 \rho_0}{V^2}  \ln\Bigg(\frac{R_{\text{max}}}{R_{\text{min}}}\Bigg) \\
    & \times \frac{1}{2}\Bigg[\Bigg(1  - \frac{C_1}{\sqrt{C_3}}\Bigg)\theta(V - c_{-})
    + \Bigg(1  + \frac{C_1}{\sqrt{C_3}}\Bigg)\theta(V - c_{+})\Bigg]
\end{split}
.\end{equation}
One feature that is clear in this limit is that $F_{\text{DF}}$ jumps from zero as $V$ becomes larger than $c_{-}$, and jumps again as it crosses $c_{+}$. It seems odd that the force should change value so dramatically when the velocity of the perturber crosses these thresholds, and indeed we find in the numerical simulations in Section \ref{sec:dynfric_numerical} that it does not. The problem is that in the steady-state case, as considered in this section, the linear over-density is symmetric upstream and downstream when the perturber moves at subsonic speeds, resulting in a zero net gravitational force at the position of the perturber. This is not an issue at supersonic speeds because the perturber moves faster than the background fluid can respond to the perturbation, which is at the speed of sound, resulting in a clear cone trailing the perturber \citep{Ostriker1999}. At subsonic speeds, on the other hand, the fluid reacts faster than the perturber moves, and with an infinite amount of time to propagate this response, the first-order perturbation of the background becomes symmetric. In order to overcome this shortcoming of steady-state linear perturbation theory, other studies have broken this symmetry by switching on the perturber for a finite time \citep{Ostriker1999,Salcedo2012}, or by going to second-order perturbations \citep{Lee2011, Shadmehri2012}. In the following section, the finite-time approach is employed for a superfluid.

\section{Dynamical friction from finite-time linear perturbation theory}
\label{sec:dynfric_linpert_finite_time}

For the finite-time calculation, Eqs. \eqref{eq:linear_pert_rho} and \eqref{eq:linear_pert_s} are used without self-gravitation, and an approach very similar to the one used by \citet{Ostriker1999} is followed.

The equations can be written in matrix form as

\begin{equation}
    \frac{\partial^2Y}{\partial t^2} + A\nabla^2 Y = F \rho_{\text{pert}},
\end{equation}
where
\begin{equation}
    Y = 
    \begin{pmatrix}
    \delta\rho\\
    \delta S
    \end{pmatrix},
\end{equation}
\begin{equation}
    A = 
    \begin{pmatrix}
    \left(\frac{\partial P}{\partial \rho}\right)_0
    & \left(\frac{\partial P}{\partial S}\right)_0 \\
    \frac{S_0}{\rho_0}\left(\frac{\partial P}{\partial \rho}\right)_0 + \frac{S_0^2}{\rho_0}\frac{\rho_{s0}}{\rho_{n0}}\left(\frac{\partial T}{\partial \rho}\right)_0 
    & \frac{S_0}{\rho_0}\left(\frac{\partial P}{\partial S}\right)_0 + \frac{S_0^2}{\rho_0}\frac{\rho_{s0}}{\rho_{n0}}\left(\frac{\partial T}{\partial S}\right)_0 
    \end{pmatrix},
\end{equation}
and
\begin{equation}
    F = 
    \begin{pmatrix}
    4\pi G \rho_0\\
    4\pi G S_0
    \end{pmatrix}.
\end{equation}
By diagonalizing matrix $A$, the coupled set of equations can be transformed into two decoupled wave equations of the form
\begin{equation}
    \frac{\partial^2 \chi_{i}}{\partial t^2} - c_{i}^2 \nabla^2 \chi_{i} = f_{i},
\end{equation}
which are solved using the retarded Green's function for the wave equation in three dimensions:
\begin{equation}
    \chi_{i} (\bm{x}, t) = \int \text{d}^3x' \int \text{d}t' \,\frac{\delta(t' - (t - |\bm{x}-\bm{x}'|/c_{i})) f_{i}(\bm{x}', t')}{4\pi c_{i}^2 |\bm{x}-\bm{x}'|}.
\end{equation}
For a point source switched on at the origin at $t=0$ and moving at speed $\bm{V}=V\hat{\bm{z}}$,
\begin{equation}
    f_{i}(\bm{x},t) = K_{i}\delta(x)\delta(y)\delta(z-Vt)H(t),
\end{equation}
where $H(x)$ is the Heaviside function, the solution of $\chi$ becomes, upon defining $s = z-Vt$, $\mathcal{M}_{i}=V/c_{i}$, and $R^2 = x^2 + y^2$,
\begin{equation}
    \chi_{i}(\bm{x},t) = \frac{K_{i}}{4\pi c_{i}^2 \sqrt{s^2 + R^2(1-\mathcal{M}_{i}^2)}}\mathcal{H}_{i},
\end{equation}
\begin{equation}
    \mathcal{H}_{i} =
    \begin{cases}
        1 & \text{for } R^2 + z^2 < (c_{i}t)^2,\\
        \multirow{2}{*}{2} & \text{for } \mathcal{M}_{i}>1, R^2 + z^2 > (c_{i}t)^2,\\
                            & s/R < - \sqrt{\mathcal{M}_{i}^2-1}, \text{and } z > c_{i}t/\mathcal{M}_{i},\\
        0 & \text{otherwise}.
        \end{cases}
\end{equation}
The resulting overdensity $\delta\rho$ is a weighted sum of $\chi_{+}$ and $\chi_{-}$, and the dynamical friction is obtained by integrating the gravitational force due to the overdensity over the
whole volume, that is,
\begin{equation}
    F_{\text{DF}} = 2\pi GM \int\text{d}s\int\text{d}R\, \frac{sR \delta\rho }{(s^2 + R^2)^{3/2}}.
\end{equation}
In spherical polar coordinates, $s = r\cos\theta = rx$ and $R = r\sin\theta = r\sqrt{1-x^2}$, we get
\begin{equation}
    \label{eq:df_finite_time}
    F_{\text{DF}} = -\frac{4\pi G^2 M^2\rho_0}{V^2}(\mathcal{I}_{+} + \mathcal{I}_{-}),
\end{equation}
\begin{equation}
    \mathcal{I}_{i} = -D_i\int_{R_{\text{min}}}^{R_{\text{max}}}\frac{\text{d}r}{2r}\int_{-1}^{1}\text{d}x\, \frac{x\mathcal{M}^2_{i}\mathcal{H}_i}{\sqrt{1 - \mathcal{M}_i^2 + x^2 \mathcal{M}_i^2}},
\end{equation}
where we have again introduced an upper and lower cutoff of scales in the integral to avoid UV- and IR divergences. The sound speeds $c_{+}$ and $c_{-}$ are the same as the ones given in Eq. \eqref{eq:dispersion_no_self-gravity}, and
\begin{equation}
    D_{+} = -\frac{S_0 \left(\frac{\partial P}{\partial S}\right)_0 \Big[S_0\frac{\rho_{s0}}{\rho_{n0}}\left(\frac{\partial T}{\partial \rho}\right)_0 + c_{+}^2 \Big]}
    {\rho_0(c_{+}^2 - c_{-}^2)\Big[\left(\frac{\partial P}{\partial \rho}\right)_0 - c_{+}^2\Big]},
\end{equation}
\begin{equation}
    D_{-} = \frac{S_0 \left(\frac{\partial P}{\partial S}\right)_0 \Big[S_0\frac{\rho_{s0}}{\rho_{n0}}\left(\frac{\partial T}{\partial \rho}\right)_0 + c_{-}^2 \Big]} 
    {\rho_0(c_{+}^2 - c_{-}^2)\Big[\left(\frac{\partial P}{\partial \rho}\right)_0 - c_{-}^2\Big]}.
\end{equation}

\begin{figure}[]
    \centering
    \subfigure[$t=0.1R_{\text{max}}/V$]{\includegraphics[width=0.98\linewidth]{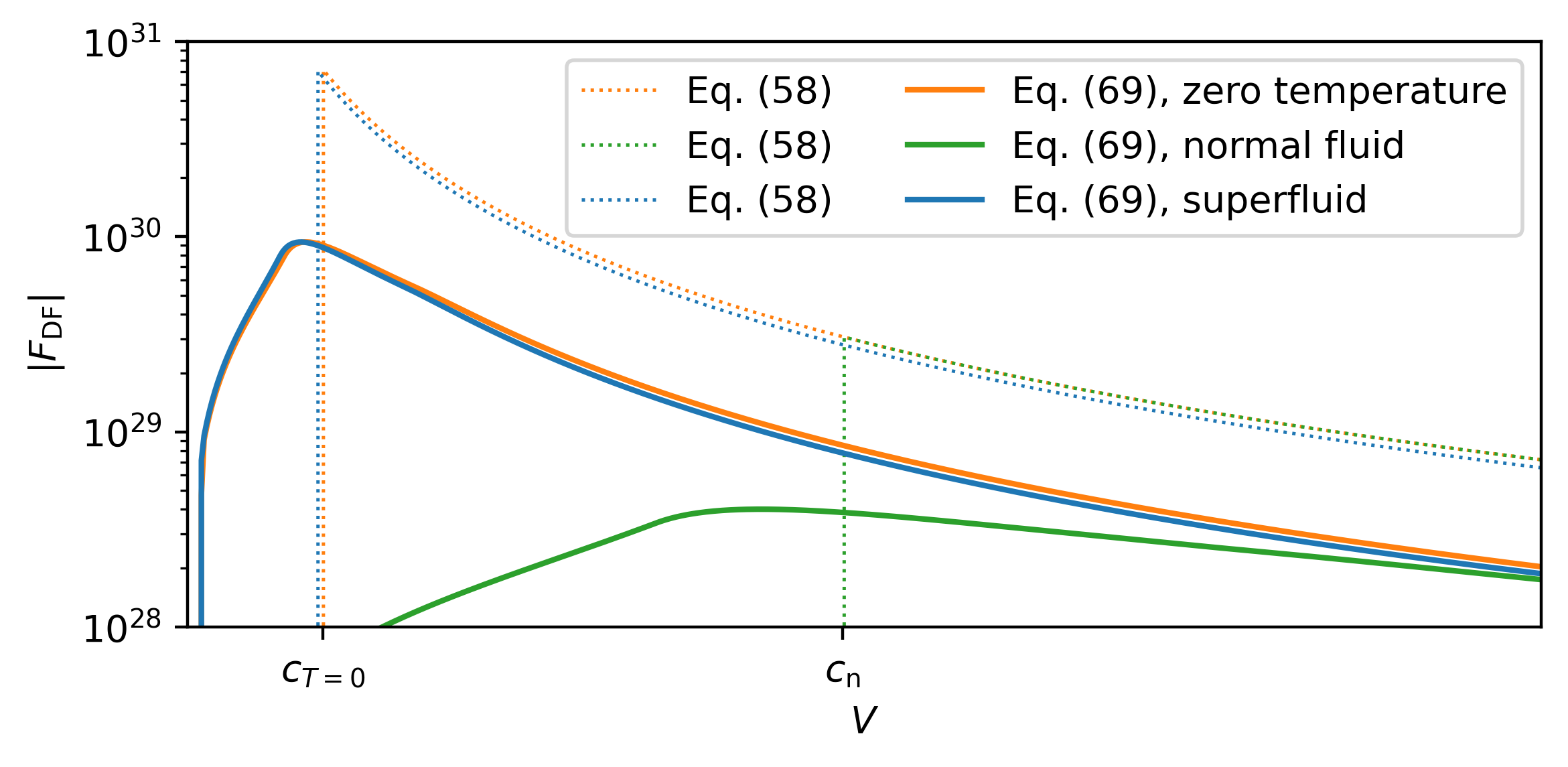}}
    \subfigure[$t=R_{\text{max}}/V$]{\includegraphics[width=0.98\linewidth]{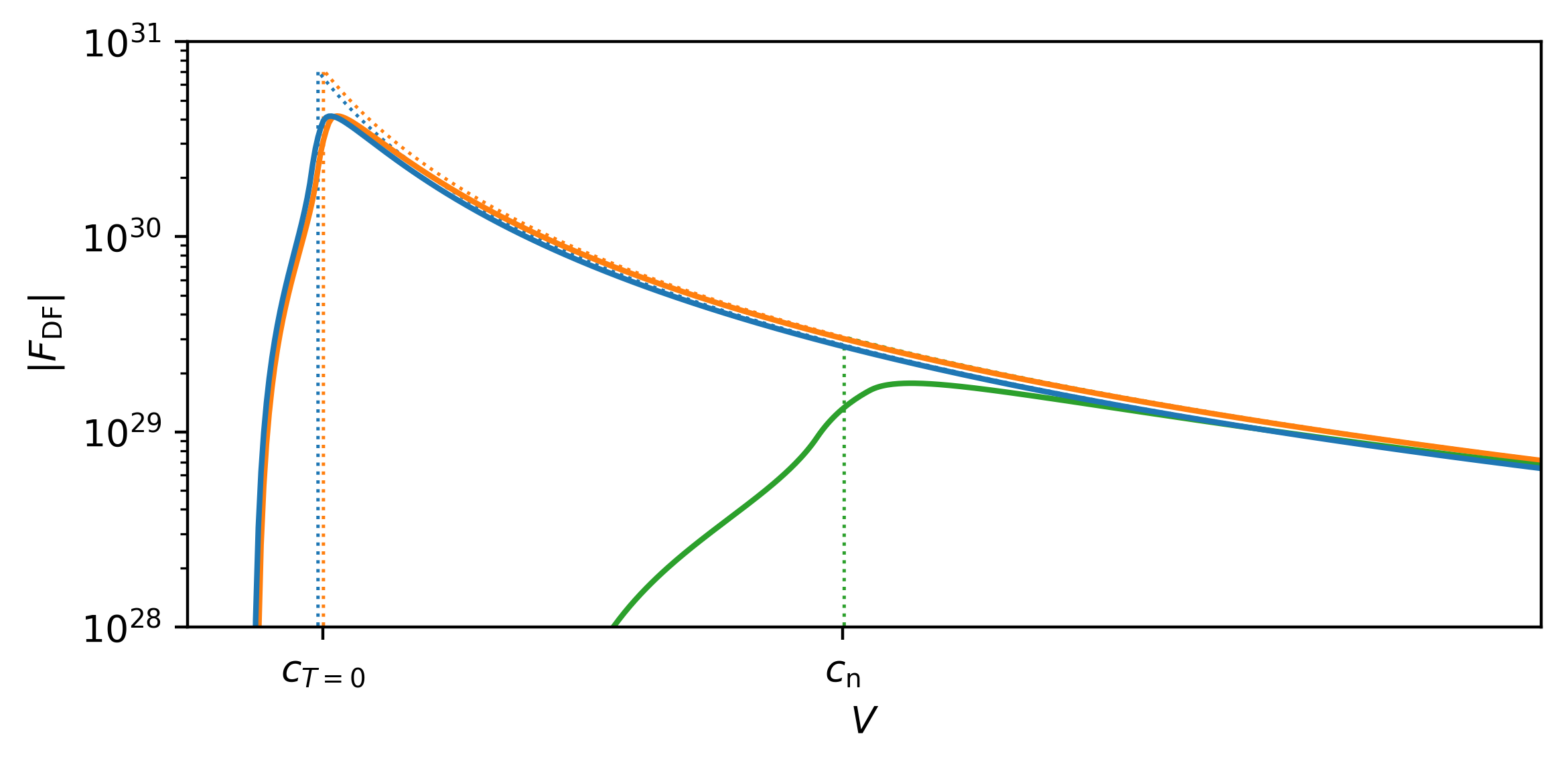}}
    \subfigure[$t=10R_{\text{max}}/V$]{\includegraphics[width=0.98\linewidth]{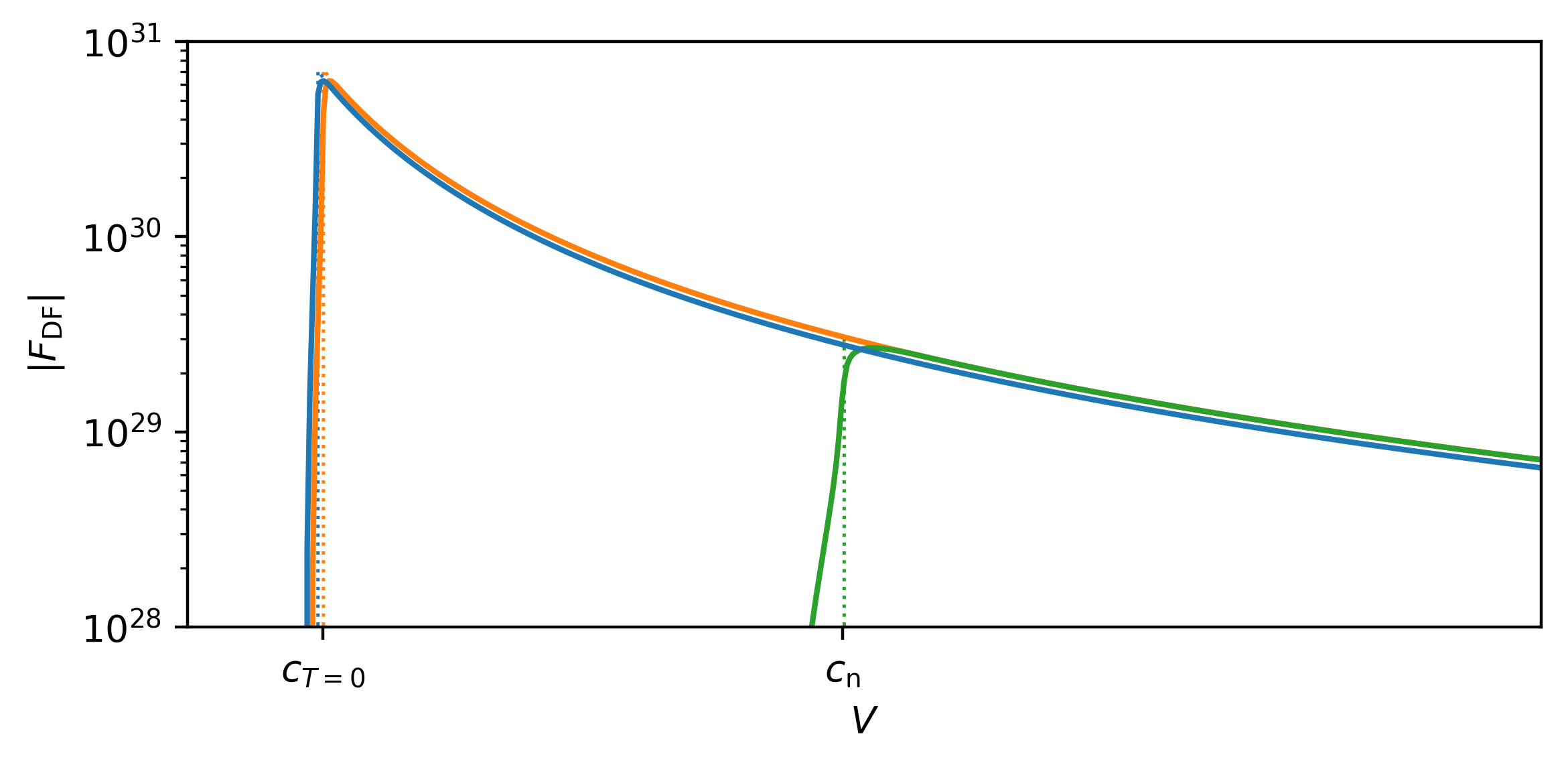}}
    \caption{Dynamical friction from linear perturbation theory using the finite-time approach (solid lines) and the steady-state approach (dotted lines) as a function of $V$. As time passes, the finite-time result, Eq. \eqref{eq:df_finite_time}, approaches the steady-state result, Eq. \eqref{eq:df_steady_state_no_self_grav}. In the zero-temperature limit, we have $T=0$, while in the normal fluid case we have $\rho_s=0$.}
\label{fig:F_DF_steady_state_vs_finite_time}
\end{figure}

The dynamical friction from the finite-time calculation is compared to the steady-state result in \figref{fig:F_DF_steady_state_vs_finite_time}. The discontinuities have been removed, with the force increasing with velocity $V$ until it reaches a maximum near the sound speed, after which the perturber becomes supersonic and the friction force decreases with the same $1/V^2$ dependence as in the steady-state result. As time passes, the finite-time result approaches the steady-state result, as expected.

Both approaches predict $F_{\text{DF}}$ in the superfluid phase to be very close to the zero-temperature value, even when thermal pressure dominates over the contribution from self-interactions. However, we must recall again that the Landau criterion is not included in linear perturbation theory, which will limit the thermal counterflow of the superfluid, making it behave more like a normal fluid, thus decreasing the dynamical friction as thermal pressure forces inhibit the growth of density perturbations. Let us therefore construct an \textit{ad hoc} scheme to include the critical velocity in the linear theory.

The critical velocity is expected to have an effect when the relative velocity is of the order of the critical velocity and larger. Therefore,  let us consider the linearized equation for the relative velocity,
\begin{equation}
    \frac{\partial \bm{w}}{\partial t} = \frac{S_0}{\rho_{n0}}\bm{\nabla}\delta T = \frac{S_0}{\rho_{n0}} \left[ \left(\frac{\partial T}{\partial S}\right)_0\bm{\nabla}\delta S + \left(\frac{\partial T}{\partial \rho}\right)_0 \bm{\nabla}\delta\rho \right].
\label{eq:linear_counterflow_evolution}
\end{equation}
The amplitude of $\delta\rho$ and $\delta S$, and hence $\delta T$, increases with $M$, driving $w$ up to the critical value faster, causing the effect of the critical velocity on the system to be more prominent. Increasing $M$ should therefore have a similar effect as lowering $v_c$ in transitioning the superfluid dynamical friction from the $T=0$ value to the fully normal fluid value.

We now assume that for an estimate of the characteristic counterflow $\bar{w}$ of the system, there is an interpolating function $f(\bar{w},v_c)$ with $f(\bar{w}\ll v_c)\rightarrow 1$, $f(\bar{w}\gg v_c)\rightarrow 0$, and a transitional region around $\bar{w}\sim v_c$, such that
\begin{equation}
\label{eq:df_interpol_vc}
    F_{\text{DF}} = f(\bar{w},v_c) F_{\text{DF}}^{\text{sf}} + [1-f(\bar{w},v_c)] F_{\text{DF}}^{\text{nf}},
\end{equation}
where $F_{\text{DF}}^{\text{sf}}$ and $F_{\text{DF}}^{\text{nf}}$ is the dynamical friction from linear theory for the superfluid and fully normal fluid, respectively. Using Eq. \eqref{eq:linear_counterflow_evolution} we can write
\begin{equation}
    \bar{w} = \frac{S_0}{\rho_{n0}} \frac{\delta T(0)}{L} \Delta t = \frac{S_0}{\rho_{n0}} \left[ \left(\frac{\partial T}{\partial S}\right)_0\frac{\delta S(0)}{L} + \left(\frac{\partial T}{\partial \rho}\right)_0 \frac{\delta \rho(0)}{L} \right] \Delta t.
\end{equation}
The length $L$ and time $\Delta t$ are characteristic scales over which the fluid attains the mass and entropy overdensity at the origin, $\delta\rho(0)$ and $\delta S(0)$. The timescale can be estimated as $\Delta t = L/v$, where $v$ is some characteristic velocity in the problem. The largest superfluid sound speed, $c_{+}$, which is essentially the fastest speed with which the superfluid can respond to disturbances, was found to work.

For the $\delta$-function perturbation, the central values for the mass and entropy overdensities diverge in the linear theory, meaning that $\delta\rho(0)$ and $\delta S(0)$ are not well-defined. Instead, these should be evaluated at some point near the origin, as was done for dynamical friction. With the equation of state used in this work, an estimate of the linear entropy contrast at $R_{\text{min}}/2$ is
\begin{equation}
    \delta S(R_{\text{min}}/2) \approx \frac{2 S_0 GM}{c^2_{+} R_{\text{min}}}.
\end{equation}
The rough estimate of the counterflow is therefore
\begin{equation}
    \bar{w} = \frac{S_0^2}{\rho_{n0}} \left(\frac{\partial T}{\partial S}\right)_0 \frac{2GM}{c_{+}^3 R_{\text{min}}}.
\end{equation}
Only the form of the interpolating function $f(\bar{w},v_c)$ remains to be specified. The simple but rather arbitrary choice,
\begin{equation}
\label{eq:vc_interpolation_form}
    f(\bar{w},v_c) = \frac{v_c}{v_c + \bar{w}} = \left[1 + \frac{S_0^2}{\rho_{n0}} \left(\frac{\partial T}{\partial S}\right)_0 \frac{2G}{c_{+}^3 R_{\text{min}}} \frac{M}{v_c} \right]^{-1}
,\end{equation}
was found to work well, as we see in Section \ref{sec:comparison_pert_theory_numeric}.

\section{Numerical simulation of dynamical friction}
\label{sec:dynfric_numerical}

To test the calculations from linear perturbation theory, the full superfluid equations are integrated numerically.
We use the frame comoving with the perturber, meaning that its gravitational field is static and centered at the origin, while the background fluid is moving. We take the perturber to be a sphere with uniform mass density $\rho_{\text{pert}} = 3M/4\pi R_{\text{pert}}^3$. The system has rotational symmetry, and so  cylindrical coordinates are employed; the axial distance is $z$, the distance along the axis of rotational symmetry, and the radial distance is $r$, the distance from the axis. The simulation volume is therefore a cylinder, and we take its domain to be $-L<z<L$ and $0<r<L$.

The superfluid is initialized as a uniform fluid moving with velocity $\bm{V} = -V\hat{z}$. More fluid is injected into the simulation volume with the same velocity at the $z = +L$ boundary. The $z = -L$ and $r = L$ boundaries are taken to have zero gradients, while the inner boundary $r=0$ has a reflecting boundary condition.

To numerically integrate the superfluid equations, a Godunov scheme similar to the one described in \citet{Hartman2020} is used. In the present work, the generation of entropy when the Landau criterion is broken is not included. Also, as we evolve the entropy instead of the energy, and we do not include any viscosity, the numerical scheme dissipates kinetic energy at shock fronts that is not converted into internal energy. In the absence of this shock heating, the total energy is not strictly conserved. Nevertheless, we have found that this inaccuracy is by and large negligible for the scenarios we consider here because the solutions are mostly in or near the linear regime.

The self-gravitation of the superfluid is neglected. The gravitational field it produces is computed only to find the resulting force on the perturber, that is, the dynamical friction. The initial fluid parameters are $\rho = 2\times 10^{7} M_\odot \text{kpc}^{-3}$, $T=0.2T_c$, $m=500\text{eV}$, and $g=2\times 10^{-3} \text{eV}^{-2}$. These parameters are chosen only to illustrate the basic features of dynamical friction in superfluids while keeping the simulation run-times reasonably short. Unless stated otherwise, the size of the perturber is taken to be $R_{\text{pert}} = 2\text{pc}$ with mass $M = 0.1M_\odot$, while the simulation size is $L=150\text{pc}$. The simulation is run until $t=10\text{pc}/V$, that is, until the background has moved $10 \text{pc}$. This is small compared to the full simulated length, but is necessary for preventing boundary effects from interfering with the results. $R_{\text{min}}$ is taken to be the size of the perturber, $R_{\text{min}} = 2 \text{pc}$, and $R_{\text{max}}$ the radius of the cylindrical simulation volume, $R_{\text{max}} = 150 \text{pc}$, when compared to linear perturbation theory. The resolution of the simulated volume is $4096\times2048$ cells, in the $z$ and $r$ directions, respectively, for the superfluid case. In the zero-temperature and normal fluid limits, for which the numerical scheme was made second-order in time and space using a MinMod slope-limited MUSCL-Hancock scheme \citep{Toro2006} without stability issues, a lower resolution of $2048\times1024$ is used.

An effective critical velocity $v_c^{\text{eff}}$, which is just $v_c$ multiplied by some factor, is used in the simulations to show the effect of varying $v_c$ without actually changing other parameters such as the particle mass and self-interaction.

\section{Comparison of perturbation theory and numerical simulation}
\label{sec:comparison_pert_theory_numeric}

\begin{figure}[]
    \centering
    \includegraphics[width=0.98\linewidth]{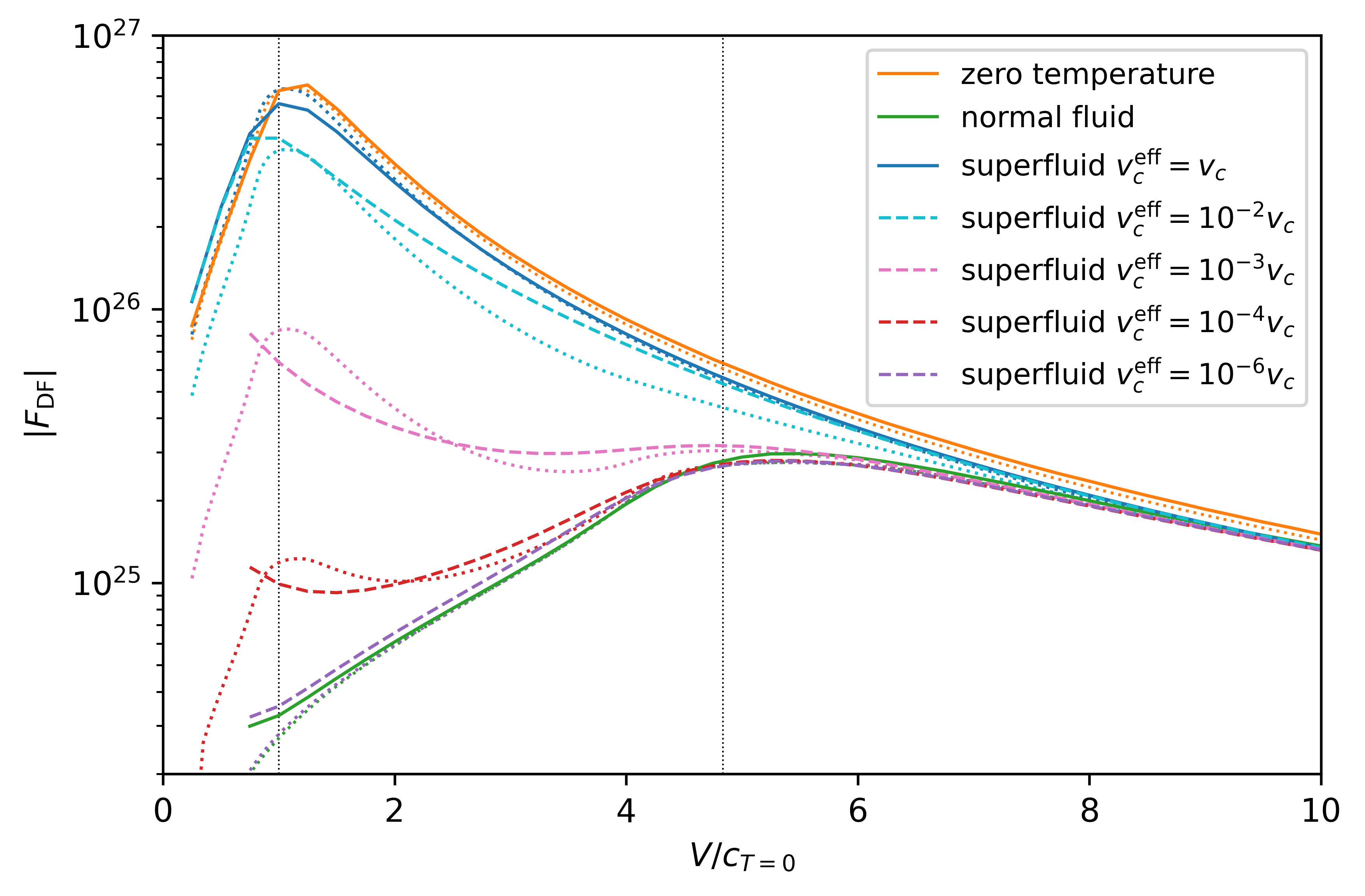}
    \caption{Dynamical friction against velocity for the superfluid with varying $v_c^{\text{eff}}$, and for the zero temperature and the fully normal fluid limits. The results from finite-time linear perturbation theory are shown with dotted lines of the same colors. Even with the critical velocity included, the superfluid case gives a dynamical friction force of the same magnitude as the zero temperature limit. When $v_c^{\text{eff}}$ is decreased, the superfluid approaches the fully normal fluid limit as thermal counterflow is increasingly limited. The sound speeds $c_{T=0}$ and $c_{n}$ are indicated by the vertical dotted lines.}
\label{fig:F_DF_vs_v_numerical}
\end{figure}

\begin{figure}[]
    \centering
    \includegraphics[width=0.98\linewidth]{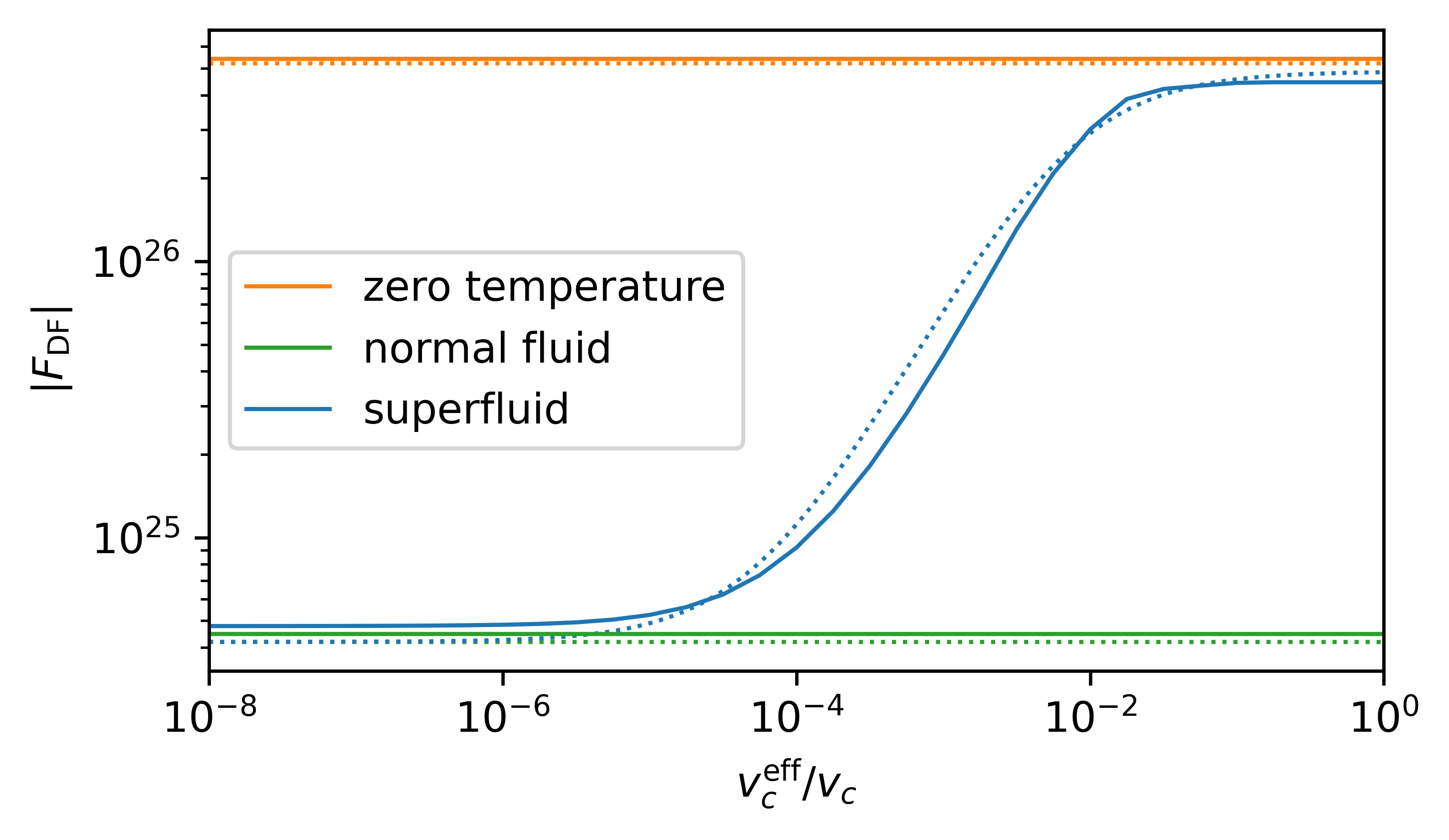}
    \caption{Dynamical friction against the effective critical velocity $v_c^{\text{eff}}$, for $V = 1.5c_{T=0}$, with the results from finite-time linear perturbation theory included with dotted lines of the same colors. As $v_c^{\text{eff}}$ is lowered, the dynamical friction goes from about the same value as the zero temperature limit to the value in the fully normal fluid limit, changing by about two orders of magnitude.}
\label{fig:F_DF_vs_vceff}
\end{figure}

\begin{figure}[]
    \centering
    \includegraphics[width=0.98\linewidth]{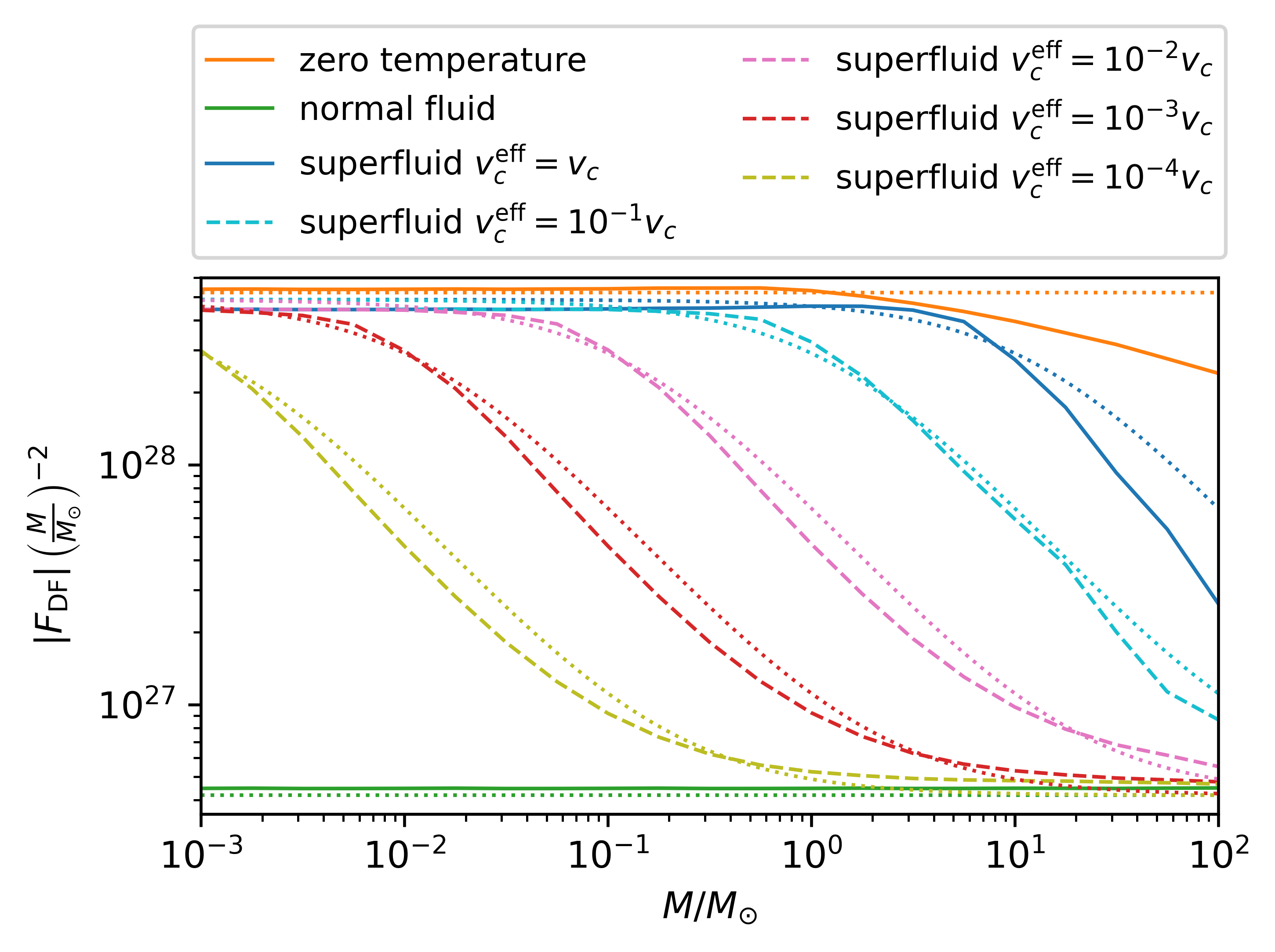}
    \caption{Dynamical friction against the perturber mass $M$, for $V = 1.5c_{T=0}$, with the results from finite-time linear perturbation theory included with dotted lines of the same colors. The departure from perturbation theory for the zero-temperature case at high $M$ is due to nonlinear effects. Increasing the mass of the perturber causes the superfluid to behave increasingly like a normal fluid, similarly to the effect of decreasing $v_c
   ^{\text{eff}}$ .}
\label{fig:F_DF_vs_M}
\end{figure}

\begin{figure*}
    \centering
    \subfigure[$T=0$ limit]{\includegraphics[width=0.35\linewidth]{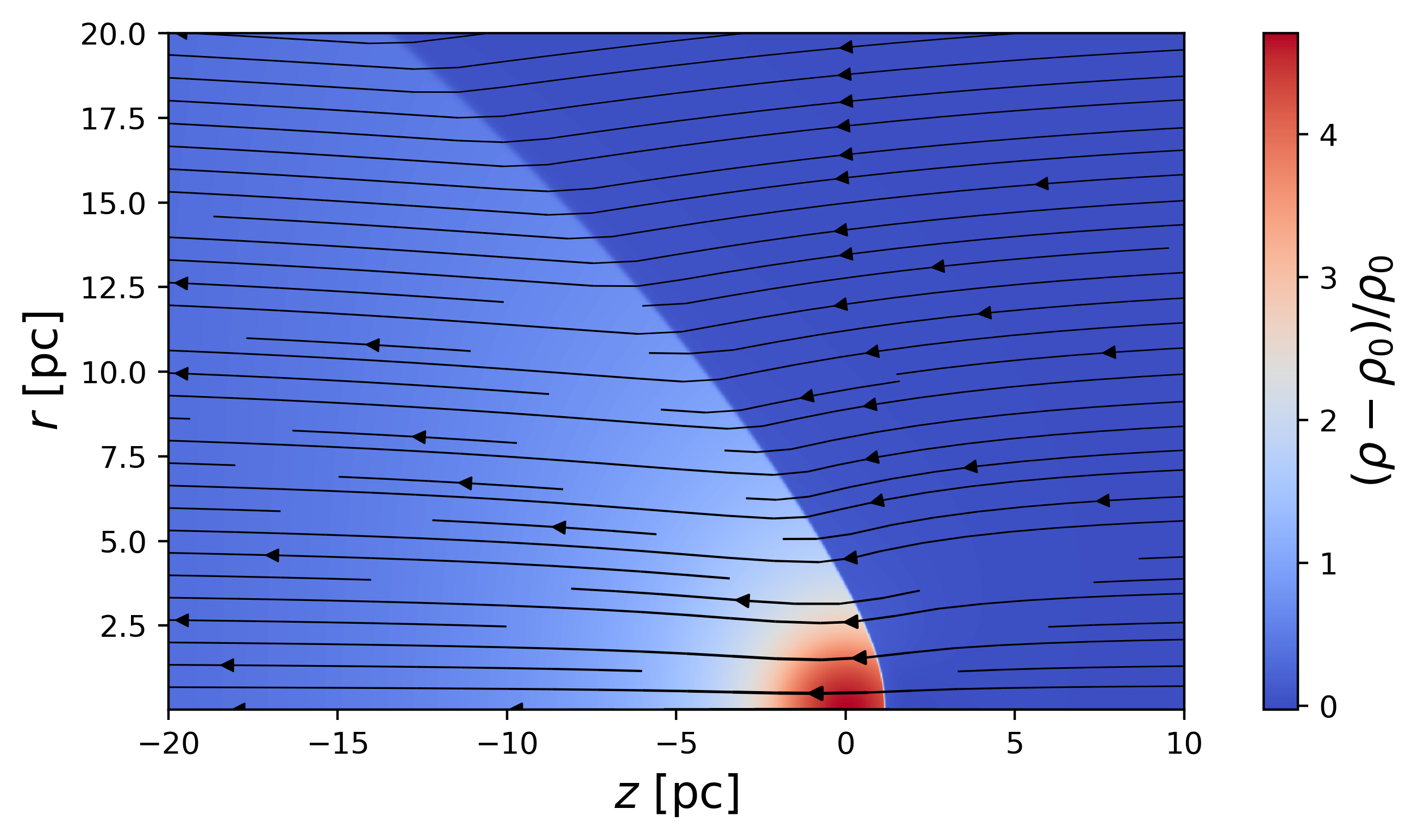}}
    \subfigure[$\rho_s=0$ limit]{\includegraphics[width=0.35\linewidth]{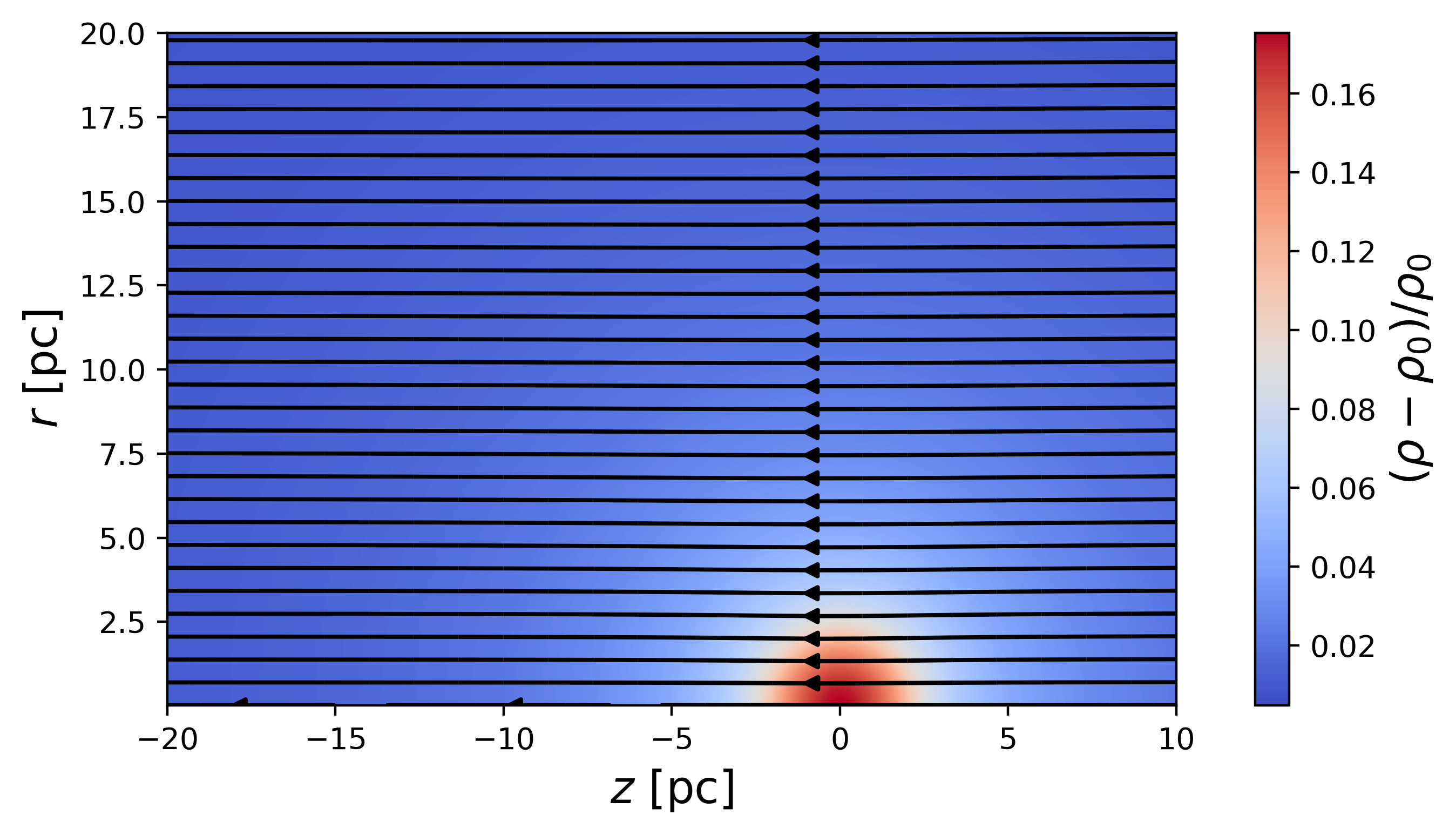}}
    \subfigure[$v_c^{\text{eff}} = v_c$]{\includegraphics[width=0.32\linewidth]{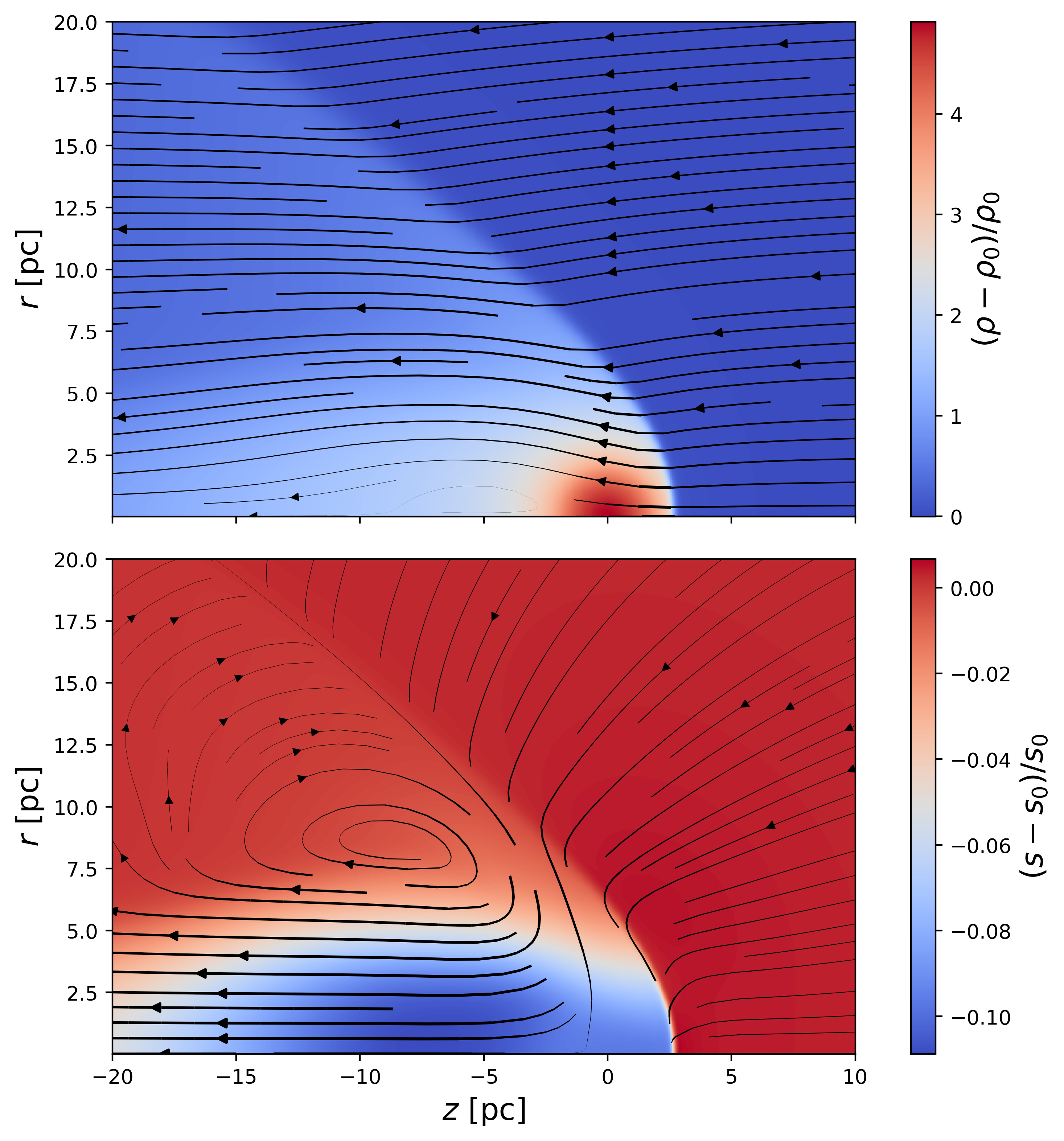}}
    \subfigure[$v_c^{\text{eff}} = 10^{-1}v_c$]{\includegraphics[width=0.32\linewidth]{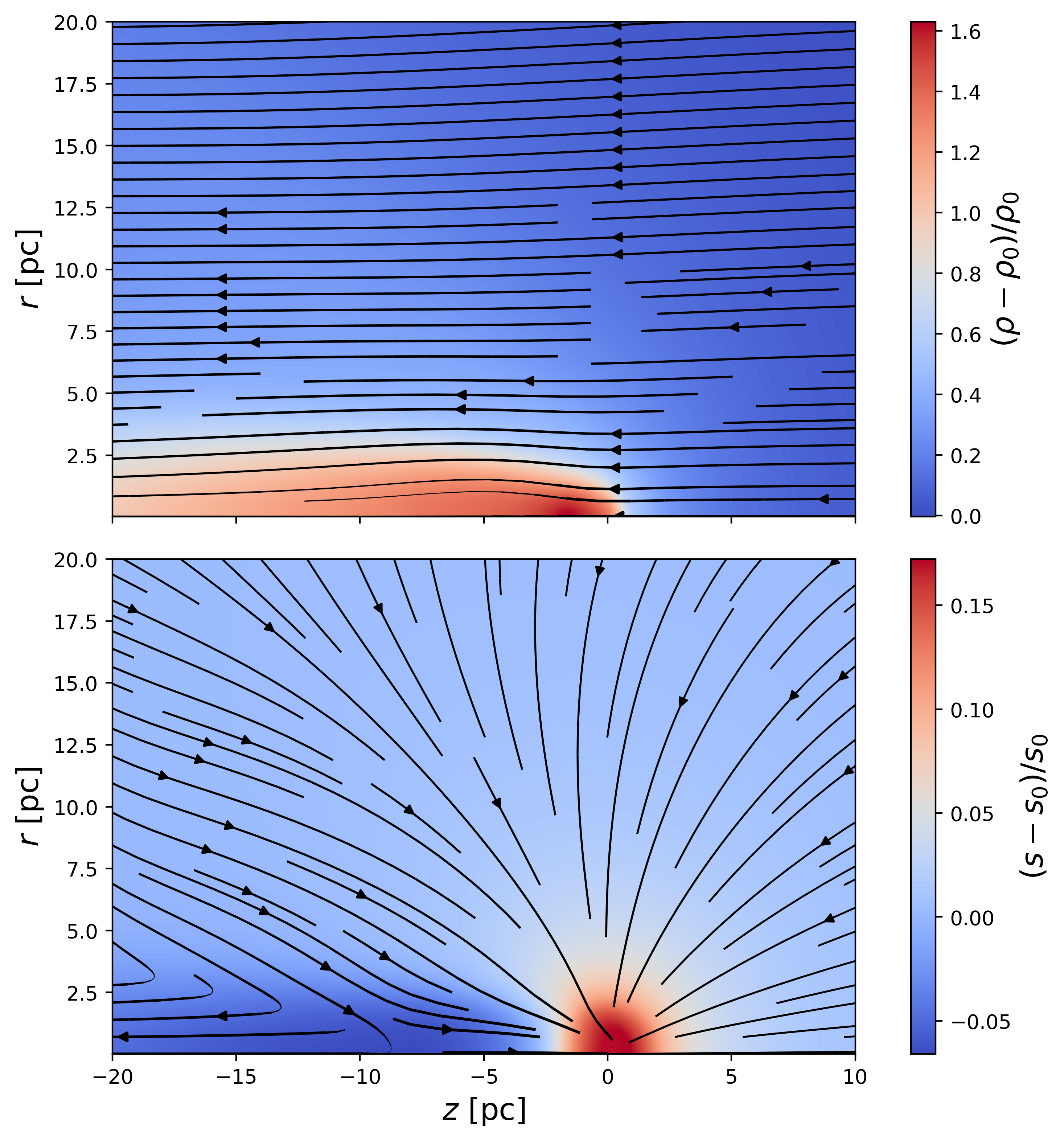}}
    \subfigure[$v_c^{\text{eff}} = 10^{-5}v_c$]{\includegraphics[width=0.32\linewidth]{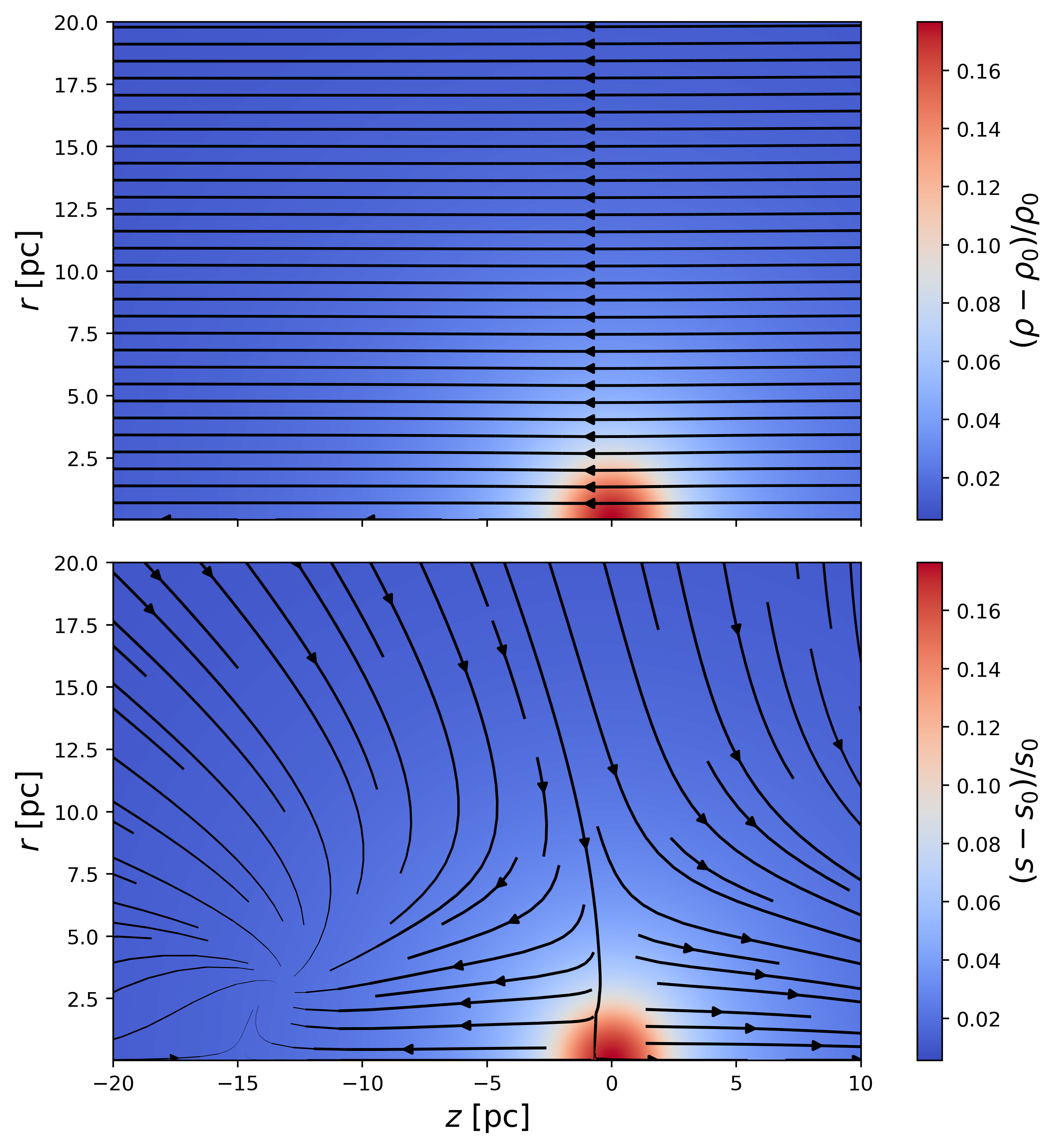}}
    \caption{Density profiles and streamlines for $V=1.5c_{T=0}$. The mass density profiles are superimposed by the net mass density velocity, $\bm{v} = \bm{j}/\rho$, while the entropy density is superimposed by the relative velocity $\bm{w} = \bm{v}_s - \bm{v}_n$. The perturber has mass $M=5M_{\odot}$, the simulation volume  is$L=75\text{pc}$, and the time is $t=50\text{pc}/V$.}
\label{fig:profiles}
\end{figure*}

In \figref{fig:F_DF_vs_v_numerical} the dynamical friction from the numerical simulations is compared to the linear result with the effect of the critical velocity included, Eqs. \eqref{eq:df_finite_time}, \eqref{eq:df_interpol_vc}, and \eqref{eq:vc_interpolation_form}. Even with the Landau criterion, given by Eq. \eqref{eq:vc}, the dynamical friction in the superfluid can be very similar to the zero-temperature limit, as was shown in the linear theory. This similarity can also be seen in the mass density profile shown in \figref{fig:profiles}. At $T=0$, for which the pertuber is supersonic with $V = 1.5c_{T=0}$, there is a well-defined supersonic cone that trails the pertuber, and the density contrast reaches about $4.5$. The finite-temperature superfluid has a similar density contrast and supersonic cone, though not as well-defined, illustrating that the superfluid behaves like the $T=0$ limit as thermal counterflow suppresses thermal perturbations. In the fully normal fluid case, the density contrast is much smaller, around $0.17$, and the perturber is instead moving at subsonic speeds, because $V=1.5c_{T=0} < c_n$, hence there is no sonic cone. As $v_c
^{\text{eff}}$ is decreased, the relative velocity becomes increasingly limited and the thermal counterflow inefficient, causing the superfluid density profile to approach the fully normal fluid limit. The dynamical friction changes accordingly, as shown in \figref{fig:F_DF_vs_vceff}. Furthermore, \figref{fig:F_DF_vs_M} shows the friction force as a function of the mass of the perturber $M$, confirming the expectation that increasing $M$ causes a transition from superfluid to normal fluid behavior in a similar manner to decreasing $v_c$.

The numerical results of \figref{fig:F_DF_vs_v_numerical}, \figref{fig:F_DF_vs_vceff}, and \figref{fig:F_DF_vs_M} show that the scheme to include $v_c$ in the linear theory (Eqs. \eqref{eq:df_interpol_vc} and \eqref{eq:vc_interpolation_form}) successfully captures the basic dependence on the perturber mass and critical velocity, though it fails at low velocities, $V < c_{T=0} \approx c_{-}$, suggesting that other factors might come into play at those speeds. However, as we see in the following section, this does not cause any problems when applied to the Fornax dSph, as in the relevant parameter space we have $\bar{w}\ll v_c$, which is far away from the transition between the superfluid and normal fluid phase, and therefore no interpolation is needed. No further attempt was therefore made to improve the scheme.

\section{Application to the Fornax system}
\label{sec:application_fornax}

So far, only the physics of dynamical friction in superfluids has been discussed, with little reference to the real world. Now, armed with the expressions derived and tested in the previous sections, the parameter space of superfluid DM---the particle mass, self-interaction, and temperature---can be explored by estimating the orbital decay time of GCs in the Fornax dSph, and checking whether the timing problem is alleviated for SIBEC-DM, or exacerbated.

\begin{figure}[]
    \centering
    \includegraphics[width=0.98\linewidth]{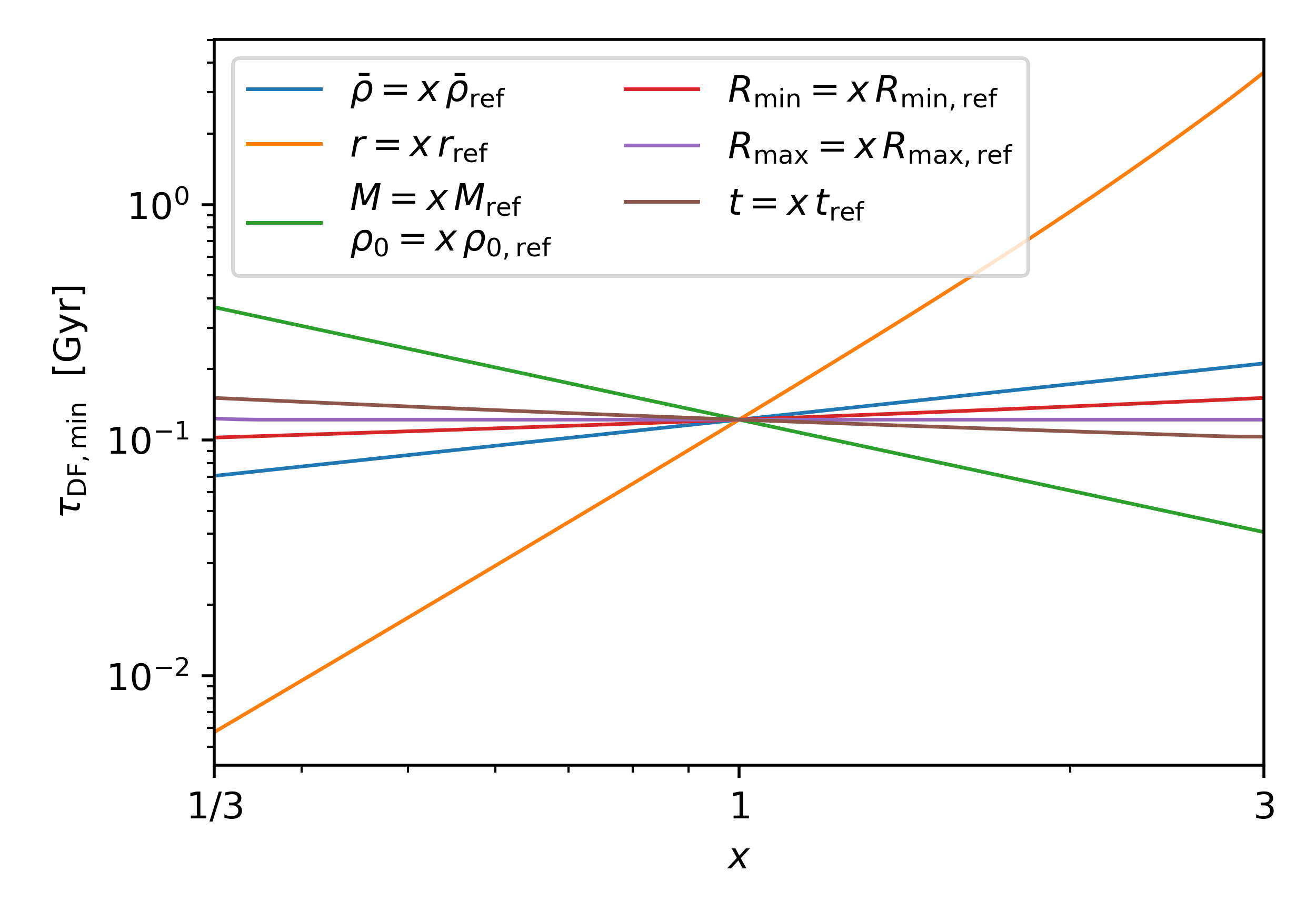}
    \caption{Change in the orbital decay time as parameters related to the modeling of the Fornax dSph, the GCs, and the dynamical friction are varied. The reference values, which are for GC3 in the LC model, are labeled with the subscript "ref".}
\label{fig:GC3_LC_tau_vary}
\end{figure}

The decay time can be defined as the time $\tau_{\text{DF}}$ it takes dynamical friction to reduce the angular momentum $L$ of the GCs to zero;
\begin{equation}
\label{eq:decay_time_estimate}
    \tau_{\text{DF}} = \frac{L}{r|F_{\text{DF}}|} = \frac{MV}{|F_{\text{DF}}|},
\end{equation}
where $M$, $V$, and $r$ are the mass, circular orbital velocity, and the orbital radius of the GCs.
The density profile of the Fornax dSph is modeled in the same way as in \citet{Cole2012}, using a spherical double-power-law profile\footnote{There appears to be a sign error in Eq. (1) in \citet{Cole2012} when comparing the resulting profiles to their own figures, as well as compared to other works that use the same type of profile \citep{Zhao1996, Walker2009, Hague2013}.} of the form
\begin{equation}
    \label{eq:fornax_density_profile}
    \rho(r) = \bar{\rho} \left(\frac{r}{r_s}\right)^{-\gamma_0} \left[1 + \left(\frac{r}{r_s}\right)^{\eta}\right]^{\frac{\gamma_0 - \gamma_{\infty}}{\eta}}.
\end{equation}
The profile parameters, still following \citet{Cole2012}, are listed in Table \ref{tab:fornax_profile}, and correspond to different models for the shape of the Fornax dSph. As SIBEC-DM, like many alternative theories for DM, is in part motivated by typically having a cored profile, we only focus on the Large core (LC) and Weak cusp (WC) models from \citet{Cole2012}. It should also be noted that the density profile Eq. \eqref{eq:fornax_density_profile} models the total mass density, that is, both stellar and DM, but as DM is the dominant component, we use it as a pure DM profile. As illustrated in \figref{fig:GC3_LC_tau_vary}, subtracting a subdominant portion of the density $\rho_0 = \rho(r)$ in the computation of the SIBEC-DM dynamical friction in order to account for the presence of stellar mass does not significantly alter the value of the orbital decay time.

\bgroup 
\def\arraystretch{1.1}
\begin{table*}[t]
\caption{Halo mass profile parameters, using Eq. \eqref{eq:fornax_density_profile}, with values from \citet{Cole2012}. The density parameter $\bar{\rho}$ is computed from the mass enclosed within $1.8\text{kpc}$.}              
\label{tab:fornax_profile}      
\centering                                      
\begin{tabular}{l l l l l l l l}          
\hline\hline
Model  & Name               & $\gamma_0$    & $\gamma_{\infty}$ & $\eta$    & $r_s$ [kpc]   & $M(1.8\text{kpc})$ [$10^8M_{\odot}$]      & $\bar{\rho}$ [$10^8 M_{\odot} \text{kpc}^{-3}$] \\
\hline
LC    & Large core          & 0.07          & 4.65              & 3.7       & 1.4           & 4.12                                      & 0.35  \\
WC    & Weak cusp           & 0.08          & 4.65              & 2.77      & 0.62          & 1.03                                      & 0.71 \\
\hline\hline
\end{tabular}
\end{table*}
\egroup 

Estimates of the masses, projected orbital radii $r_{\perp}$, and core radii $r_c$ of the GCs, which we use as $R_{\text{min}}$ in perturbation theory, are listed in Table \ref{tab:fornax_GC}. As in \citet{Lancaster2020} and \citet{Hui2017}, $r=2r_{\perp}/\sqrt{3}$ is used as the "true" radial distance from the Fornax center. This larger radial distance leads to a longer estimate of the decay time $\tau_{\text{DF}}$, as illustrated in \figref{fig:GC3_LC_tau_vary}. The orbital velocities of the  GCs are assumed to be circular, determined by the total halo mass enclosed by their orbits, $M_{\text{encl}}$,
\begin{equation}
\label{eq:orbital_velocity_vs_radius}
    V=\sqrt{\frac{G M_{\text{encl}}}{r}}.
\end{equation}

Inside $r_s$ the density profile of the Fornax dSph is approximately constant and cored for the LC and WC models. Hence, we use $r_s$ as the "core size" of the Fornax, $R_c$, and the upper length scale when we use perturbation theory, $R_{\text{max}}$. The DM density is determined at the position of the GCs using Eq. \eqref{eq:fornax_density_profile}.

\bgroup 
\def\arraystretch{1.1}
\begin{table}
\caption{Projected radial distances, masses, and core radii of the GCs (not including the sixth found by \citet{Wang2019a}) in the Fornax dSph, taken from \citet{Mackey2003}.}              
\label{tab:fornax_GC}      
\centering                                      
\begin{tabular}{l l l l}          
\hline\hline
GC label & \makecell{Projected radial \\ distance $r_{\perp}$ [kpc]} & \makecell{GC core \\radius $r_c$ [pc]} & \makecell{Mass \\$M$ [$10^5M_{\odot}$]}
\\ \hline
GC1      & 1.6                                                       & 10.03                                   & 0.37      \\
GC2      & 1.05                                                      & 5.81                                    & 1.82      \\
GC3      & 0.43                                                      & 1.60                                    & 3.63      \\
GC4      & 0.24                                                      & 1.75                                    & 1.32      \\
GC5      & 1.43                                                      & 1.38                                    & 1.78      \\
\hline\hline
\end{tabular}
\end{table}
\egroup 

There is a limited region of parameter space that is both physically relevant, and may provide a reasonable estimate of $\tau_{\text{DF}}$. This region should satisfy the following:
\begin{itemize}
    \item The core radius of the halo obtained from hydrostatic equilibrium should not exceed the core radius of the dSph as modeled by Eq. \eqref{eq:fornax_density_profile}.
    \item The DM mass and self-interaction should satisfy constraints from observations of the deceleration of DM in cluster collisions \citep{Harvey2015}.
    \item The relaxation rate of DM should be higher than the rate of dynamical changes in the dSph, so that the system can thermalize and form a superfluid.
    \item Perturbation theory is only properly valid for $\delta \rho / \rho \ll 1$.
    \item The ad hoc scheme to include the critical velocity introduced at the end of Section \ref{sec:dynfric_linpert_finite_time} failed to reproduce the numerical results of Section \ref{sec:comparison_pert_theory_numeric} for velocities $V<c_{-}$. Therefore, we cannot trust the dynamical friction obtained from linear perturbation theory for these velocities. However, this should only be a problem near the transition $\bar{w}\sim v_c$, where the form of interpolation between the superfluid and normal fluid result is important.
\end{itemize}
While this list is not exhaustive, it provides a minimum set of criteria that should be fulfilled. Due to our ignorance of the general behavior of superfluid DM in a number of situations, we enforce relaxed variants of the above constraints.

\begin{figure}[]
    \centering
    \includegraphics[width=0.98\linewidth]{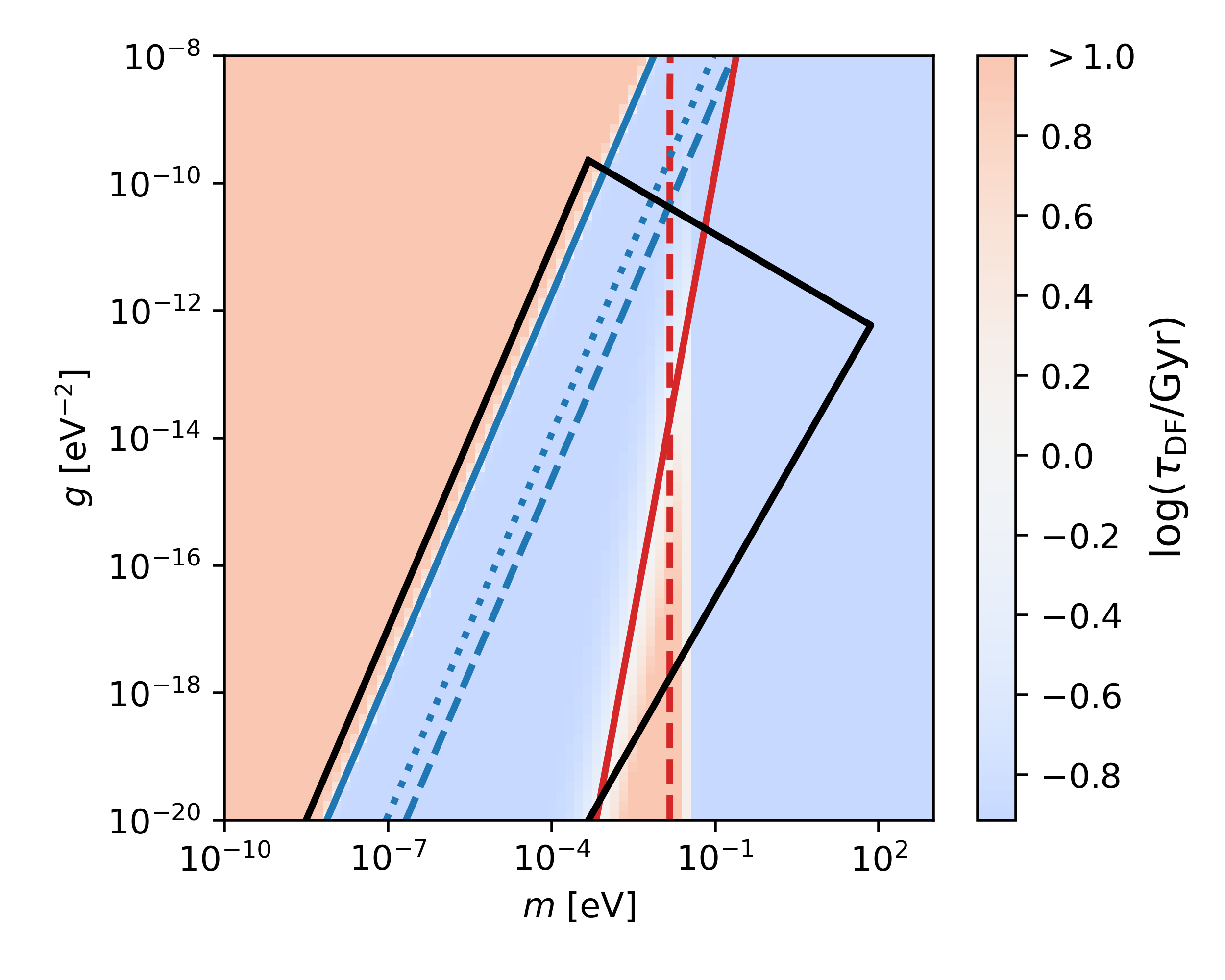}
    \caption{Criteria listed in the text, and the orbital decay timescale for GC3 at $T/T_c = 10^{-4}$ in the LC model for reference. 
    (\textit{solid black line}) Permitted parameter space; the left side is from the constraint on the halo core radius in hydrostatic equilibrium, Eq. \eqref{eq:halo_size_criterion}; the upper right side from the constraint from galaxy cluster collisions, Eq. \eqref{eq:cluster_constraint}; and the lower right side from the minimum relaxation rate needed to thermalize the fluid across the halo, Eq. \eqref{eq:thermalize_constraint}. 
    (\textit{solid blue line}) $V=c_{-}$, with $V<c_{-}$ on the left side. 
    (\textit{dotted blue line}) Criterion for linear perturbation theory to be properly valid, with $\delta\rho/\rho_0<1$ satisfied on the left side. 
    (\textit{dashed blue line}) Supplementary criterion for the $T=0$ treatment of the hydrostatic halo size, with Eq. \eqref{eq:halo_size_suppl_criterion} satisfied on the left side. 
    (\textit{solid red line}) $\bar{w}=v_c$, where the superfluid dynamical friction transitions from superfluid on the left side, to normal fluid on the right. 
    (\textit{dashed red line}) Constraint on the halo core radius in hydrostatic equilibrium with thermal pressure included, with halo cores smaller than the core as modeled by Eq. \eqref{eq:fornax_density_profile} on the right side.}
\label{fig:parameter_space_criteria}
\end{figure}

As seen in the previous sections, and shown in an earlier work \citep{Hartman2020}, counterflow can effectively redistribute thermal energy in a superfluid. Therefore, the shape of the temperature profile of a realistic superfluid DM halo is unknown. The least constraining assumption is therefore made; that the counterflow has washed out any significant thermal differences, so that only the interaction part of the pressure (the only pressure present at $T=0$) determines the hydrostatic profile. Demanding that the core radius of the halo be larger than the core radius obtained from hydrostatic equilibrium, which we define as $\rho(R_c) \approx \rho(0)/2$, gives
\begin{equation}
\label{eq:halo_size_criterion}
    g \lesssim \pi GR_c^2 m^2.
\end{equation}
This relaxed constraint is only possible if it is physically feasible for the counterflow to transport a significant portion of the thermal energy away from the halo core. A supplementary criterion can be derived by demanding that the total entropy flux due to thermal counterflow at the core edge $R_c$, with $w = v_c$, be of the same order as the total entropy enclosed in $R_c$. This leads to
\begin{equation}
\label{eq:halo_size_suppl_criterion}
    g \gtrsim \frac{m^2}{9\rho}\frac{R_c^2}{\Delta t^2},
\end{equation}
where $\Delta t$ should be smaller than the age of the dSph, for example $\Delta t \sim 1 \text{Gyr}$. As shown in \figref{fig:parameter_space_criteria}, the difference between the $T=0$ and the finite temperature treatment of the hydrostatic halo size can be very large, and we do not expect a realistic superfluid halo to be able to completely remove thermal differences, even if upper estimates of the thermal counterflow suggest it could. A realistic superfluid core radius is therefore expected to be larger than the zero-temperature estimate used to derive Eq. \eqref{eq:halo_size_criterion}.

By measuring the spatial offset of stars, gas, and DM in colliding galaxy clusters, a constraint on the self-interaction cross section of DM, $\sigma$, can be established \citep{Harvey2015}. The lack of deceleration of DM and its proximity to the collisionless stars in these collisions places an upper limit of $\sigma/m \lesssim 0.5 \text{cm}^2/\text{g}$. In terms of the self-interaction parameter $g$, this constraint reads \citep{Pitaevskii2016}
\begin{equation}
\label{eq:cluster_constraint}
    g = \frac{\sqrt{4\pi\sigma}}{m} \lesssim 5\times 10^{-12}\left(\frac{1\,\text{eV}}{m}\right)^{1/2} \,\text{eV}^{-2}.
\end{equation}

While the above places upper limits on $g$, there is also a lower limit that must be considered given by the criterion that the DM superfluid should be thermalized across much of the halo core. For this we require the relaxation rate of DM, $\Gamma_{\text{DM}}$, to be higher than the rate of dynamical changes in the halo core, $\Gamma_{\text{grav}} \sim \sqrt{G\rho}$. For two-body interactions, the relaxation rate is $\Gamma \sim n\sigma\delta v$, where $\sigma$ is the scattering cross-section and $\delta v$ the velocity dispersion of the particles. In terms of $g$, as above, the cross-section is $\sigma=m^2g^2/4\pi$. However, for a condensed Bose gas, the relaxation rate is enhanced, that is, $\Gamma \sim \mathcal{N}n\sigma\delta v$, where
\begin{equation}
    \mathcal{N} = n\frac{(2\pi)^3}{\frac{4\pi}{3}(m\delta v)^3},
\end{equation}
because of the high occupation number of the ground state \citep{Sikivie2009}. Using $\delta v \sim V$, that is, that the DM velocity dispersion is of the same order as the GC orbital velocity, the criterion $\Gamma_{\text{DM}} > \Gamma_{\text{grav}}$ becomes 
\begin{equation}
\label{eq:thermalize_constraint}
    g \gtrsim \sqrt{\frac{2}{3\pi}}\frac{m^{3/2} G^{1/4} V}{\rho^{3/4}}.
\end{equation}
It should be noted that the enhancement factor is included in this criterion, but not in the constraint from cluster collisions. This is another example of a relaxed constraint due to our ignorance of how the superfluid properties might change in the various situations. The characteristic speeds of cluster collisions are usually much higher than inside halos, which might result in a much larger disruption of the DM BEC. Furthermore, the DM fluid may not even be condensed throughout most of the cluster, only inside dense structures. We therefore choose the least restrictive constraint by including $\mathcal{N}$ inside the dSph DM halo, but not outside.

The remaining constraints due to $\delta \rho / \rho \ll 1$ and $V<c_{-}$ are readily obtained from perturbation theory and Eq. \eqref{eq:cs_m}. The result from the finite-time approach, Eq. \eqref{eq:df_finite_time}, with our proposed scheme for including the critical velocity, Eqs. \eqref{eq:df_interpol_vc} and \eqref{eq:vc_interpolation_form}, is used to compute the dynamical friction. The characteristic timescale $t=r/V$ is used as the finite time the perturber has been active, though the results are not sensitive to this choice. A deficiency of the finite-time formalism is the lack of self-gravitation in the background fluid, and it might seem that a better choice is to instead use the steady-state expression, Eq. \eqref{eq:F_k_int_remaining}, which includes this property. However, that result assumes the perturber has acted on an otherwise static background for an infinite amount time, and does not take into account that the background can be rotationally supported, and therefore resist the large-scale gravitational collapse induced by the perturber. Furthermore, numerical studies of dynamical friction in realistic halos show that linear theory can provide reasonable estimates of the gravitational drag force even with self-gravitation neglected if the mass of the perturber is significantly smaller than the mass of the host halo \citep{Fujii2006,Aceves2007,Binney2008,Chapon2013,Antonini2011,Tamfal2020}, as is the case here. However, because linear perturbation theory assumes a uniform background with an upper cutoff of scales to take into account the finite extent of the background, we focus on the GCs located inside $r_s$, where the density profile of the LC and WC models are approximately flat, which are GC3 and GC4. These are also the ones that the timing-problem usually applies too \citep{Cole2012,Hui2017,Arca-Sedda2017}.

\begin{figure*}
    \centering   
    \subfigure[GC3 LC, $T/T_c=10^{-2}$]{\includegraphics[width=0.32\linewidth]{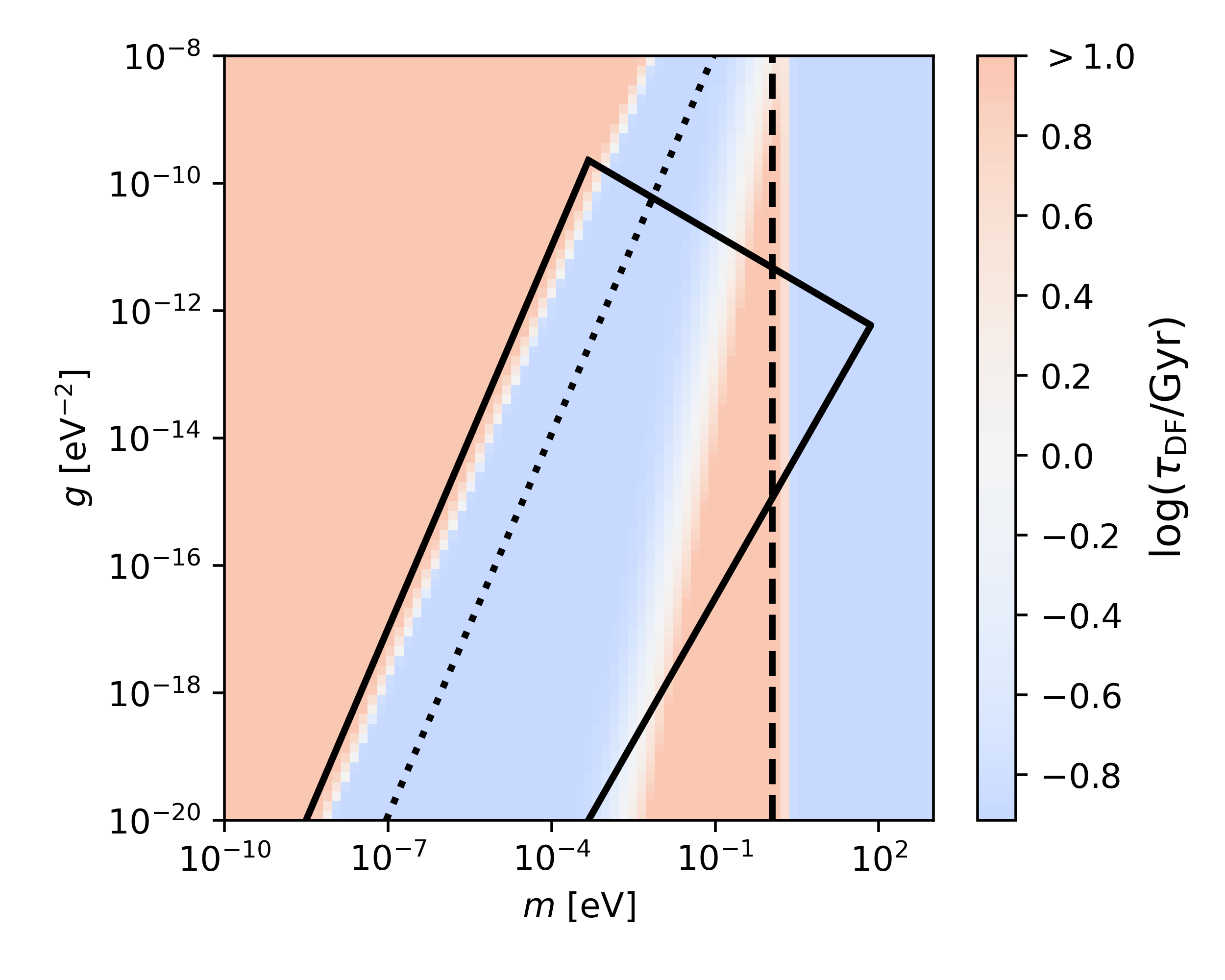}}
    \subfigure[GC3 LC, $T/T_c=10^{-4}$]{\includegraphics[width=0.32\linewidth]{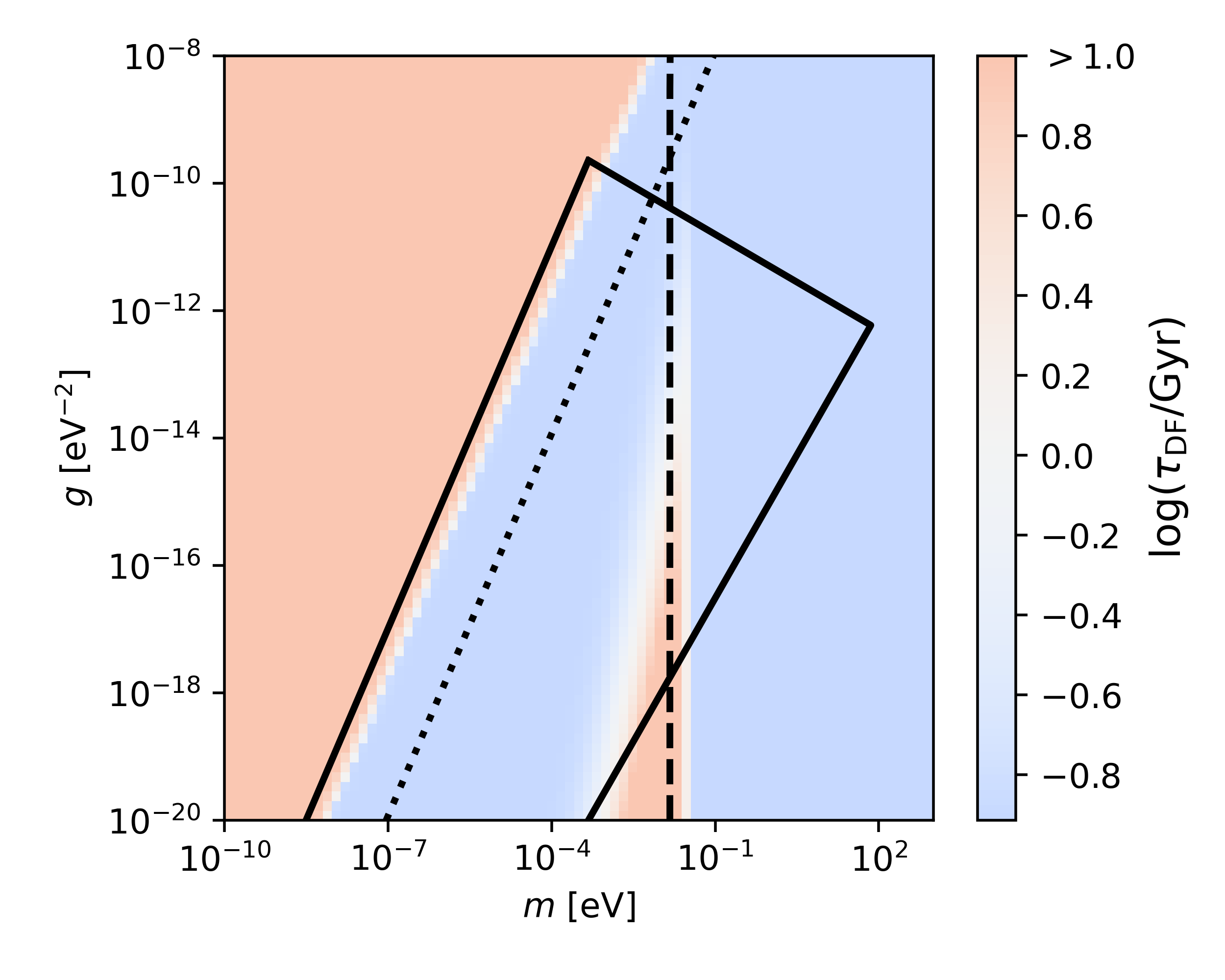}}
    \subfigure[GC3 LC, $T/T_c=10^{-6}$]{\includegraphics[width=0.32\linewidth]{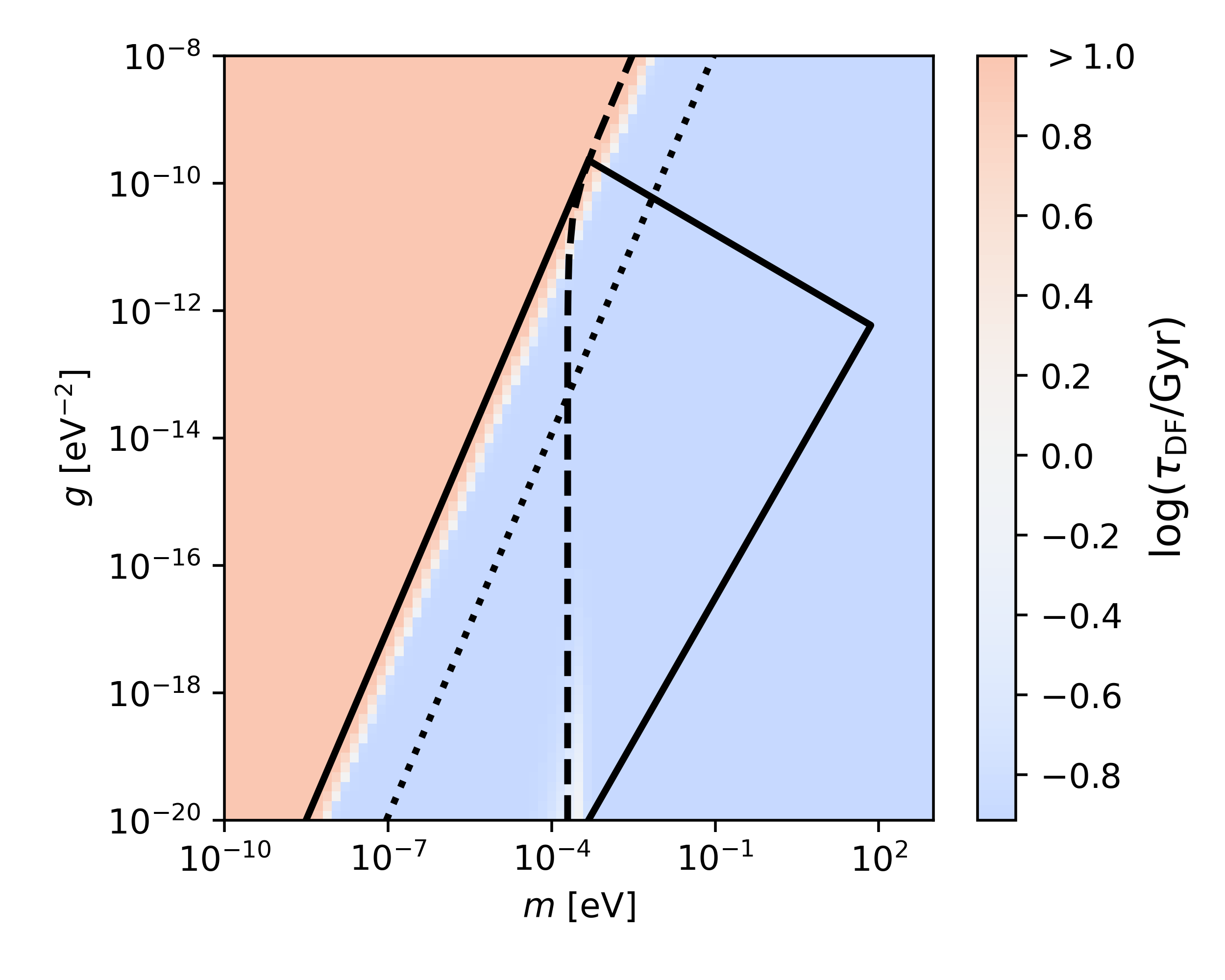}}
    \subfigure[GC3 WC, $T/T_c=10^{-2}$]{\includegraphics[width=0.32\linewidth]{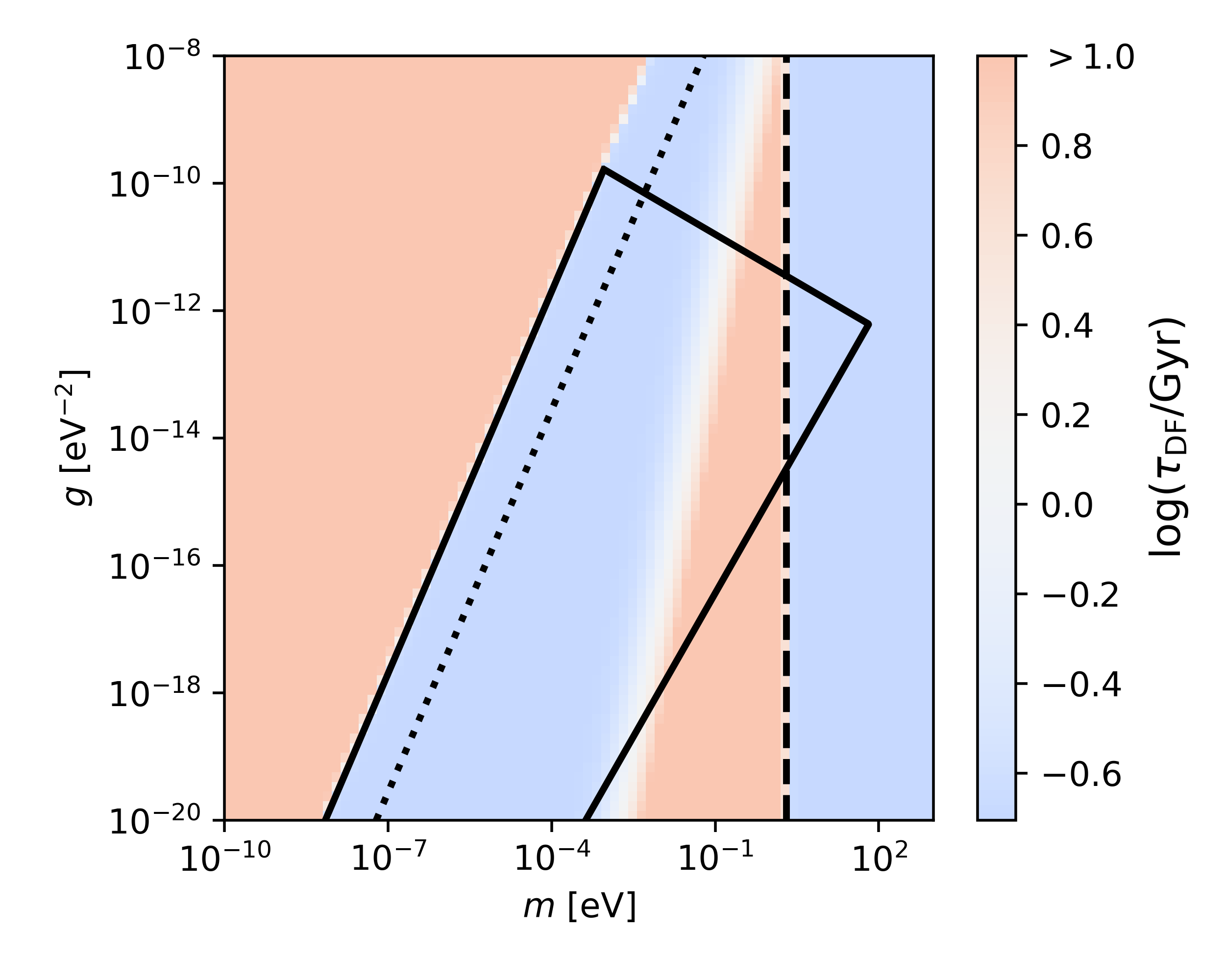}}
    \subfigure[GC3 WC, $T/T_c=10^{-4}$]{\includegraphics[width=0.32\linewidth]{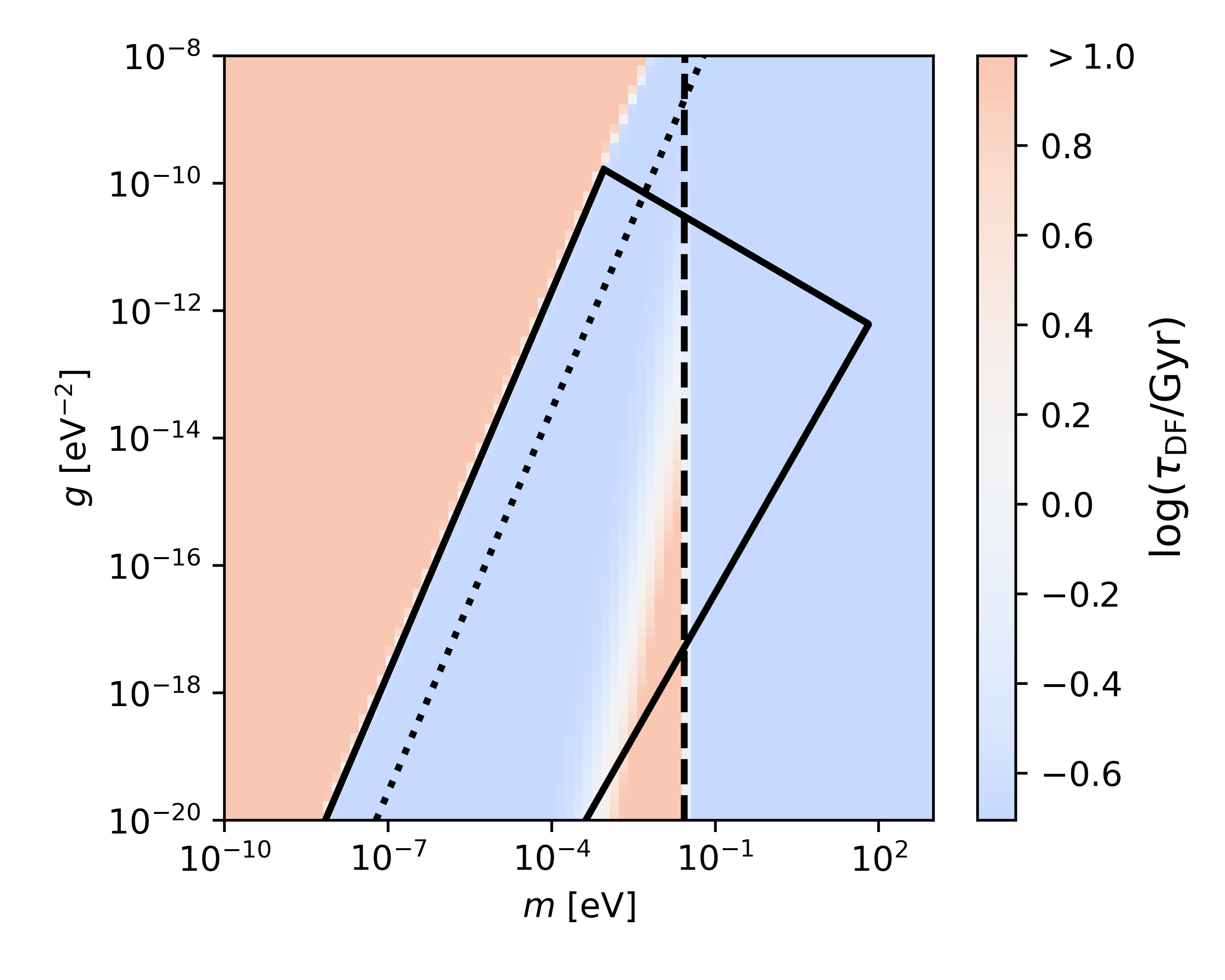}}
    \subfigure[GC3 WC, $T/T_c=10^{-6}$]{\includegraphics[width=0.32\linewidth]{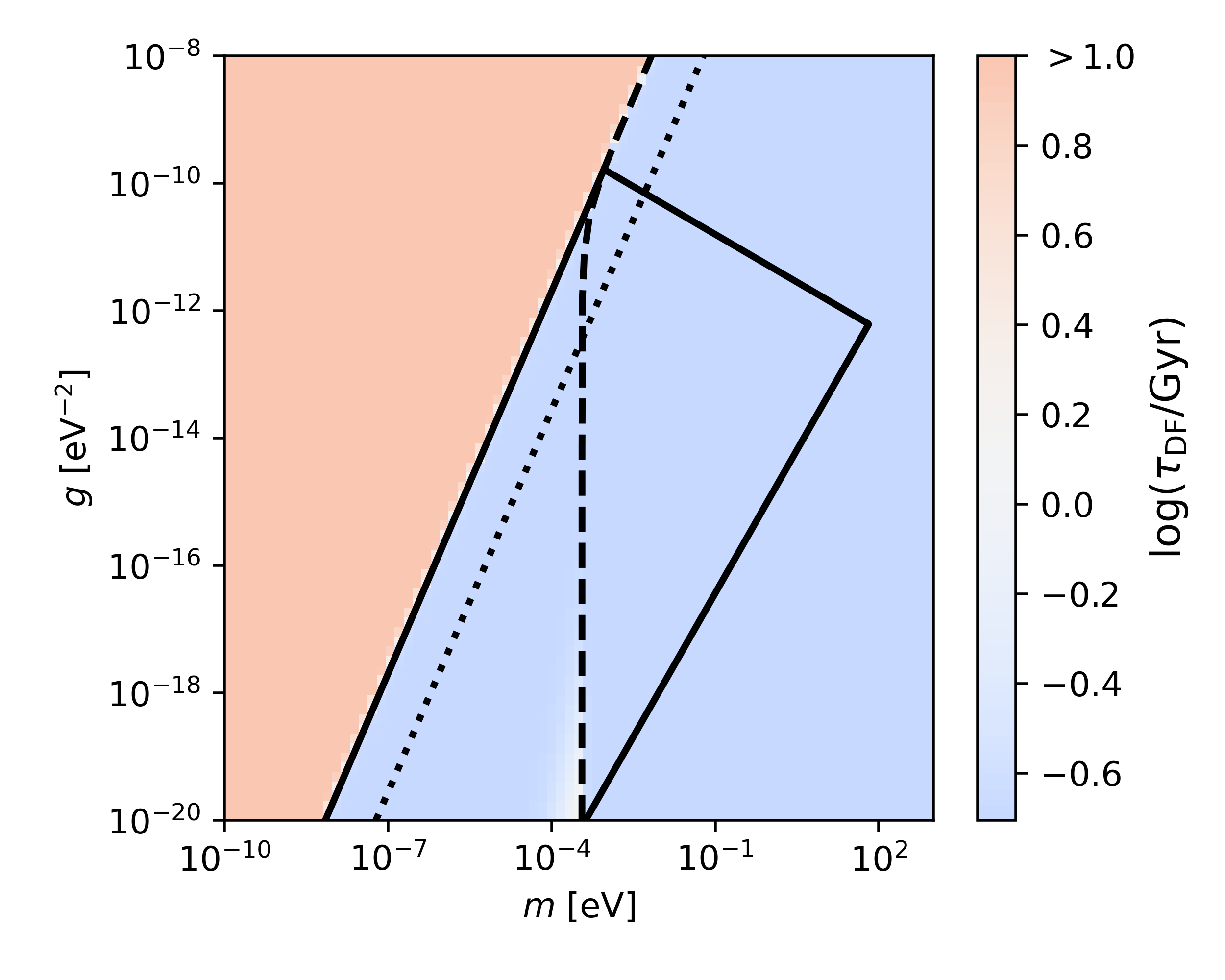}}
    \subfigure[GC4 LC, $T/T_c=10^{-2}$]{\includegraphics[width=0.32\linewidth]{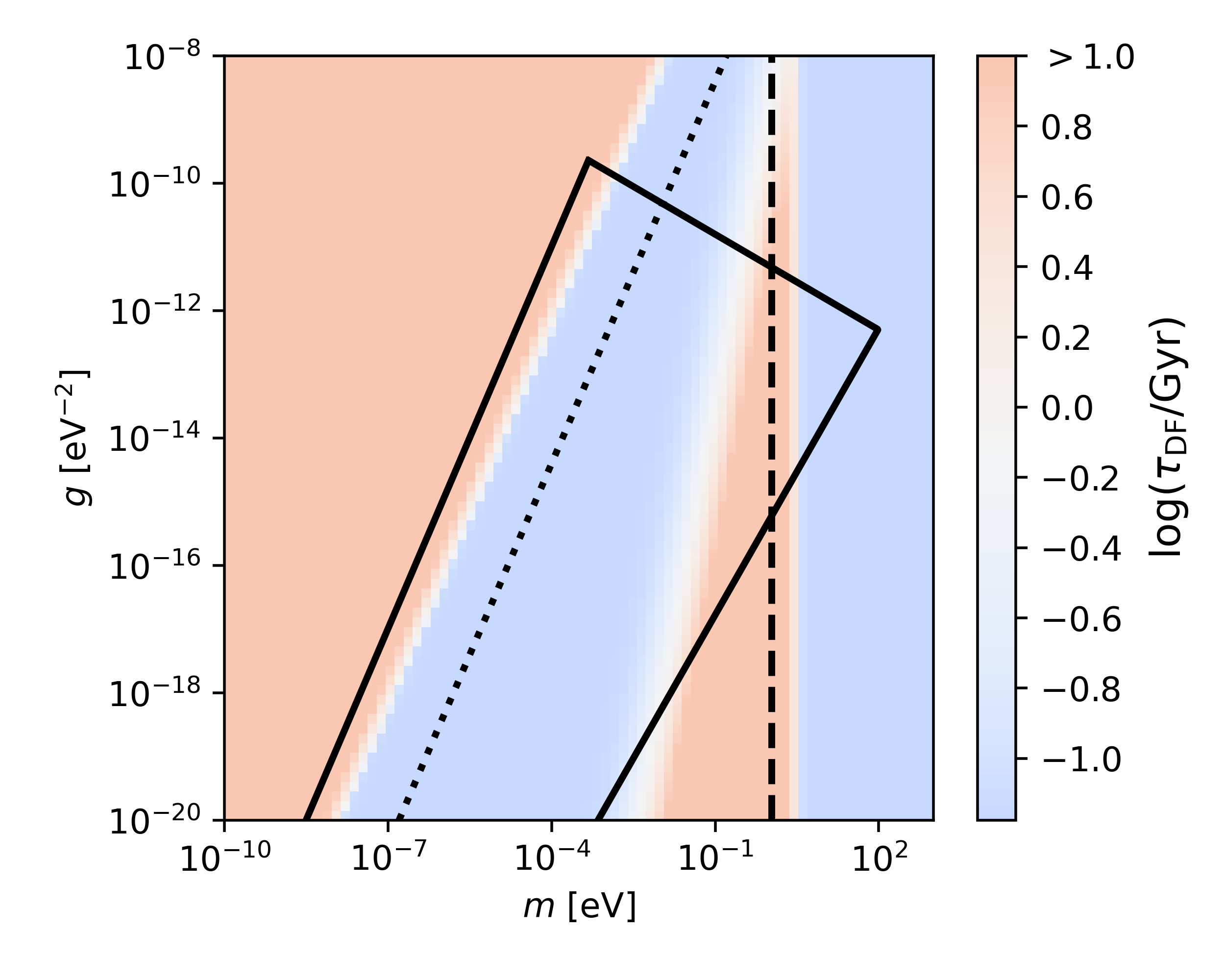}}
    \subfigure[GC4 LC, $T/T_c=10^{-4}$]{\includegraphics[width=0.32\linewidth]{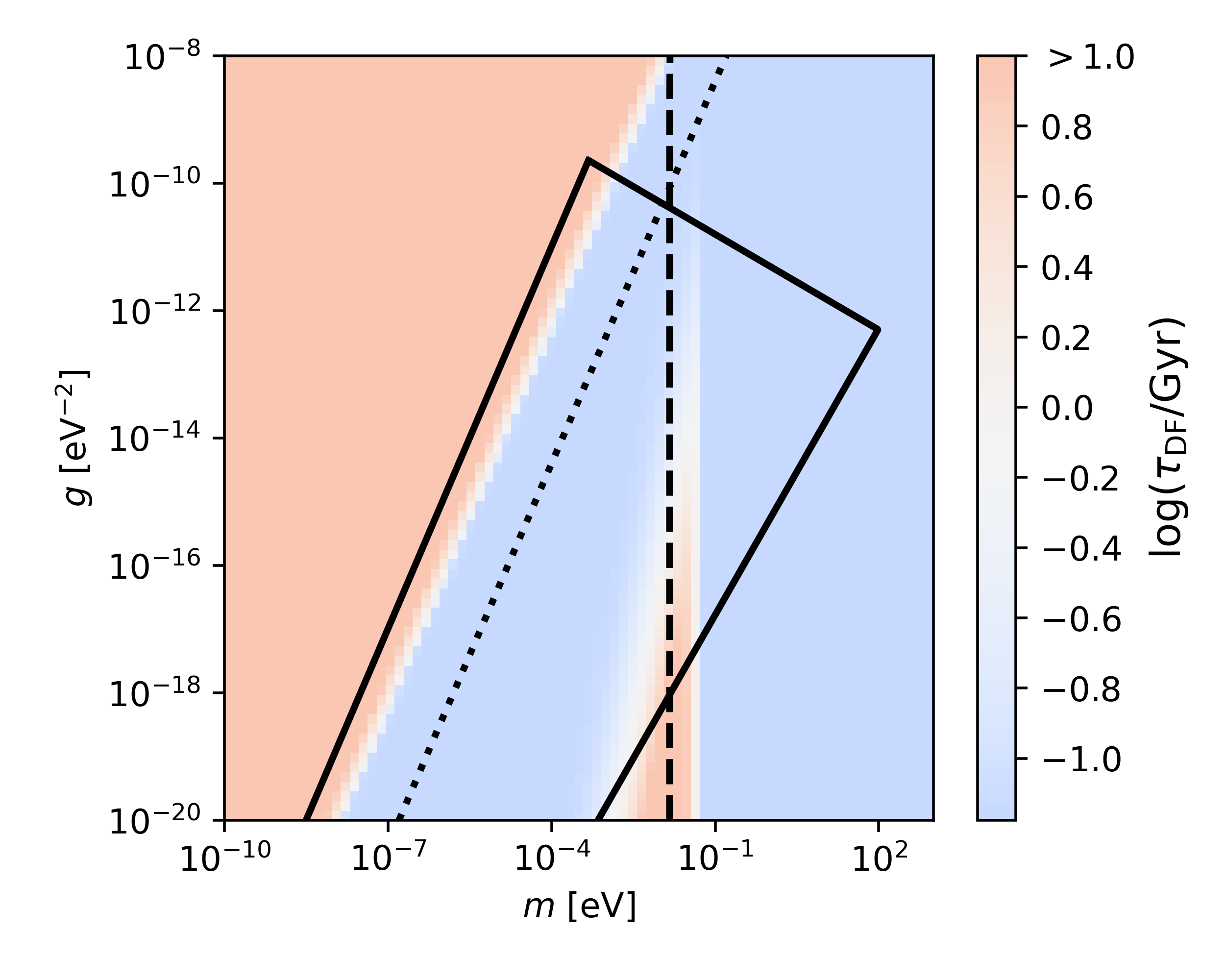}}
    \subfigure[GC4 LC, $T/T_c=10^{-6}$]{\includegraphics[width=0.32\linewidth]{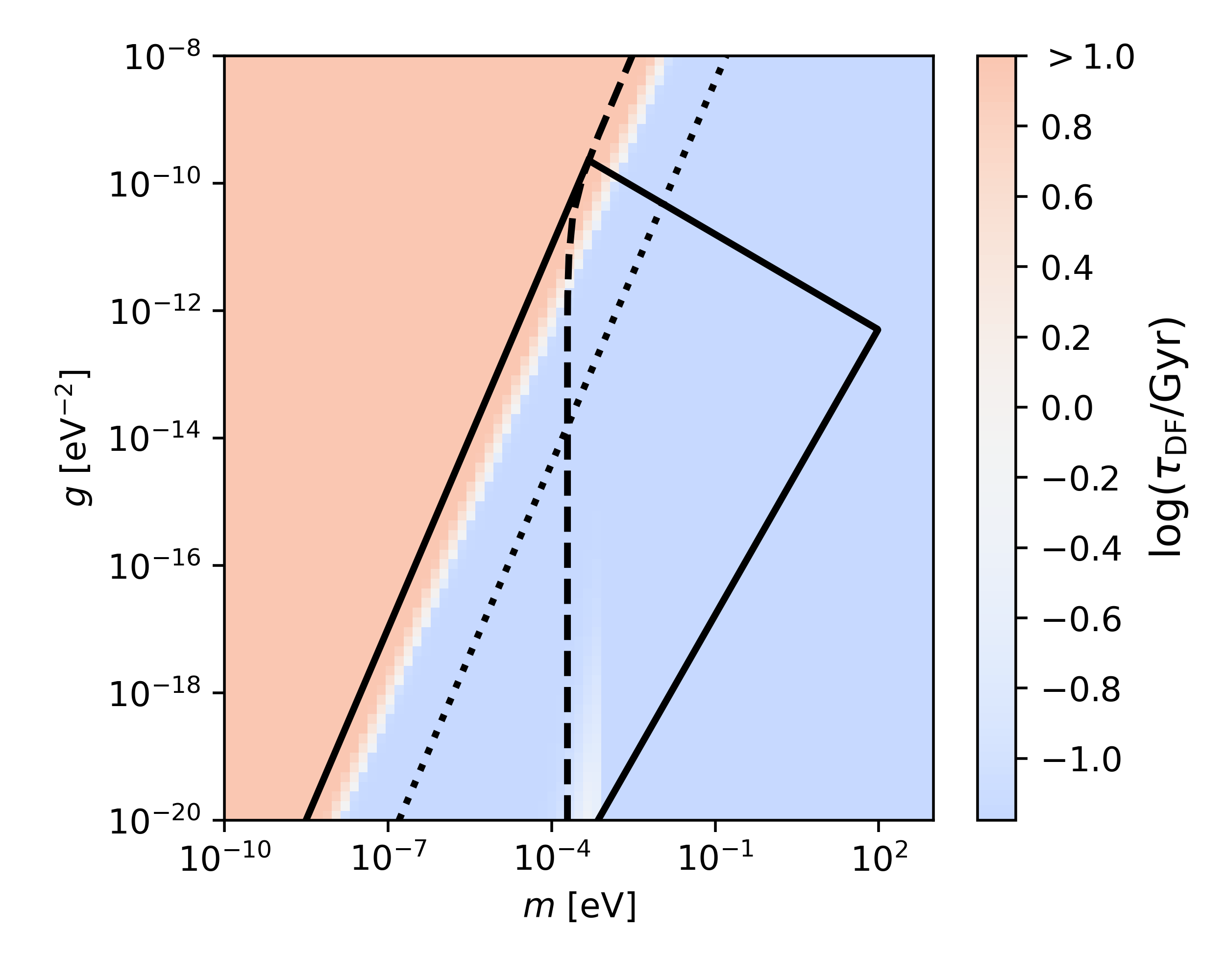}}
    \subfigure[GC4 WC, $T/T_c=10^{-2}$]{\includegraphics[width=0.32\linewidth]{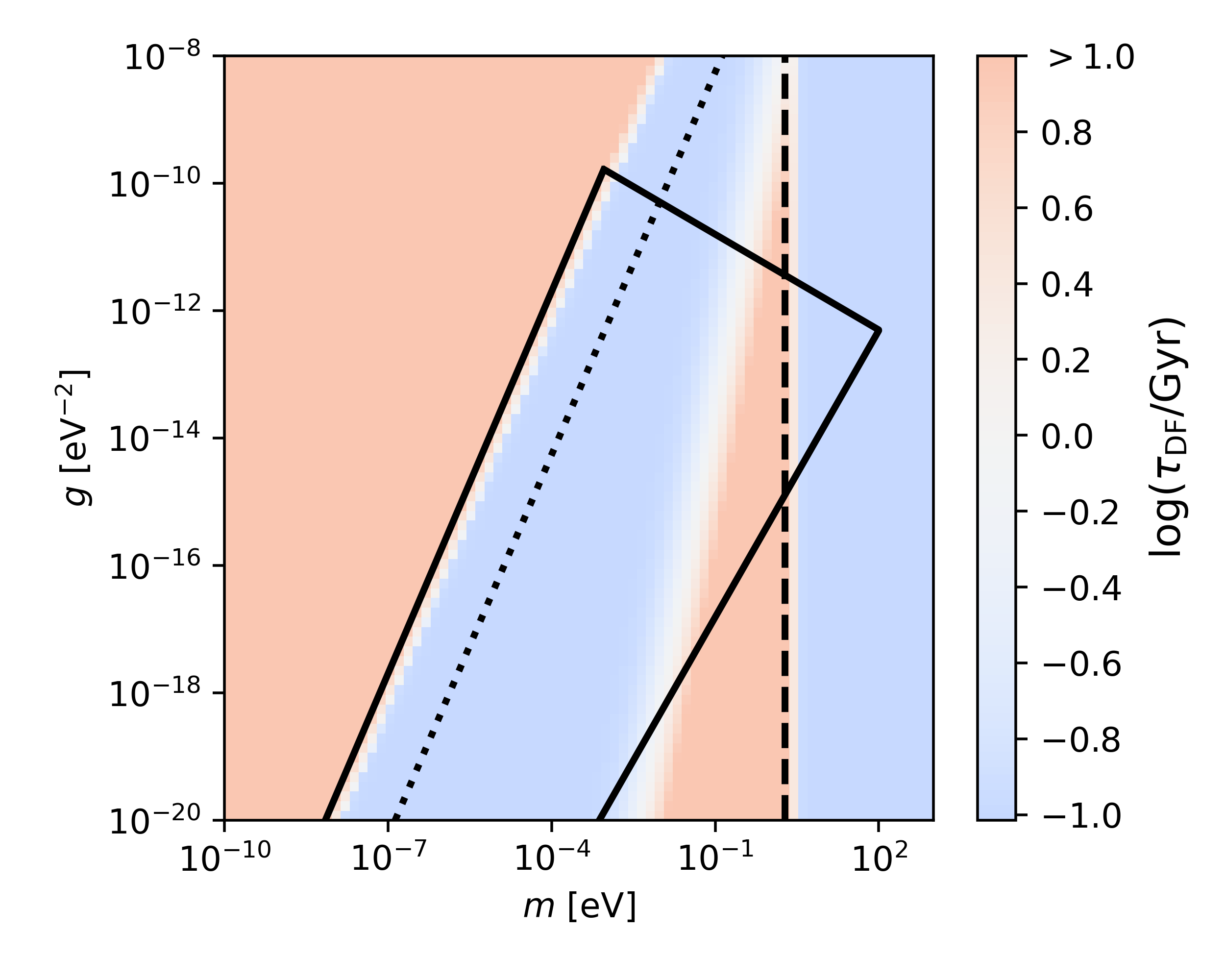}}
    \subfigure[GC4 WC, $T/T_c=10^{-4}$]{\includegraphics[width=0.32\linewidth]{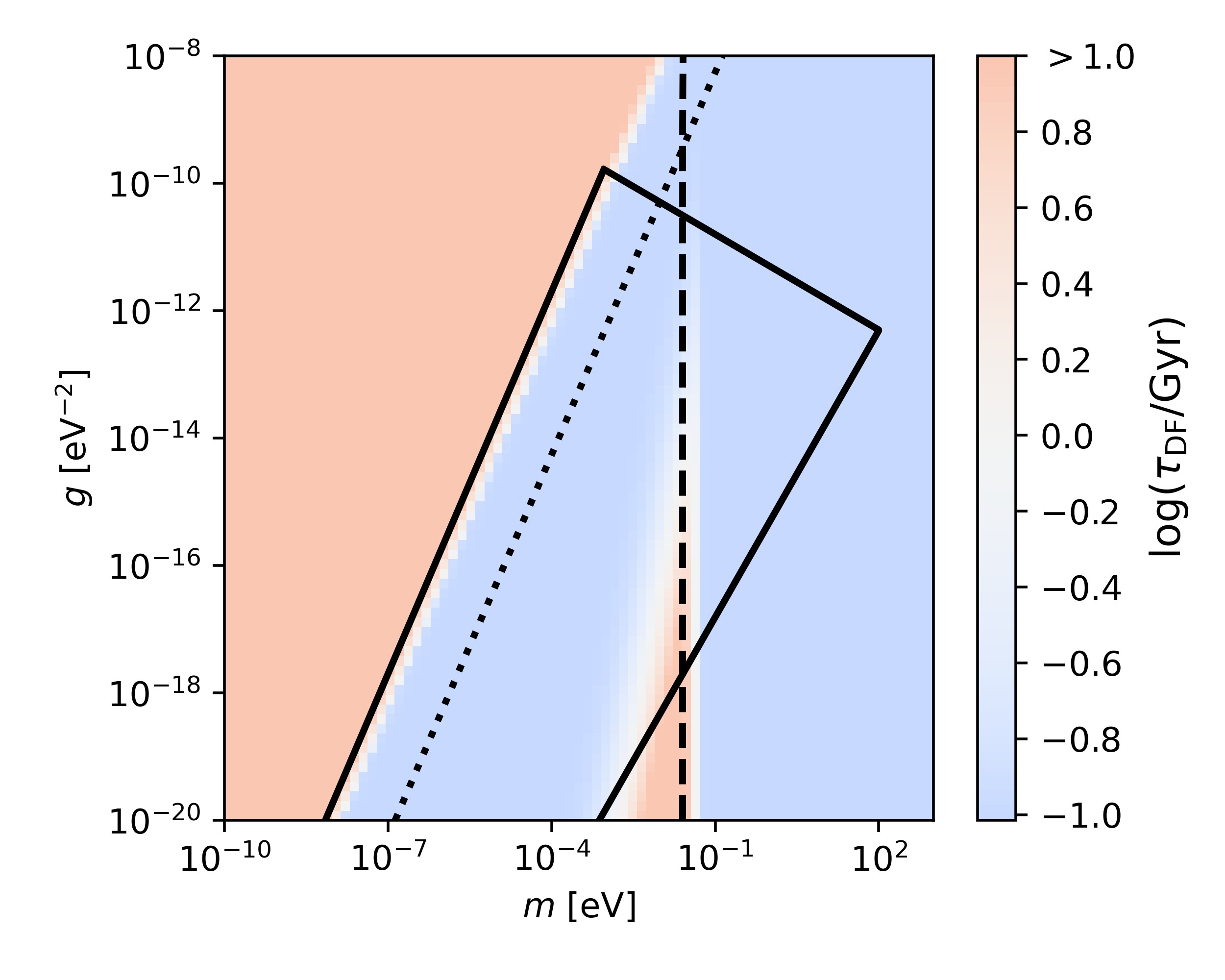}}
    \subfigure[GC4 WC, $T/T_c=10^{-6}$]{\includegraphics[width=0.32\linewidth]{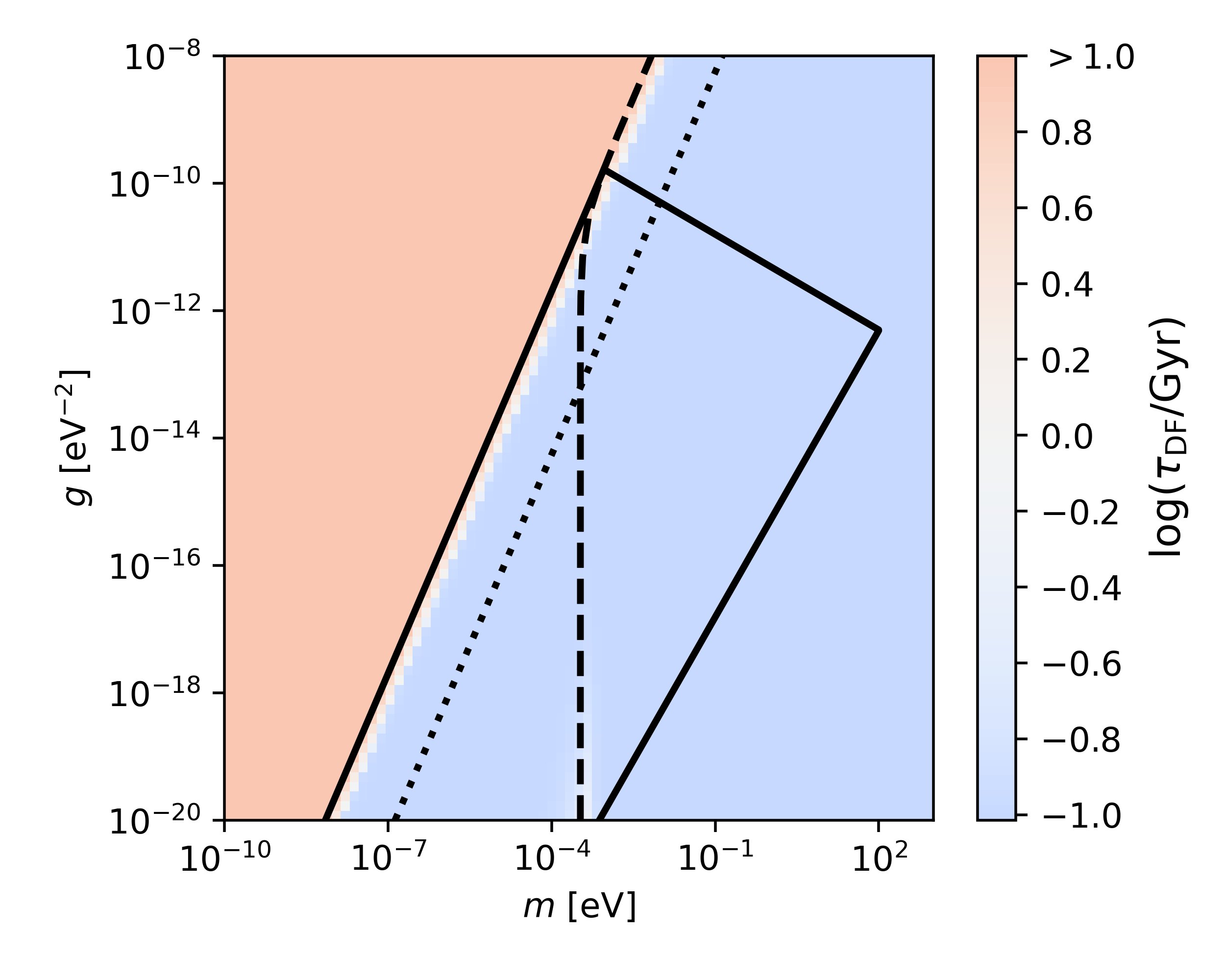}}
    \caption{Decay time of GC3 and GC4, as listed in Table \ref{tab:fornax_GC}, in the LC and WC models for the Fornax dSph density profile from Table \ref{tab:fornax_profile}. 
    (\textit{solid line}) Permitted parameter space; the left side is from the constraint on the halo core radius in hydrostatic equilibrium, Eq. \eqref{eq:halo_size_criterion}; the upper right side from the constraint from galaxy cluster collisions, Eq. \eqref{eq:cluster_constraint}; and the lower right side from the minimum relaxation rate needed to thermalize the fluid across the halo, Eq. \eqref{eq:thermalize_constraint}.
    (\textit{dotted line}) Criterion for linear perturbation theory to be properly valid, with $\delta\rho/\rho_0<1$ satisfied on left side.
    (\textit{dashed line}) Limit due to a hydrostatic halo with thermal pressure included, with resulting core radii smaller than the core of the Fornax dSph as modeled by Eq. \eqref{eq:fornax_density_profile} to the right. Changing the temperature only changes the decay time of the normal fluid phase, as well as the crossover from superfluid to normal fluid. However, for the temperatures shown and lower, the normal phase is well outside the parameter space where perturbation theory is valid.}
\label{fig:fornax_parameter_space}
\end{figure*}

The criteria on the parameter space listed above are illustrated in \figref{fig:parameter_space_criteria} for GC3 in the LC profile, with the estimated orbital decay time for reference. Some features that also hold for the other cases shown in \figref{fig:fornax_parameter_space} are worth pointing out. First, the transition point between superfluid and normal fluid behavior, $\bar{w} = v_c$, lies far away from the region $\delta\rho/\rho_0 < 1$ where perturbation theory is valid, meaning that we do not need to worry about the accuracy of the interpolation scheme described in Section \ref{sec:dynfric_linpert_finite_time}. Second, the decay time becomes very large for $V<c_{-}$, because, as we have seen in the previous sections, the dynamical friction vanishes quickly for velocities below the lowest sound speed.

\bgroup 
\def\arraystretch{1.1}
\begin{table*}[t]
\caption{Minimum orbital decay time $\tau_{\text{DF, min}}$ of GC3 and GC4 in the LC and WC models for the Fornax dSph, found in the region $V>c_{-}$ and $\delta\rho / \rho_0 < 1$ (for which perturbation theory is properly valid), with the CDM result given by Eq. \eqref{eq:df_cdm} for comparison. The listed values are essentially constant for all temperatures for which the assumptions made are valid, $T/T_c \lesssim 0.1$. For $V<c_{-}$, perturbation theory instead predicts the dynamical friction to quickly vanish, causing $\tau_{\text{DF}}$ to become infinite. The criterion on $g/m^2$ due to the hydrostatic core radius, Eq. \eqref{eq:halo_size_criterion}, is also listed. The values for $g/m^2$ are given in units of $\text{eV}^{-4}$.}              
\label{tab:GC_decay_times}      
\centering                                      
\begin{tabular}{l l l l l l}          
\hline\hline
GC label \& model & $\tau_{\text{DF, min}}$ [Myr] & $\tau_{\text{DF, CDM}}$ [Myr] & $V>c_{-}$                           & $\delta\rho / \rho_0 < 1$   & Eq. \eqref{eq:halo_size_criterion}
\\ \hline
GC3 \& LC         & 122                           & 883                           & $g/m^2 < 1.8\times 10^{-4}$         & $g/m^2 > 1.2\times 10^{-6}$ & $g/m^2 < 1.0\times 10^{-3}$    \\
GC3 \& WC         & 197                           & 1327                          & $g/m^2 < 2.5\times 10^{-4}$         & $g/m^2 > 2.7\times 10^{-6}$ & $g/m^2 < 2.0\times 10^{-4}$    \\
GC4 \& LC         & 67                            & 515                           & $g/m^2 < 5.4\times 10^{-5}$         & $g/m^2 > 4.0\times 10^{-7}$ & $g/m^2 < 1.0\times 10^{-3}$    \\
GC4 \& WC         & 97                            & 635                           & $g/m^2 < 2.5\times 10^{-4}$         & $g/m^2 > 5.5\times 10^{-7}$ & $g/m^2 < 2.0\times 10^{-4}$    \\
\hline\hline
\end{tabular}
\end{table*}
\egroup 

The orbital decay time for a wide range of parameters is shown in \figref{fig:fornax_parameter_space} for the two GCs inside the core radius of the Fornax dSph, GC3 and GC4, in the two density profiles considered. $\tau_{\text{DF}}$ generally either attains a minimum value, $\tau_{\text{DF, min}}$, or approaches infinity. The minimum values in the region $V>c_{-}$ and $\delta\rho/\rho_0 < 1$ are summarized in Table \ref{tab:GC_decay_times}, with $\tau_{\text{DF}}$ in the range $67 \text{Myr}$ -- $197 \text{Myr}$. These timescales are considerably smaller than the CDM result assuming the same density profiles, $515 \text{Myr}$ -- $1327 \text{Myr}$, with the dynamical friction given by \citep{Binney2008}
\begin{equation}
\label{eq:df_cdm}
    F_{\text{DF, CDM}} = -\frac{4\pi M^2 G^2 \rho_0 \ln\Lambda}{V^2}\left[\text{erf}(X) - \frac{2X}{\sqrt{\pi}}e^{-X^2}\right],
\end{equation}
where $\Lambda \approx r\delta v^2/GM$, $X = V/\sqrt{2}\delta v$, $\delta v$ is the velocity dispersion of CDM particles, taken to be $\delta v\approx V$, and $\text{erf}$ is the error function.

The decay time remains small even if parameters used to model the Fornax dSph, the GCs, and the dynamical friction are varied, as illustrated in \figref{fig:GC3_LC_tau_vary}. A notable exception is the position of the GC, for which $\tau_{\text{DF}}$ is considerably shorter when closer to the halo center, and likewise longer when further away. This implies that the value for $\tau_{\text{DF}}$ obtained from Eq. \eqref{eq:decay_time_estimate} overestimates the time it takes the GC to fully decay from its current position, but it also implies that the migration towards the halo center was slower in the past when the GCs were at larger radial distances. Indeed, estimates of $\tau_{\text{DF, min}}$ for GC1, GC2, and GC5, all of which are located at $r \gtrsim 1\text{kpc}$, give decay times in excess of $4 \text{Gyr}$. In the CDM case, the decay times for these GCs are even longer: $17 \text{Gyr}$ and more. These estimates do not suggest a timing problem for the the outer GCs, even if their decay times are considerably shorter for SIBEC-DM compared to CDM. However, we note that these GCs are near or outside the radius $r_s$, where the density profile of the dSph falls sharply, and therefore we do not expect the result for the dynamical friction, nor Eq. \eqref{eq:decay_time_estimate}, to necessarily provide a reasonable estimate of $\tau_{\text{DF}}$. 
Nonetheless, the present results show that for a large region of the relevant parameter space of the SIBEC-DM model considered here, GC3 and GC4 are currently racing towards the center of their host halo in a SIBEC-DM universe.

Let us now consider $\tau_{\text{DF}}$ in light of constraints on the SIBEC-DM model from the literature. By fitting rotation curves of slowly rotating SIBEC-DM halos in hydrostatic equilibrium in 173 nearby galaxies from the Spitzer Photomery \& Accurate Rotation Curves (SPARC) data \citep{Lelli2016}, \citet{Craciun2020} estimated the properties of SIBEC-DM halos at $T=0$, and found the best fit values for $g/m^2$ to be between $2.7\times 10^{-4} \text{eV}^{-4}$ and $5.0\times 10^{-2} \text{eV}^{-4}$. For reference, the estimated limit from hydrostatic equilibrium using Eq. \eqref{eq:halo_size_criterion} gives $g/m^2$ of less than about $2\times 10^{-4} \text{eV}^{-4}$ or $10^{-3} \text{eV}^{-4}$, depending on the profile used for the dSph. As the preferred values obtained by \citet{Craciun2020} for zero-temperature SIBEC-DM satisfy $V<c_{-}$ for the GCs and dSph profiles considered, leading to a vanishing dynamical friction, the $T=0$ case does not have a timing-problem, a result that could also have been found using heuristic arguments; if the halo is largely supported by hydrostatic pressure, that is, its Jeans' length $R_{J} \sim c_s/\sqrt{G\rho}$ is of the order of the DM halo core radius $R_c$, then density perturbations on smaller scales inside the halo will be highly suppressed, resulting in very weak dynamical friction, and therefore long decay times.

In a finite-temperature SIBEC-DM halo---for which we expect the preferred values for $g/m^2$ obtained from fitting rotation curves to be lowered, as it provides additional pressure forces to support DM halos---the present results instead suggest that overly large orbital decay rates due to strong dynamical friction may arise. This is the opposite of what one would naively expect if the superfluid had been treated as a conventional thermal fluid, because an increased pressure generally leads to a smaller maximum dynamical friction. Instead, the superfluid essentially ignores the thermal contribution, and responds to a perturber as if it were at $T=0$, which can yield a much stronger friction force.

\section{Conclusion}
\label{sec:Conclusion}
We investigated the dynamical friction acting on an object due to a superfluid background, starting with steady-state linear perturbation theory. The well-known issue of discontinuities in the friction force as the perturber's velocity crosses the fluid sound speed was encountered. We
therefore also employed a finite-time formalism, which removed these discontinuities, agreeing with previous studies that the dynamical friction increases with the velocity of the  perturber until the sound speed is reached, after which the force decreases with the same $V^{-2}$ dependence as the steady-state result. Both approaches predict the force in the superfluid phase to be very similar to the $T=0$ limit, even when there are large thermal contributions, yielding a much stronger friction force than one might naively expect when compared to a conventional fluid at the same temperature. This happens because counterflow conspires against thermal perturbations, allowing the superfluid to respond to a perturbation as if it were at zero temperatures. However, the counterflow is only effective as long as it does not exceed the critical velocity $v_c$, which acts as an upper limit. For flows where the counterflow would normally exceed, but is limited by, the critical velocity, the superfluid instead behaves as a normal fluid. Therefore, decreasing $v_c$ essentially causes a transition from a superfluid to a normal fluid, interpolating the dynamical friction from about the value at $T=0$ to the value of the normal fluid, which can differ by several orders of magnitude. Numerical simulations were also used to investigate dynamical friction, confirming the general dependence of the force on the critical velocity and the mass of the  perturber, which was found using linear perturbation theory. However, the linear theory failed to reproduce the shape of the superfluid-normal fluid transition for velocities smaller than the smallest sound speed, $V<c_{-}$.

Finally, the superfluid dynamical friction was applied to the Fornax dSph and two of its GCs. It was found that the relevant parameter space in which, among other things, perturbation theory is valid gives orbital decay times for these GCs that are much smaller than the age of the dSph, except in the region preferred in the literature \citep{Craciun2020}. The present work therefore suggests that there is no timing problem for Fornax GCs in the SIBEC-DM model for the values of $g/m^2$ obtained by \citet{Craciun2020} by fitting rotation curves at $T=0$. For a finite-temperature SIBEC-DM, on the other hand, for which the preferred parameter space of $g/m^2$ is likely lowered, very large decay rates of Fornax GCs pose a problem.

The use of linear perturbation theory made it possible to probe a large region of parameter space that is difficult to explore with numerical simulations. The main limitations of the numerical scheme used in this work are the low order of the Godunov scheme used; the absence of entropy production, both when the critical velocity was enforced and in shock waves, which leads to the total energy not being strictly conserved; and the large difference between the superfluid sound speeds and dynamics, which results in very small time-stepping and hence excessive diffusion of the numerical solution. All of these limit the parameters for which we can be confident that the numerical solution is correct, and therefore limits the range within which perturbation theory can be tested. Ideally, superfluid dynamical friction would have also been explored using simulations with realistic models for both the DM halo and perturber, as has been done for galaxies with standard CDM and gas \citep{Chapon2013,Tamfal2020}, but such a study requires an improved scheme for solving the superfluid hydrodynamics equations.

\begin{acknowledgements}
We thank the Research Council of Norway for their support, and Benjamin Elder for the discussions that initiated this work. We also thank the anonymous referee for their helpful comments and suggestions that greatly improved this manuscript.
\end{acknowledgements}


\begin{thebibliography}{111}
\expandafter\ifx\csname natexlab\endcsname\relax\def\natexlab#1{#1}\fi

\bibitem[{Aceves \& Colosimo(2007)}]{Aceves2007}
Aceves, H. \& Colosimo, M. 2007, American Journal of Physics, 75, 139,
  publisher: American Association of Physics Teachers

\bibitem[{Andersen(2004)}]{Andersen2004}
Andersen, J.~O. 2004, Rev. Mod. Phys., 76, 599

\bibitem[{Antonini \& Merritt(2011)}]{Antonini2011}
Antonini, F. \& Merritt, D. 2011, The Astrophysical Journal, 745, 83,
  publisher: IOP Publishing

\bibitem[{Arca-Sedda \& Capuzzo-Dolcetta(2017)}]{Arca-Sedda2017}
Arca-Sedda, M. \& Capuzzo-Dolcetta, R. 2017, Monthly Notices of the Royal
  Astronomical Society, 464, 3060, publisher: Oxford Academic

\bibitem[{Bar-Or {et~al.}(2019)Bar-Or, Fouvry, \& Tremaine}]{Or2019}
Bar-Or, B., Fouvry, J.-B., \& Tremaine, S. 2019, The Astrophysical Journal,
  871, 28, publisher: American Astronomical Society

\bibitem[{Barausse(2007)}]{Barausse2007}
Barausse, E. 2007, Monthly Notices of the Royal Astronomical Society, 382, 826,
  publisher: Oxford Academic

\bibitem[{Barenghi {et~al.}(2014)Barenghi, Skrbek, \&
  Sreenivasan}]{Barenghi2014}
Barenghi, C.~F., Skrbek, L., \& Sreenivasan, K.~R. 2014, Proceedings of the
  National Academy of Sciences, 111, 4647

\bibitem[{Battaglia {et~al.}(2013)Battaglia, Helmi, \&
  Breddels}]{Battaglia2013}
Battaglia, G., Helmi, A., \& Breddels, M. 2013, New Astronomy Reviews, 57, 52

\bibitem[{Berezhiani {et~al.}(2019)Berezhiani, Elder, \&
  Khoury}]{Berezhiani2019}
Berezhiani, L., Elder, B., \& Khoury, J. 2019, Journal of Cosmology and
  Astroparticle Physics, 2019, 074

\bibitem[{Berezhiani \& Khoury(2015)}]{Berezhiani2015}
Berezhiani, L. \& Khoury, J. 2015, Phys. Rev. D, 92, 103510

\bibitem[{{Binney} \& {Tremaine}(2008)}]{Binney2008}
{Binney}, J. \& {Tremaine}, S. 2008, {Galactic Dynamics: Second Edition}
  (Princeton Univ. Press)

\bibitem[{Boldrini {et~al.}(2020)Boldrini, Miki, Wagner, Mohayaee, Silk, \&
  Arbey}]{Boldrini2020}
Boldrini, P., Miki, Y., Wagner, A.~Y., {et~al.} 2020, Monthly Notices of the
  Royal Astronomical Society, 492, 5218, publisher: Oxford Academic

\bibitem[{Boldrini {et~al.}(2019)Boldrini, Mohayaee, \& Silk}]{Boldrini2019}
Boldrini, P., Mohayaee, R., \& Silk, J. 2019, Monthly Notices of the Royal
  Astronomical Society, 485, 2546, arXiv: 1903.00354

\bibitem[{Boylan-Kolchin {et~al.}(2008)Boylan-Kolchin, Ma, \&
  Quataert}]{Kolchin2008}
Boylan-Kolchin, M., Ma, C.-P., \& Quataert, E. 2008, Monthly Notices of the
  Royal Astronomical Society, 383, 93, publisher: Oxford Academic

\bibitem[{Bullock \& Boylan-Kolchin(2017)}]{Bullock2017}
Bullock, J.~S. \& Boylan-Kolchin, M. 2017, Annual Review of Astronomy and
  Astrophysics, 55, 343

\bibitem[{Chandrasekhar(1943)}]{Chandrasekhar1943}
Chandrasekhar, S. 1943, The Astrophysical Journal, 97, 255

\bibitem[{Chapman {et~al.}(2014)Chapman, Hoyos, \& Oz}]{Chapman2014}
Chapman, S., Hoyos, C., \& Oz, Y. 2014, Journal of High Energy Physics, 2014,
  27

\bibitem[{Chapon {et~al.}(2013)Chapon, Mayer, \& Teyssier}]{Chapon2013}
Chapon, D., Mayer, L., \& Teyssier, R. 2013, Monthly Notices of the Royal
  Astronomical Society, 429, 3114, publisher: Oxford Academic

\bibitem[{Clesse \& García-Bellido(2018)}]{Clesse2018}
Clesse, S. \& García-Bellido, J. 2018, Physics of the Dark Universe, 22, 137

\bibitem[{Cole {et~al.}(2012)Cole, Dehnen, Read, \& Wilkinson}]{Cole2012}
Cole, D.~R., Dehnen, W., Read, J.~I., \& Wilkinson, M.~I. 2012, Monthly Notices
  of the Royal Astronomical Society, 426, 601

\bibitem[{Colpi {et~al.}(1999)Colpi, Mayer, \& Governato}]{Colpi1999}
Colpi, M., Mayer, L., \& Governato, F. 1999, The Astrophysical Journal, 525,
  720, publisher: IOP Publishing

\bibitem[{Cowsik {et~al.}(2009)Cowsik, Wagoner, Berti, \& Sircar}]{Cowsik2009}
Cowsik, R., Wagoner, K., Berti, E., \& Sircar, A. 2009, The Astrophysical
  Journal, 699, 1389

\bibitem[{Crăciun \& Harko(2020)}]{Craciun2020}
Crăciun, M. \& Harko, T. 2020, arXiv:2007.12222 [astro-ph, physics:gr-qc,
  physics:hep-th], arXiv: 2007.12222

\bibitem[{Darve {et~al.}(2012)Darve, Bottura, Patankar, \&
  Van~Sciver}]{Darve2012}
Darve, C., Bottura, L., Patankar, N.~A., \& Van~Sciver, S. 2012, AIP Conference
  Proceedings, 1434, 247

\bibitem[{Davis {et~al.}(1985)Davis, Efstathiou, Frenk, \& White}]{Davis1985}
Davis, M., Efstathiou, G., Frenk, C.~S., \& White, S. D.~M. 1985, The
  Astrophysical Journal, 292, 371

\bibitem[{Debattista \& Sellwood(2000)}]{Debattista2000}
Debattista, V.~P. \& Sellwood, J.~A. 2000, The Astrophysical Journal, 543, 704,
  publisher: IOP Publishing

\bibitem[{del Pino {et~al.}(2013)del Pino, Hidalgo, Aparicio, Gallart, Carrera,
  Monelli, Buonanno, \& Marconi}]{delPino2013}
del Pino, A., Hidalgo, S.~L., Aparicio, A., {et~al.} 2013, Monthly Notices of
  the Royal Astronomical Society, 433, 1505, publisher: Oxford Academic

\bibitem[{Del~Popolo \& Le~Delliou(2017)}]{DelPopolo2017}
Del~Popolo, A. \& Le~Delliou, M. 2017, Galaxies, 5

\bibitem[{Doi {et~al.}(2008)Doi, Shirai, \& Shiotsu}]{Doi2008}
Doi, D., Shirai, Y., \& Shiotsu, M. 2008, AIP Conference Proceedings, 985, 648

\bibitem[{Dosopoulou \& Antonini(2017)}]{Dosopoulou2017}
Dosopoulou, F. \& Antonini, F. 2017, The Astrophysical Journal, 840, 31,
  publisher: American Astronomical Society

\bibitem[{Elbert {et~al.}(2015)Elbert, Bullock, Garrison-Kimmel, Rocha,
  Oñorbe, \& Peter}]{Elbert2015}
Elbert, O.~D., Bullock, J.~S., Garrison-Kimmel, S., {et~al.} 2015, Monthly
  Notices of the Royal Astronomical Society, 453, 29

\bibitem[{Famaey \& McGaugh(2012)}]{Famaey2012}
Famaey, B. \& McGaugh, S.~S. 2012, Living Reviews in Relativity, 15, 10

\bibitem[{Fujii {et~al.}(2006)Fujii, Funato, \& Makino}]{Fujii2006}
Fujii, M., Funato, Y., \& Makino, J. 2006, Publications of the Astronomical
  Society of Japan, 58, 743, publisher: Oxford Academic

\bibitem[{Goerdt {et~al.}(2010)Goerdt, Moore, Read, \& Stadel}]{Goerdt2010}
Goerdt, T., Moore, B., Read, J.~I., \& Stadel, J. 2010, The Astrophysical
  Journal, 725, 1707, publisher: IOP Publishing

\bibitem[{Goerdt {et~al.}(2006)Goerdt, Moore, Read, Stadel, \&
  Zemp}]{Goerdt2006}
Goerdt, T., Moore, B., Read, J.~I., Stadel, J., \& Zemp, M. 2006, Monthly
  Notices of the Royal Astronomical Society, 368, 1073, publisher: Oxford
  Academic

\bibitem[{Gómez \& Rueda(2017)}]{Gomez2017}
Gómez, L.~G. \& Rueda, J. 2017, Physical Review D, 96, 063001, publisher:
  American Physical Society

\bibitem[{Hague \& Wilkinson(2013)}]{Hague2013}
Hague, P.~R. \& Wilkinson, M.~I. 2013, Monthly Notices of the Royal
  Astronomical Society, 433, 2314, publisher: Oxford Academic

\bibitem[{Harko {et~al.}(2015)Harko, Liang, Liang, \& Mocanu}]{Harko2015}
Harko, T., Liang, P., Liang, S.-D., \& Mocanu, G. 2015, Journal of Cosmology
  and Astroparticle Physics, 2015, 027, publisher: IOP Publishing

\bibitem[{Harko \& Mocanu(2012)}]{Harko2012}
Harko, T. \& Mocanu, G. 2012, Physical Review D, 85, 084012

\bibitem[{Hartman {et~al.}(2020)Hartman, Winther, \& Mota}]{Hartman2020}
Hartman, S. T.~H., Winther, H.~A., \& Mota, D.~F. 2020, Astronomy \&
  Astrophysics, 639, A90, publisher: EDP Sciences

\bibitem[{Harvey {et~al.}(2015)Harvey, Massey, Kitching, Taylor, \&
  Tittley}]{Harvey2015}
Harvey, D., Massey, R., Kitching, T., Taylor, A., \& Tittley, E. 2015, Science,
  347, 1462

\bibitem[{Hu {et~al.}(2000)Hu, Barkana, \& Gruzinov}]{Hu2000}
Hu, W., Barkana, R., \& Gruzinov, A. 2000, Phys. Rev. Lett., 85, 1158

\bibitem[{Hui {et~al.}(2017)Hui, Ostriker, Tremaine, \& Witten}]{Hui2017}
Hui, L., Ostriker, J.~P., Tremaine, S., \& Witten, E. 2017, Physical Review D,
  95, 043541

\bibitem[{Jiang {et~al.}(2008)Jiang, Jing, Faltenbacher, Lin, \&
  Li}]{Jiang2007}
Jiang, C.~Y., Jing, Y.~P., Faltenbacher, A., Lin, W.~P., \& Li, C. 2008, The
  Astrophysical Journal, 675, 1095, publisher: IOP Publishing

\bibitem[{Just {et~al.}(2011)Just, Khan, Berczik, Ernst, \& Spurzem}]{Just2011}
Just, A., Khan, F.~M., Berczik, P., Ernst, A., \& Spurzem, R. 2011, Monthly
  Notices of the Royal Astronomical Society, 411, 653, publisher: Oxford
  Academic

\bibitem[{Katz {et~al.}(2019)Katz, Kurkela, \& Soloviev}]{Katz2019}
Katz, A., Kurkela, A., \& Soloviev, A. 2019, Journal of Cosmology and
  Astroparticle Physics, 2019, 017, publisher: IOP Publishing

\bibitem[{Kaur \& Sridhar(2018)}]{Kaur2018}
Kaur, K. \& Sridhar, S. 2018, The Astrophysical Journal, 868, 134, publisher:
  American Astronomical Society

\bibitem[{Khoury(2016)}]{Khoury2016}
Khoury, J. 2016, Phys. Rev. D, 93, 103533

\bibitem[{Lancaster {et~al.}(2020)Lancaster, Giovanetti, Mocz, Kahn, Lisanti,
  \& Spergel}]{Lancaster2020}
Lancaster, L., Giovanetti, C., Mocz, P., {et~al.} 2020, Journal of Cosmology
  and Astroparticle Physics, 2020, 001, publisher: IOP Publishing

\bibitem[{Landau(1941)}]{Landau1941}
Landau, L. 1941, Phys. Rev., 60, 356

\bibitem[{Lee \& Stahler(2011)}]{Lee2011}
Lee, A.~T. \& Stahler, S.~W. 2011, Monthly Notices of the Royal Astronomical
  Society, 416, 3177, publisher: Oxford Academic

\bibitem[{Lee \& Stahler(2014)}]{Lee2014}
Lee, A.~T. \& Stahler, S.~W. 2014, Astronomy \& Astrophysics, 561, A84,
  publisher: EDP Sciences

\bibitem[{Lelli {et~al.}(2016)Lelli, McGaugh, \& Schombert}]{Lelli2016}
Lelli, F., McGaugh, S.~S., \& Schombert, J.~M. 2016, The Astronomical Journal,
  152, 157, publisher: American Astronomical Society

\bibitem[{Leung {et~al.}(2020)Leung, Leaman, van de Ven, \&
  Battaglia}]{Leung2020}
Leung, G. Y.~C., Leaman, R., van de Ven, G., \& Battaglia, G. 2020, Monthly
  Notices of the Royal Astronomical Society, 493, 320, publisher: Oxford
  Academic

\bibitem[{Lovell {et~al.}(2014)Lovell, Frenk, Eke, Jenkins, Gao, \&
  Theuns}]{Lovell2014}
Lovell, M.~R., Frenk, C.~S., Eke, V.~R., {et~al.} 2014, Monthly Notices of the
  Royal Astronomical Society, 439, 300, publisher: Oxford Academic

\bibitem[{Mackey \& Gilmore(2003)}]{Mackey2003}
Mackey, A.~D. \& Gilmore, G.~F. 2003, Monthly Notices of the Royal Astronomical
  Society, 340, 175, publisher: Oxford Academic

\bibitem[{Madelung(1926)}]{Madelung1926}
Madelung, E. 1926, Naturwissenschaften, 14, 1004

\bibitem[{Meadows {et~al.}(2020)Meadows, Navarro, Santos-Santos,
  Benítez-Llambay, \& Frenk}]{Meadows2020}
Meadows, N., Navarro, J.~F., Santos-Santos, I., Benítez-Llambay, A., \& Frenk,
  C. 2020, Monthly Notices of the Royal Astronomical Society, 491, 3336,
  publisher: Oxford Academic

\bibitem[{Milgrom(1983{\natexlab{a}})}]{Milgrom1983b}
Milgrom, M. 1983{\natexlab{a}}, \apj, 270, 371

\bibitem[{Milgrom(1983{\natexlab{b}})}]{Milgrom1983c}
Milgrom, M. 1983{\natexlab{b}}, \apj, 270, 384

\bibitem[{Milgrom(1983{\natexlab{c}})}]{Milgrom1983a}
Milgrom, M. 1983{\natexlab{c}}, \apj, 270, 365

\bibitem[{Mina {et~al.}(2020{\natexlab{a}})Mina, Mota, \& Winther}]{Mina2020}
Mina, M., Mota, D.~F., \& Winther, H.~A. 2020{\natexlab{a}}, Astronomy \&
  Astrophysics, 641, A107, publisher: EDP Sciences

\bibitem[{Mina {et~al.}(2020{\natexlab{b}})Mina, Mota, \& Winther}]{Mina2020b}
Mina, M., Mota, D.~F., \& Winther, H.~A. 2020{\natexlab{b}}, arXiv:2007.04119
  [astro-ph, physics:gr-qc], arXiv: 2007.04119

\bibitem[{Mocz {et~al.}(2017)Mocz, Vogelsberger, Robles, Zavala,
  Boylan-Kolchin, Fialkov, \& Hernquist}]{Mocz2017}
Mocz, P., Vogelsberger, M., Robles, V.~H., {et~al.} 2017, Monthly Notices of
  the Royal Astronomical Society, 471, 4559

\bibitem[{{Mulder}(1983)}]{Mulder1983}
{Mulder}, W.~A. 1983, \aap, 117, 9

\bibitem[{Navez \& Graham(2006)}]{Navez2006}
Navez, P. \& Graham, R. 2006, Physical Review A, 73, 043612, publisher:
  American Physical Society

\bibitem[{Nori \& Baldi(2018)}]{Nori2018}
Nori, M. \& Baldi, M. 2018, Monthly Notices of the Royal Astronomical Society,
  478, 3935, publisher: Oxford Academic

\bibitem[{Nori \& Baldi(2020)}]{Nori2020}
Nori, M. \& Baldi, M. 2020, arXiv:2007.01316 [astro-ph], arXiv: 2007.01316

\bibitem[{Oh {et~al.}(2000)Oh, Lin, \& Richer}]{Oh2000}
Oh, K.~S., Lin, D. N.~C., \& Richer, H.~B. 2000, The Astrophysical Journal,
  531, 727, publisher: IOP Publishing

\bibitem[{Ostriker(1999)}]{Ostriker1999}
Ostriker, E.~C. 1999, The Astrophysical Journal, 513, 252, publisher: IOP
  Publishing

\bibitem[{Pani(2015)}]{Pani2015}
Pani, P. 2015, Physical Review D, 92, 123530, publisher: American Physical
  Society

\bibitem[{Percival {et~al.}(2001)Percival, Baugh, Bland-Hawthorn, Bridges,
  Cannon, Cole, Colless, Collins, Couch, Dalton, De~Propris, Driver,
  Efstathiou, Ellis, Frenk, Glazebrook, Jackson, Lahav, Lewis, Lumsden, Maddox,
  Moody, Norberg, Peacock, Peterson, Sutherland, \& Taylor}]{Percival2001}
Percival, W.~J., Baugh, C.~M., Bland-Hawthorn, J., {et~al.} 2001, Monthly
  Notices of the Royal Astronomical Society, 327, 1297, publisher: Oxford
  Academic

\bibitem[{Pitaevskii \& Stringari(2016)}]{Pitaevskii2016}
Pitaevskii, L.~P. \& Stringari, S. 2016, Bose-Einstein Condensation and
  Superfluidity (Great Clarendon Street, Oxford, United Kingdom: Oxford
  University Press)

\bibitem[{{Planck Collaboration} {et~al.}(2016){Planck Collaboration}, {Ade},
  {Aghanim}, {Arnaud}, {Ashdown}, {Aumont}, {Baccigalupi}, {Banday},
  {Barreiro}, {Bartlett}, {Bartolo}, {Battaner}, {Battye}, {Benabed},
  {Beno{\^\i}t}, {Benoit-L{\'e}vy}, {Bernard}, {Bersanelli}, {Bielewicz},
  {Bock}, {Bonaldi}, {Bonavera}, {Bond}, {Borrill}, {Bouchet}, {Boulanger},
  {Bucher}, {Burigana}, {Butler}, {Calabrese}, {Cardoso}, {Catalano},
  {Challinor}, {Chamballu}, {Chary}, {Chiang}, {Chluba}, {Christensen},
  {Church}, {Clements}, {Colombi}, {Colombo}, {Combet}, {Coulais}, {Crill},
  {Curto}, {Cuttaia}, {Danese}, {Davies}, {Davis}, {de Bernardis}, {de Rosa},
  {de Zotti}, {Delabrouille}, {D{\'e}sert}, {Di Valentino}, {Dickinson},
  {Diego}, {Dolag}, {Dole}, {Donzelli}, {Dor{\'e}}, {Douspis}, {Ducout},
  {Dunkley}, {Dupac}, {Efstathiou}, {Elsner}, {En{\ss}lin}, {Eriksen},
  {Farhang}, {Fergusson}, {Finelli}, {Forni}, {Frailis}, {Fraisse},
  {Franceschi}, {Frejsel}, {Galeotta}, {Galli}, {Ganga}, {Gauthier}, {Gerbino},
  {Ghosh}, {Giard}, {Giraud-H{\'e}raud}, {Giusarma}, {Gjerl{\o}w},
  {Gonz{\'a}lez-Nuevo}, {G{\'o}rski}, {Gratton}, {Gregorio}, {Gruppuso},
  {Gudmundsson}, {Hamann}, {Hansen}, {Hanson}, {Harrison}, {Helou},
  {Henrot-Versill{\'e}}, {Hern{\'a}ndez-Monteagudo}, {Herranz}, {Hildebrand t},
  {Hivon}, {Hobson}, {Holmes}, {Hornstrup}, {Hovest}, {Huang}, {Huffenberger},
  {Hurier}, {Jaffe}, {Jaffe}, {Jones}, {Juvela}, {Keih{\"a}nen}, {Keskitalo},
  {Kisner}, {Kneissl}, {Knoche}, {Knox}, {Kunz}, {Kurki-Suonio}, {Lagache},
  {L{\"a}hteenm{\"a}ki}, {Lamarre}, {Lasenby}, {Lattanzi}, {Lawrence}, {Leahy},
  {Leonardi}, {Lesgourgues}, {Levrier}, {Lewis}, {Liguori}, {Lilje},
  {Linden-V{\o}rnle}, {L{\'o}pez-Caniego}, {Lubin}, {Mac{\'\i}as-P{\'e}rez},
  {Maggio}, {Maino}, {Mandolesi}, {Mangilli}, {Marchini}, {Maris}, {Martin},
  {Martinelli}, {Mart{\'\i}nez-Gonz{\'a}lez}, {Masi}, {Matarrese}, {McGehee},
  {Meinhold}, {Melchiorri}, {Melin}, {Mendes}, {Mennella}, {Migliaccio},
  {Millea}, {Mitra}, {Miville-Desch{\^e}nes}, {Moneti}, {Montier}, {Morgante},
  {Mortlock}, {Moss}, {Munshi}, {Murphy}, {Naselsky}, {Nati}, {Natoli},
  {Netterfield}, {N{\o}rgaard-Nielsen}, {Noviello}, {Novikov}, {Novikov},
  {Oxborrow}, {Paci}, {Pagano}, {Pajot}, {Paladini}, {Paoletti}, {Partridge},
  {Pasian}, {Patanchon}, {Pearson}, {Perdereau}, {Perotto}, {Perrotta},
  {Pettorino}, {Piacentini}, {Piat}, {Pierpaoli}, {Pietrobon}, {Plaszczynski},
  {Pointecouteau}, {Polenta}, {Popa}, {Pratt}, {Pr{\'e}zeau}, {Prunet},
  {Puget}, {Rachen}, {Reach}, {Rebolo}, {Reinecke}, {Remazeilles}, {Renault},
  {Renzi}, {Ristorcelli}, {Rocha}, {Rosset}, {Rossetti}, {Roudier},
  {Rouill{\'e} d'Orfeuil}, {Rowan-Robinson}, {Rubi{\~n}o-Mart{\'\i}n},
  {Rusholme}, {Said}, {Salvatelli}, {Salvati}, {Sandri}, {Santos},
  {Savelainen}, {Savini}, {Scott}, {Seiffert}, {Serra}, {Shellard}, {Spencer},
  {Spinelli}, {Stolyarov}, {Stompor}, {Sudiwala}, {Sunyaev}, {Sutton},
  {Suur-Uski}, {Sygnet}, {Tauber}, {Terenzi}, {Toffolatti}, {Tomasi},
  {Tristram}, {Trombetti}, {Tucci}, {Tuovinen}, {T{\"u}rler}, {Umana},
  {Valenziano}, {Valiviita}, {Van Tent}, {Vielva}, {Villa}, {Wade}, {Wandelt},
  {Wehus}, {White}, {White}, {Wilkinson}, {Yvon}, {Zacchei}, \&
  {Zonca}}]{Planck2015}
{Planck Collaboration}, {Ade}, P.~A.~R., {Aghanim}, N., {et~al.} 2016, \aap,
  594, A13

\bibitem[{Pontzen \& Governato(2012)}]{Pontzen2012}
Pontzen, A. \& Governato, F. 2012, Monthly Notices of the Royal Astronomical
  Society, 421, 3464, publisher: Oxford Academic

\bibitem[{Read {et~al.}(2019)Read, Walker, \& Steger}]{Read2019}
Read, J.~I., Walker, M.~G., \& Steger, P. 2019, Monthly Notices of the Royal
  Astronomical Society, 484, 1401, publisher: Oxford Academic

\bibitem[{Riess {et~al.}(2016)Riess, Macri, Hoffmann, Scolnic, Casertano,
  Filippenko, Tucker, Reid, Jones, Silverman, Chornock, Challis, Yuan, Brown,
  \& Foley}]{Riess2016}
Riess, A.~G., Macri, L.~M., Hoffmann, S.~L., {et~al.} 2016, The Astrophysical
  Journal, 826, 56, publisher: American Astronomical Society

\bibitem[{Sales {et~al.}(2016)Sales, Navarro, Oman, Fattahi, Ferrero, Abadi,
  Bower, Crain, Frenk, Sawala, Schaller, Schaye, Theuns, \& White}]{Sales2016}
Sales, L.~V., Navarro, J.~F., Oman, K., {et~al.} 2016, Monthly Notices of the
  Royal Astronomical Society, 464, 2419

\bibitem[{{S{\'a}nchez-Salcedo} \& {Brandenburg}(1999)}]{Salcedo1999}
{S{\'a}nchez-Salcedo}, F.~J. \& {Brandenburg}, A. 1999, \apjl, 522, L35

\bibitem[{Santos-Santos {et~al.}(2015)Santos-Santos, Brook, Stinson, Di~Cintio,
  Wadsley, Domínguez-Tenreiro, Gottlöber, \& Yepes}]{Santos-Santos2015}
Santos-Santos, I.~M., Brook, C.~B., Stinson, G., {et~al.} 2015, Monthly Notices
  of the Royal Astronomical Society, 455, 476

\bibitem[{Sawala {et~al.}(2016)Sawala, Frenk, Fattahi, Navarro, Bower, Crain,
  Vecchia, Furlong, Helly, Jenkins, Oman, Schaller, Schaye, Theuns, Trayford,
  \& White}]{Sawala2016}
Sawala, T., Frenk, C.~S., Fattahi, A., {et~al.} 2016, Monthly Notices of the
  Royal Astronomical Society, 457, 1931

\bibitem[{Schive {et~al.}(2014)Schive, Chiueh, \& Broadhurst}]{Schive2014}
Schive, H.-Y., Chiueh, T., \& Broadhurst, T. 2014, Nature Physics, 10, 496

\bibitem[{Schwabe {et~al.}(2016)Schwabe, Niemeyer, \& Engels}]{Schwabe2016}
Schwabe, B., Niemeyer, J.~C., \& Engels, J.~F. 2016, Phys. Rev. D, 94, 043513

\bibitem[{Sellwood(2014)}]{Sellwood2014}
Sellwood, J.~A. 2014, Reviews of Modern Physics, 86, 1, publisher: American
  Physical Society

\bibitem[{Shadmehri \& Khajenabi(2012)}]{Shadmehri2012}
Shadmehri, M. \& Khajenabi, F. 2012, Monthly Notices of the Royal Astronomical
  Society, 424, 919, publisher: Oxford Academic

\bibitem[{Shao {et~al.}(2013)Shao, Gao, Theuns, \& Frenk}]{Shao2013}
Shao, S., Gao, L., Theuns, T., \& Frenk, C.~S. 2013, Monthly Notices of the
  Royal Astronomical Society, 430, 2346, publisher: Oxford Academic

\bibitem[{Sharma {et~al.}(2019)Sharma, Khoury, \& Lubensky}]{Sharma2019}
Sharma, A., Khoury, J., \& Lubensky, T. 2019, Journal of Cosmology and
  Astroparticle Physics, 2019, 054

\bibitem[{Sikivie \& Yang(2009)}]{Sikivie2009}
Sikivie, P. \& Yang, Q. 2009, Physical Review Letters, 103, 111301

\bibitem[{Skrbek(2011)}]{Skrbek2011}
Skrbek, L. 2011, Journal of Physics: Conference Series, 318, 012004

\bibitem[{Skrbek \& Sreenivasan(2012)}]{Skrbek2012}
Skrbek, L. \& Sreenivasan, K.~R. 2012, Physics of Fluids, 24, 011301

\bibitem[{Slepian \& Goodman(2012)}]{Slepian2012}
Slepian, Z. \& Goodman, J. 2012, Monthly Notices of the Royal Astronomical
  Society, 427, 839, publisher: Oxford Academic

\bibitem[{Soulaine {et~al.}(2017)Soulaine, Quintard, Baudouy, \&
  Van~Weelderen}]{Soulaine2017}
Soulaine, C., Quintard, M., Baudouy, B., \& Van~Weelderen, R. 2017, Physical
  Review Letters, 118, 074506

\bibitem[{Spergel \& Steinhardt(2000)}]{Spergel2000}
Spergel, D.~N. \& Steinhardt, P.~J. 2000, Phys. Rev. Lett., 84, 3760

\bibitem[{Strigari(2018)}]{Strigari2018}
Strigari, L.~E. 2018, Reports on Progress in Physics, 81, 056901, publisher:
  IOP Publishing

\bibitem[{Sánchez-Salcedo(2012)}]{Salcedo2012}
Sánchez-Salcedo, F.~J. 2012, The Astrophysical Journal, 745, 135, publisher:
  IOP Publishing

\bibitem[{Tamfal {et~al.}(2020)Tamfal, Mayer, Quinn, Capelo, Kazantzidis,
  Babul, \& Potter}]{Tamfal2020}
Tamfal, T., Mayer, L., Quinn, T.~R., {et~al.} 2020, Revisiting dynamical
  friction: the role of global modes and local wakes

\bibitem[{Taylor \& Griffin(2005)}]{Taylor2005}
Taylor, E. \& Griffin, A. 2005, Phys. Rev. A, 72, 8739

\bibitem[{Tegmark {et~al.}(2004)Tegmark, Blanton, Strauss, Hoyle, Schlegel,
  Scoccimarro, Vogeley, Weinberg, Zehavi, Berlind, Budavari, Connolly,
  Eisenstein, Finkbeiner, Frieman, Gunn, Hamilton, Hui, Jain, Johnston, Kent,
  Lin, Nakajima, Nichol, Ostriker, Pope, Scranton, Seljak, Sheth, Stebbins,
  Szalay, Szapudi, Verde, Xu, Annis, Bahcall, Brinkmann, Burles, Castander,
  Csabai, Loveday, Doi, Fukugita, III, Hennessy, Hogg, Ivezi{\'{c}}, Knapp,
  Lamb, Lee, Lupton, McKay, Kunszt, Munn, O'Connell, Peoples, Pier, Richmond,
  Rockosi, Schneider, Stoughton, Tucker, Berk, Yanny, \& and}]{Tegmark2004}
Tegmark, M., Blanton, M.~R., Strauss, M.~A., {et~al.} 2004, The Astrophysical
  Journal, 606, 702

\bibitem[{Thun {et~al.}(2016)Thun, Kuiper, Schmidt, \& Kley}]{Thun2016}
Thun, D., Kuiper, R., Schmidt, F., \& Kley, W. 2016, Astronomy \& Astrophysics,
  589, A10, publisher: EDP Sciences

\bibitem[{Toro(2006)}]{Toro2006}
Toro, E. 2006, Applied Numerical Mathematics, 56, 1464

\bibitem[{Trujillo-Gomez {et~al.}(2011)Trujillo-Gomez, Klypin, Primack, \&
  Romanowsky}]{Trujillo-Gomez2011}
Trujillo-Gomez, S., Klypin, A., Primack, J., \& Romanowsky, A.~J. 2011, The
  Astrophysical Journal, 742, 16, publisher: American Astronomical Society

\bibitem[{Tulin \& Yu(2018)}]{Tulin2018}
Tulin, S. \& Yu, H.-B. 2018, Physics Reports, 730, 1 , dark matter
  self-interactions and small scale structure

\bibitem[{Vogelsberger {et~al.}(2014)Vogelsberger, Genel, Springel, Torrey,
  Sijacki, Xu, Snyder, Bird, Nelson, \& Hernquist}]{Vogelsberger2014}
Vogelsberger, M., Genel, S., Springel, V., {et~al.} 2014, Nature, 509, 177,
  number: 7499 Publisher: Nature Publishing Group

\bibitem[{Walker(2013)}]{Walker2013}
Walker, M. 2013, in Planets, {Stars} and {Stellar} {Systems}: {Volume} 5:
  {Galactic} {Structure} and {Stellar} {Populations}, ed. T.~D. Oswalt \&
  G.~Gilmore (Dordrecht: Springer Netherlands), 1039--1089

\bibitem[{Walker {et~al.}(2009)Walker, Mateo, Olszewski, Peñarrubia, Evans, \&
  Gilmore}]{Walker2009}
Walker, M.~G., Mateo, M., Olszewski, E.~W., {et~al.} 2009, The Astrophysical
  Journal, 704, 1274, publisher: IOP Publishing

\bibitem[{Wang {et~al.}(2019{\natexlab{a}})Wang, Boer, Pieres, Li,
  Drlica-Wagner, Koposov, Vivas, Pace, Santiago, Walker, Tucker, Strigari,
  Marshall, Yanny, DePoy, Bechtol, Roodman, Abbott, Abdalla, Allam, Annis,
  Avila, Bertin, Brooks, Burke, Rosell, Kind, Cunha, D'Andrea, Costa, Vicente,
  Desai, Eifler, Estrada, Flaugher, Frieman, García-Bellido, Gerdes, Gruen,
  Gruendl, Gutierrez, Hollowood, Honscheid, James, Kuehn, Kuropatkin, Lahav,
  Maia, Miquel, Sanchez, Scarpine, Sevilla-Noarbe, Smith, Smith, Sobreira,
  Suchyta, Swanson, \& and}]{Wang2019b}
Wang, M.~Y., Boer, T.~d., Pieres, A., {et~al.} 2019{\natexlab{a}}, The
  Astrophysical Journal, 881, 118, publisher: American Astronomical Society

\bibitem[{Wang {et~al.}(2019{\natexlab{b}})Wang, Koposov, Drlica-Wagner,
  Pieres, Li, Boer, Bechtol, Belokurov, Pace, Bacon, Abbott, Annis, Bertin,
  Brooks, Buckley-Geer, Burke, Rosell, Kind, Carretero, Costa, Vicente, Desai,
  Diehl, Doel, Estrada, Flaugher, Fosalba, Frieman, García-Bellido, Gerdes,
  Gruen, Gruendl, Gschwend, Gutierrez, Hollowood, Honscheid, Hoyle, James,
  Kent, Kuehn, Kuropatkin, Maia, Marshall, Menanteau, Miquel, Plazas, Sanchez,
  Santiago, Scarpine, Schindler, Schubnell, Serrano, Sevilla-Noarbe, Smith,
  Smith, Sobreira, Suchyta, Swanson, Tarle, Thomas, Tucker, \& and}]{Wang2019a}
Wang, M.~Y., Koposov, S., Drlica-Wagner, A., {et~al.} 2019{\natexlab{b}}, The
  Astrophysical Journal, 875, L13, publisher: American Astronomical Society

\bibitem[{Weinberg {et~al.}(2015)Weinberg, Bullock, Governato, Naray, \&
  Peter}]{Weinberg2015}
Weinberg, D.~H., Bullock, J.~S., Governato, F., Naray, R. K.~d., \& Peter, A.
  H.~G. 2015, Proceedings of the National Academy of Sciences, 112, 12249,
  publisher: National Academy of Sciences Section: Colloquium Paper

\bibitem[{Weinberg(1985)}]{Weinberg1985}
Weinberg, M.~D. 1985, Monthly Notices of the Royal Astronomical Society, 213,
  451, publisher: Oxford Academic

\bibitem[{Zhao(1996)}]{Zhao1996}
Zhao, H. 1996, Monthly Notices of the Royal Astronomical Society, 278, 488,
  publisher: Oxford Academic

\bibitem[{Zhu {et~al.}(2016)Zhu, Marinacci, Maji, Li, Springel, \&
  Hernquist}]{Zhu2016}
Zhu, Q., Marinacci, F., Maji, M., {et~al.} 2016, Monthly Notices of the Royal
  Astronomical Society, 458, 1559

\end{thebibliography}



\end{document}